\documentclass[final,12pt,authoryear]{elsarticle}

\usepackage{amssymb}
\usepackage{amsmath}
\usepackage{lipsum}
\usepackage{rotating}
\usepackage{graphicx}
\usepackage{natbib}
\usepackage{amsthm}
\usepackage{lscape}
\usepackage{url}
\usepackage{float}
\usepackage{hyperref}
\usepackage{booktabs}
\usepackage{caption}
\usepackage{appendix}
\usepackage[utf8]{inputenc}

\journal{Astronomy and Computing}

\begin{document}

\begin{frontmatter}



\title{Unveiling the Power of Uncertainty: A Journey into Bayesian Neural Networks for Stellar dating}


\author[first]{Víctor Tamames Rodero}
\author[first]{Andrés Moya Bedón}
\affiliation[first]{organization={Departament d’Astronomia i Astrofísica, Universitat de València},
            addressline={C/ Dr. Moliner, 50}, 
            city={Burjassot},
            postcode={46100}, 
            state={Valencia},
            country={Spain}}

\author[third]{Luis Manuel Sarro Baro}
\affiliation[third]{organization={Departamento de Inteligencia Artificial, ETSI Informática, UNED},
            addressline={Juan del Rosal}, 
            city={E-16},
            postcode={28040}, 
            state={Madrid},
            country={Spain}}

\author[fourth]{Roberto Javier López Sastre}
\affiliation[fourth]{organization={GRAM Research Group, Department of Signal Theory and Communications, University of Alcalá},
            city={Alcalá de Henares},
            postcode={28805}, 
            state={Madrid},
            country={Spain}}

\begin{abstract}

\noindent\textbf{Context}: Astronomy and astrophysics demand rigorous handling of uncertainties to ensure the credibility of outcomes. The growing integration of artificial intelligence offers a novel avenue to address this necessity. This convergence presents an opportunity to create advanced models capable of quantifying diverse sources of uncertainty and automating complex data relationship exploration.

\noindent\textbf{What:} We introduce a hierarchical Bayesian architecture whose probabilistic relationships are modeled by neural networks, designed to forecast stellar attributes such as mass, radius, and age (our main target). This architecture handles both observational uncertainties stemming from measurements and epistemic uncertainties inherent in the predictive model itself. As a result, our system generates distributions that encapsulate the potential range of values for our predictions, providing a comprehensive understanding of their variability and robustness.

\noindent\textbf{Methods:}  Our focus is on dating main sequence stars using a technique known as Chemical Clocks, which serves as both our primary astronomical challenge and a model prototype. In this work, we use hierarchical architectures to account for correlations between stellar parameters and optimize information extraction from our dataset. We also employ Bayesian neural networks for their versatility and flexibility in capturing complex data relationships.

\noindent\textbf{Results:} By integrating our machine learning algorithm into a Bayesian framework, we have successfully propagated errors consistently and managed uncertainty treatment effectively, resulting in predictions characterized by broader uncertainty margins. This approach facilitates more conservative estimates in stellar dating. Our architecture achieves age predictions with a mean absolute error of less than 1 Ga for the stars in the test dataset.

\end{abstract}



\end{frontmatter}




\section{Introduction}
\label{introduction}

Nowadays, in many different fields, we see how Artificial Intelligence (AI) approaches emerge; new machine learning (ML), deep learning (DL) algorithms, and other statistical models come to the stage to help us automatize and extract as much information from our datasets as possible. These new tools are changing our workflow and the way we treat our data, promising to be a potential solution for many tasks in the near future.

In the dynamic field of AI, a major challenge has been the ability to clearly express the uncertainties associated with their predictions. Notwithstanding these limitations, significant efforts have been made to enhance and elucidate the functioning of these algorithms, with the objective of developing architectures that can address uncertainties.

Our objective is to apply this capability to stellar dating, with the aim of improving the accuracy and confidence of stellar age estimates. Employing varied methods, from traditional isochrone fitting \citep[e.g.][]{2014Valls, 2017Rodrigues} to more modern techniques like gyrochronology \citep[e.g.][]{2003Barnes, 2015Angus, 2023Mathur}, asteroseismology \citep[e.g.][]{2010Moya,2015Silva, 2015Silva2,2023Mathur2} and chemical abundance analysis \citep[e.g.][]{2012Silva, 2015Nissen, 2016Nissen,2020Nissen, Moya2022}, accurate stellar dating in astrophysics demands rigorous uncertainty treatment for reliable results. When estimating stellar ages one is confronted with multifaceted challenges in accurately estimating the parameter. Inherent uncertainties within theoretical models describing stellar evolution pose a significant hurdle, as these models rely on assumptions about intricate physical processes, introducing potential inaccuracies. Limitations in observational data quality, characterized by incomplete or sparse data, particularly for distant or faint stars, hinder precise age determinations. Moreover, degeneracies in stellar properties used for obtaining ages, along with calibration intricacies, can introduce biases, complicating age estimates. Furthermore, age dispersions observed within stellar populations challenge the uniformity of stellar ages within clusters, impeding a comprehensive understanding of stellar formation and evolution. Addressing these challenges necessitates refined models, improved observational capabilities, meticulous calibration techniques, and innovative, interdisciplinary approaches to refine our grasp of stellar ages and cosmic chronologies. 

Modern advancements in computational models and observational techniques have substantially enhanced contemporary stellar dating methodologies. Notably, the advent of missions such as Corot (France, ESA), Kepler, and TESS (NASA) has revolutionized age estimations through the application of asteroseismology \citep[e.g.][]{2010Moya,2013Mathur, 2015Silva, 2015Silva2, 2015Soderblom2, 2021Morel, 2023Mathur2, 2024Goupil}. These missions have played a transformative role in refining our understanding of stellar ages. Moreover, the anticipated contributions of missions like Plato 2.0 (ESA) hold substantial promise for delving deeper into stellar age determinations, offering the potential for significant advancements in our comprehension of cosmic chronologies. 

Overall, in the field of stellar dating techniques, it's essential to acknowledge that no single method universally applies to all types of stars, except for isochrones model fitting. Depending on observations, this approach holds the potential to yield age predictions with very tight uncertainties. However, this could represent a limitation in broader applicability, as it might be deemed unrealistic or overly constrained in certain contexts. Work has been done in this direction, using Bayesian statistical approaches to manage and better propagate uncertainties for more realistic age estimates \citep[e.g.][]{2012Guede, 2013Serenelli, 2015Brandt, 2015Angus,2016Hippel,2016Jeffery, 2018Almeida, 2018Lin, 2020Kiman, 2021Kiman, 2022Kiman, Moya2022, 2023Mathur, 2023Mathur2, 2024Lu}, some of them use hierarchical architectures \citep[e.g.][]{2015Angus, 2017olivaresromero, Moya2022}, and other ones using ML techniques \citep[e.g.][]{2020basu, 2020bu, 2021Moya, 2023Lane}.

Currently, using clusters for stellar dating \citep[e.g.][]{2022Viscasillas} and asteroseismology emerge as promising alternatives, each with its distinct strengths and constraints. Calculating the age of stars through cluster characterization, while effective in certain contexts, often yields quantized databases, constraining their utility in broader astronomical applications. On the other hand, asteroseismology, renowned for its precision and accuracy, has predominantly been applied to giants and a select few hundred main sequence (MS) stars of specific spectral types. Despite their advancements, these methodologies encounter challenges in achieving comprehensive coverage across stellar populations.

In this paper, we present a novel approach designed for dating main sequence stars by leveraging machine learning and statistical methods. Our model aims to provide a more accurate and conservative approach to stellar dating by quantifying uncertainties in the predictions. By utilizing machine learning, we can automatically capture intricate relationships within the data, thus potentially enhancing the quality of our estimates. Our contributions are rooted in embedding this approach within a Bayesian framework, known for its robust and well-established methodologies. By employing a hierarchical architecture, our approach not only quantifies uncertainties but also maximizes the extraction of information from available datasets, leading to more reliable and comprehensive stellar age determinations.

\subsection{Scope \& case}
In this study, we adopt a hierarchical Bayesian framework integrated with machine learning algorithms, specifically multilayer perceptrons, to model the relationships embedded within our dataset and be able to derive robust distributions of potential stellar ages. In this work we decided to use the chemical clocks (CCs) technique for calculating stellar ages, which exploits the unaltered surface chemical compositions of stars as proxies for their formation epochs.  
MS stars are particularly well-suited for this approach because their photospheric abundances remain largely unchanged throughout their lifetimes (e.g. \citealt{2019Delgado}, from now on DM19, \citealt{2022Viscasillas}; \citealt{Moya2022}). Once established at formation, the chemical composition of an MS star effectively serves as a relic of the interstellar medium (ISM) or formation cloud at that time. Aside from minor effects such as atomic diffusion, the surface layers of these stars do not experience significant alterations, preserving the original chemical signature. This characteristic stability is central to the chemical clocks method, as it enables the inference of stellar ages directly from the observed abundance patterns.
The foundation of the CCs approach lies in the framework of Galactic chemical evolution. The ISM is progressively enriched over time by a variety of nucleosynthetic sources that operate on different timescales. By empirically calibrating the abundance ratios of elements produced on these disparate timescales (e.g., [Y/Mg] or [Ba/Al]), chemical clocks effectively timestamp the epoch at which a star's birth environment was chemically characterized. Thus, a star’s immutable chemical signature encodes vital information about the temporal evolution of the formation cloud.
It is important to note that while the majority of heavy elements in MS stars remain stable, certain species do undergo evolutionary changes. However, the reliability of the chemical clock method hinges on the systematic variation of abundance ratios, primarily governed by the nucleosynthetic history of the ISM rather than by intrinsic stellar evolution. This distinction is critical, as it ensures that the observed abundance ratios serve as faithful indicators of stellar age.
The applicability of the chemical clock is one of its main advantages since it does not depend on the evolution and structure of the stars, but the method varies across different stellar types.  The technique is most effective for low- and intermediate-mass stars, which possess convective envelopes that do not facilitate extensive mixing during the MS phase. In contrast, high-mass stars, characterized by radiative envelopes and rapid evolution may experience convective dredge-up events that transport nuclear-processed material to the surface, thereby altering their chemical abundances and compromising the chemical clock signal. In these cases, either alternative chemical tracers or additional corrective calibrations are necessary to obtain reliable age estimates.
CCs offer a valuable complement to other established age-dating methodologies, including gyrochronology, isochrone fitting, and asteroseismology. Gyrochronology, which correlates stellar rotation periods with age, is most effective for cool dwarfs but relies on precise rotational data. Isochrone fitting, based on luminosity and effective temperature, can suffer from degeneracies in certain regions of the Hertzsprung–Russell diagram (e.g. regions where isochrones are tightly packed). Although asteroseismology provides highly precise age determinations, it demands observational resources that may not be available for large samples. In contrast, chemical clocks can be readily applied to extensive spectroscopic surveys, thereby facilitating robust statistical analyses of stellar populations and extending age estimates to stars that are challenging to date using other techniques.

\section{Tools \& data sample}\label{datasams}

\subsection{Tools}\label{toools}
Inside the vast field of AI, neural networks (NNs) stand out from the rest of the algorithms thanks to their potential versatility to adapt to data, tasks, or domains. However, the limitations of such networks became evident over time and it wasn't until the last decades when we witnessed a remarkable transformation in NNs, driven by advancements in Gradient-based learning \citep{Lecun98}, the availability of powerful hardware such as graphics processing units (GPUs), and large datasets. This has led to the development of deep neural networks, including other architectures like deep convolutional or recurrent neural networks \citep[e.g.][]{Krizhevsky12, Silver17}. 

These models have improved a lot in the last few years, but they often require large amounts of labeled, high-quality data for training, which is not always available. In addition, NNs lack interpretability, making it challenging to understand the reasoning behind their predictions, especially when they make errors or biased decisions \citep[e.g.][]{2017Dong, 2017Ancona,2022Rauker}. Furthermore, traditional deep learning models were not initially designed for quantifying uncertainty and generally focus on providing point estimates as predictions. Although, there are algorithms like Bayesian Neural Networks (BNNs) \citep{Tishby1989},  Monte Carlo Dropout  \citep{2015Gal}, Deep Ensembles \citep{Krizhevsky12} or Masksembles \citep{2020Durasov} that can produce approximations of uncertainties.  In this context, probability theory has become a cornerstone for modeling uncertainty in machine learning, seamlessly integrating probabilistic modeling with the predictive capabilities of ML algorithms. In our work, we leverage BNNs, a powerful framework for effectively incorporating and managing uncertainty in AI. This approach represents a significant step toward building reliable and interpretable systems.

BNNs offer a powerful approach for incorporating uncertainty into predictions by modeling their weights and biases as probability distributions rather than fixed values. Unlike traditional neural networks that yield point estimates, BNNs quantify uncertainty in both model parameters and outputs, which is essential for applications demanding reliable decision-making and interoperability. In particular, by modeling uncertainty, BNNs mitigate overfitting: they capture epistemic uncertainty and avoid overconfident predictions through implicit regularization. This capability is especially important in our application, where understanding the confidence of predictions is critical.

Bayesian statistics underpins this methodology by providing a formal framework to represent uncertainty in model parameters and observations through probability distributions. This framework not only enables the calculation of credible intervals and distributions for parameter estimates, contrasting with the frequentist approach, which assigns probabilities to data rather than hypotheses, but also allows the integration of prior knowledge. Priors in the Bayesian framework discourage extreme weight values, while the posterior distribution naturally favors simpler models when data is scarce. Moreover, by marginalizing over multiple weight configurations instead of relying on a single optimized set, BNNs reduce sensitivity to noise in the training data. As new data is gathered, Bayesian models update the probability of hypotheses, a feature that is fundamental for our stellar characterization model, where handling uncertainties at every step is critical.

In this context, hierarchical models emerge as a particularly valuable tool. These architectures have the ability to capture and account for variations at multiple levels, encompassing both within-group and between-group variations simultaneously. This capability is especially beneficial when dealing with correlated data or where different levels may exhibit distinct sources of variability, providing a more nuanced and accurate analysis. Hierarchical models are often applied within a Bayesian framework because they facilitate the integration of prior knowledge and enable the quantification of uncertainty in parameter estimates. By including prior knowledge and data likelihood, the posterior distribution offers a comprehensive depiction of parameter information, providing a holistic understanding of the model's nuances post-data observation \citep[e.g.][]{Gelman06, Gelman17}.  They excel in handling situations with limited data, providing stability and informativeness in parameter estimates by borrowing information from other levels of the hierarchy \citep[e.g.][]{2002Stephen,gelmanbda04, Gelman06}. 
Considering these advantages and inspired by the promising results presented in \cite{Moya2022}, we have adapted and extended its Hierarchical Bayesian Model (HBM) to suit the specific requirements of our current experimentation. Technically, in this work, we built upon the HBM architecture proposed in \cite{Moya2022}, which models probabilistic relationships using multi-linear regressions. However, we enhanced the model by replacing the multi-linear components with a machine learning module, specifically neural networks (NNs).

In our hierarchical model, we use Markov Chain Monte Carlo (MCMC) methods for parameter inference, leveraging their ability to estimate posterior distributions in Bayesian frameworks. MCMC methods are a class of computational algorithms that rely on repeated random sampling to approximate these distributions, making them particularly suited to complex, high-dimensional models where traditional methods fall short. The need for such techniques has driven advances in probabilistic programming, with languages like Stan, PyMC, and Pyro making MCMC more accessible and applicable to sophisticated models.

However, MCMC methods are also computationally demanding, which can limit their scalability, especially in large models or time-sensitive applications. Addressing these limitations is an active area of research, with current work in BNNs and deep learning–based approaches focused on making these methods more efficient and scalable. These innovations are helping to broaden the practical use of Bayesian inference for uncertainty quantification in machine learning.

\subsection{Training sample} \label{training}

The training dataset utilized in our study is thoroughly documented in \citep{2019Delgado2}. It comprises observations of 1059 stars from the HARPS-GTO planetary search program. These stars were selected from a volume-limited area within roughly 70 parsecs of the Sun, with few stars located at greater distances.

Details about derived effective temperature ($T_\mathrm{eff}$), metallicity ([Fe/H]), surface gravity ($\log$ g), and chemical abundances ([X/Fe], where X represents the different chemical elements) are presented in DM19. Stellar ages (from now on \textit{t}) were computed utilizing the PARAM v1.3 tool, employing PARSEC isochrones \citep{2012Bressan} in conjunction with a Bayesian estimation method \citep{2006Silva}. However, it is pertinent to highlight that not all age determinations among these 1059 stars are deemed reliable. DM19 established reliable age estimates as those presenting an age uncertainty smaller than 1.5 billion years (Ga). Stars with age uncertainties below 0.2 Ga were omitted because we regard them as unrealistic for a standard isochrone fitting process, having the potential to introduce bias into the final model \citep{Moya2022}. Adopting a uniform threshold was a practical decision to avoid introducing biases that might arise from a non-uniform error selection. Consequently, this resulted in a refined dataset comprising 328 stars extracted from the original set of 1059 stars for our research. Further elucidation regarding the primary characteristics of this subset can be found in DM19.

In summary, our investigation was centered on categorizing 244 thin-disk stars, 14 high-$\alpha$ metal-rich stars, 68 thick-disk stars, and 2 halo stars, following the classification methodology delineated by \cite{2011Adibekyan,2012Adibekyan}. The effective temperature ranged from 5010 to 6788 K, with 95\% of the values falling between 5271 and 6416 K. The surface gravity spanned from 3.73 to 4.71 dex, with 95\% of the measurements ranging between 3.93 and 4.58 dex. Additionally, the metallicity exhibited variation from -1.15 to 0.55 dex, with 95\% of values within the range of -0.81 to 0.33 dex. Each parameter measurement was accompanied by its corresponding observational error. We chose to use the chemical abundances of Mg, Al, Si, Zn, Ti, Sr, and Y for these stars.

\subsection{Testing sample}

A distinct testing set comprising stars not included in the training was employed to assess the accuracy of the HBM predictions, produced using the training dataset detailed in Section \ref{training}. The specific composition of this dataset, as referenced in \cite{Moya2022}, is twenty-three stars designated with 'reliable' age estimates. They are delineated below:

\begin{itemize}
    \item Twenty stars whose age determination was derived through asteroseismology. Specific characteristics of eighteen Kepler Input Catalog (KIC) stars, as well as 16 Cygni A and 16 Cygni B, were acquired from \cite{2017Nissen} and \cite{2021Morel}. Their chemical abundances were acquired via specialized observations. For data sourced from \cite{2017Nissen}, $log(g)$ uncertainties were not explicitly provided; thus, a standard and conservative uncertainty of 0.05 dex was imposed. The ages utilized in this context were primarily sourced from \cite{Aguirre_2017}.
    
    \item Two stars belong to the M67 open cluster (this is one of the best-characterized clusters). Details about these stars were sourced from \cite{2016Liu}, which are identified as solar twins. The age of the cluster was obtained from \cite{2008Yadav}.
    
    \item The Sun, whose general characteristics were taken from \cite{2016Prvsa}. To minimize biases and facilitate direct comparisons with previous works, we adopt the same abundance reference as in prior studies \cite{2007Sousa,2008Sousa}.
    
\end{itemize}

We consider them more reliable than other selections because they were obtained through asteroseismology and cluster characterization, which nowadays represent two of the most accurate methods for dating MS stars. These stars exhibit well-determined effective temperature, surface gravity, metallicity, age, and chemical abundances of Mg, Al, Si, Zn, Ti, Sr, and Y relative to iron, along with their associated uncertainties. It is noteworthy that not all stars have available data for all chemical abundances, contributing to the realistic nature of our testing dataset. In the case of the Sun, it was treated as a standard field star within the testing dataset. For doing so we intentionally adjusted its uncertainties to align with those of the remaining stars.

\section{Inference technique: HBM}

\begin{figure}
    \centering
   \rotatebox{90}{\includegraphics[width=1.18\linewidth]{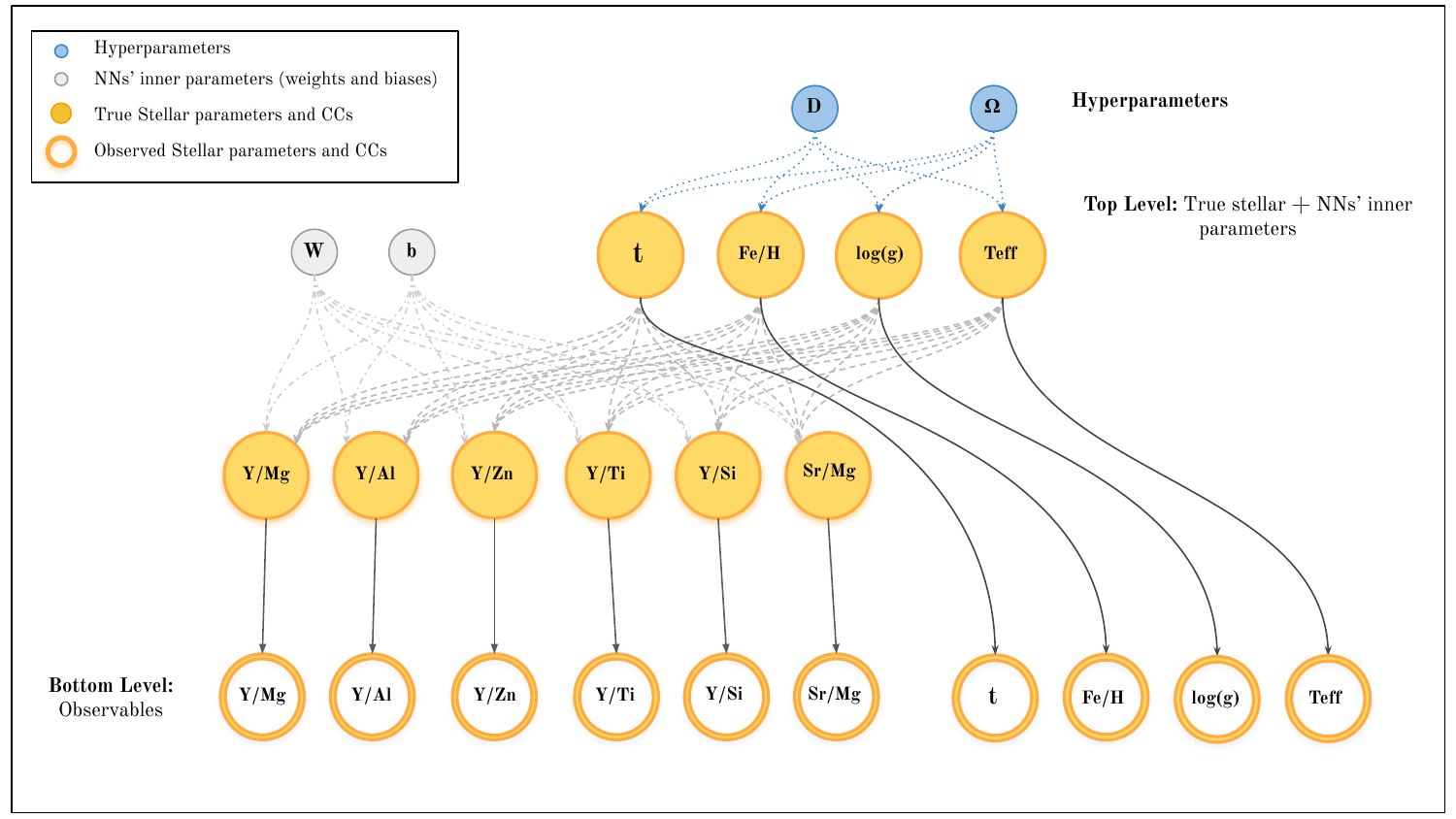}}
    \caption{Structure of the hierarchical Bayesian model used in this work, comprising two levels: the \textit{Top} and \textit{Bottom}. Hyperparameters \textit{$\Omega$} (correlation matrix) and \textit{D} (scaling matrix), which are used to define the covariance matrices for the stellar parameters represent the model's hyperpriors. Additionally, \(\textit{\textbf{W}}\) and \(\textit{\textbf{b}}\) are the weights matrices and biases vectors, respectively, of the neural networks. \textit{Filled circles} are the true stellar parameters and CCs, and \textit{empty circles} represent the observed ones.}
    \label{fig:hierarchical}
\end{figure}

As it was previously introduced in Section \ref{toools}, in this work we decided to inherit the HBM architecture used in \cite{Moya2022}, whose probabilistic relationships are modeled by multi-linear regressions (from now on HBM-MLR). Thanks to this decision we are able to stratify information for hyperprior specification, account for observational uncertainties, and propagate the errors through the model. In this work, we replaced the multi-dimensional linear models, with a ML module, specifically, NNs (from now on HBM-NNs). This new approach eliminates the need for manual exploration of internal data relationships, as previously suggested in \cite{2019Delgado}. The schematic of the architecture is depicted in Figure \ref{fig:hierarchical}.

\subsection{Bayesian Neural Networks}

Drawing upon insights from \cite{Dotter_2017} and \cite{2021Gavel}, we acknowledge the proposition that all potential predictor variables may carry pertinent physical information. In pursuit of optimal data treatment, several strategic decisions were undertaken. Firstly, our methodology aligns with the principles advocated by \cite{gelmanbda04}, wherein all available predictors are incorporated into our BNN model, with priors for the neural network weights and biases centered at zero. This decision serves as a regularization mechanism, ensuring that the posterior distributions of neural network coefficients diverge significantly from zero solely when supported by empirical evidence \citep{gelmanbda04}. Additionally, an arrangement was imposed on the bias vectors of hidden layers (HLs) in the BNNs. The reordering of a bias vector $\textit{b}$ with components $b_k$ is done in ascending order (\(b_k < b_{k+1}\) for \(k \in \{1, \ldots, k-1\}\)). This deliberate ordering aims to break weight-space symmetries, thereby enhancing general identifiability of BNNs. The organized bias vectors contribute to the network's discernment of relevant patterns and dependencies, reducing the risk of underfitting or oversimplification. This, in turn, augments the network's learning capacity, promoting accurate predictions and improving generalization \citep{Pourzanjani2017ImprovingTI}. Furthermore, the chosen activation function for the BNNs' HLs, namely the classical sigmoid, is applied within nodes after the linear combination of the inputs with the weights and biases, effectively constraining the domain of internal neural network parameters. The output of the hidden units is defined as

\begin{figure}
	\centering 
	\includegraphics[width=\textwidth]{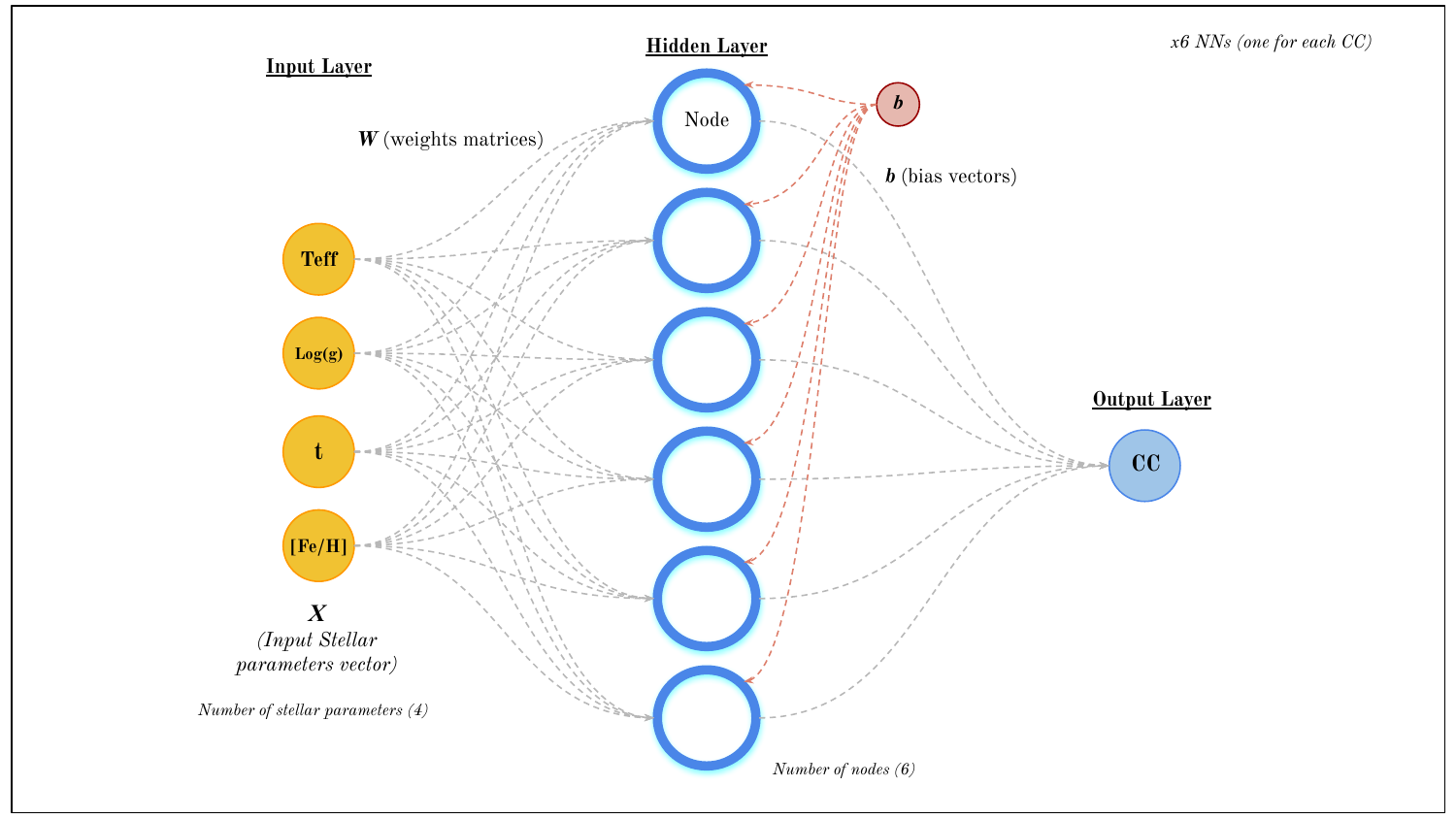}	
	\caption{An example of a single-output NN architecture with one hidden layer, matching the best BNN schematics used for single-output configurations during experimentation. \(\textit{\textbf{W}}\) and \(\textit{\textbf{b}}\) represent the weights matrices and biases vectors, illustrated through the dashed gray and red lines, respectively, depicting the information flow within the NN. Additionally, \textit{Orange filled circles} are the stellar parameters ($X$) as input, and \textit{Blue filled circle} is the single CC output.} 
	\label{fig_mom0}%
\end{figure}

\begin{equation}
    h_i = f \left( \sum_{j=1}^{n} w_{ji} x_j + b_i \right),
    \label{eq-NN}
\end{equation}

\noindent where $i$ indexes neurons in the hidden layer from 1 to $m$ (the number of neurons in the hidden layer). Index \textit{j} indexes the input features. It ranges from 1 to \textit{n}, where \textit{n} is the total number of input features. $x_j$ denotes the \textit{j}-th input feature and $w_{ji}$ stands for the weight connecting the \textit{j}-th input feature to the \textit{i}-th neuron in the hidden layer. Each input feature $x_j$ is weighted by $w_{ji}$ before being summed up and passed through the activation function \textit{f} (sigmoid). $b_i$ represents the bias term for the \textit{i}-th neuron in the HL.  The output value of the neural network, $y$, is calculated as 

\begin{equation}
    y = \sum_{i=1}^{m} w_i h_i + b .
    \label{eq-NN2}
\end{equation}

In the primary case of a single CC per BNN (Figure \ref{fig_mom0}), in the output layer the weight $w_{ji}$ is represented as a vector, while the bias $b_i$ is represented as a scalar. Our model would then be defined by the inner operations of the neural networks, in this case, the feed-forward multilayer perceptron.

The dimensions of the weights and biases in NN architectures depend on the number of inputs, nodes, and HLs. To maintain clarity and simplify nomenclature, we define \(\textit{\textbf{W}}\) and \(\textit{\textbf{b}}\) as the sets encompassing all weight matrices and bias vectors, respectively, used in any given experimental configuration, including all 6 NNs (one for each CC). Specific details will be provided as necessary for single NN with multi-output experiments.

\subsubsection{Posterior sampling techniques} \label{ref: sms}

Bayesian inference updates priors using observed data to compute the posterior distribution, encapsulating the model's uncertainty (epistemic). Specifically, within BNNs, approximation techniques such as MCMC or Variational Inference (VI) \citep[e.g.][]{gelmanbda04, Gelman06, 2016Liu2, 2015Gal2} are employed due to the intractable nature of the posterior distribution over neural network parameters \citep[e.g.][]{2020Wilson, arbel2023primer, wiese2023efficient, papamarkou2021challenges}. In our case, we used PyStan \citep{Gelman17}, which is the Python version of the probabilistic work-frame Stan \citep{Carpenter2017}.  Among the different sampling methods offered by this language, we used the NUTS version of the Hamiltonian Monte Carlo (HMC) \citep[e.g.][]{gelmanbda04, Gelman06, 2011Hoffman} sampler to obtain the desired samples from the posterior distribution. This approach better fits our needs in the sense that even if the computing time is generally longer than typical VI ones, we sample directly from the posterior distribution and not from an approximation of it.

Our Bayesian model was build using the 2.19.1.1 PyStan version, which just support CPU. Experiments were launched on Intel®  i9-13905H and Intel® Xeon® E5-2695 v4 processors, both ranging a clock speed between 2.6-3.3 GHz.

\subsection{Hierarchical Architecture} \label{hierararq}

As depicted in Figure \ref{fig:hierarchical}, our model architecture is defined by two layers or groups. The first layer in which we can find the true stellar parameters $\Theta$ ($\log$ g, $T_\mathrm{eff}$, $t$, and [Fe/H]) and the inner parameters of the NNs used for predictions (Figure \ref{fig_mom0}), and the final layer, where the observables are. The latter are characterized as random variables, modeled as a normal distribution centered around the actual values. Their variability is defined by the uncertainties inherent in the measurements detailed in the datasets section. In our context, \textit{observed} parameters refer to the direct measurements from different techniques such as spectroscopy or photometry, which inherently include uncertainties. \textit{True} parameters, on the other hand, correspond to the latent stellar properties inferred through our hierarchical Bayesian model, which accounts for observational errors and propagates uncertainties.

Measurement uncertainties of the stellar parameters were introduced in the model through multivariate Gaussian distributions centered at the mean value of the observations and constrained by the uncertainties themselves through a covariance matrix \citep[for deeper insight please refer to][]{Moya2022}.  This way, we were able to take into account the two sources of uncertainty: random noise due to the measurement process; and the epistemic uncertainty that affects the parameters of our model.

Our model is hierarchical in the sense that the prior on the stellar parameters is learned from the data. Figure \ref{fig:hierarchical} shows the two information levels of our hierarchical model, where the hyperparameters $\Omega$ and $D$ represent the correlation and scaling matrices needed to define the covariance matrix of the distribution of stellar parameters. \textbf{\textit{W}} and \textbf{\textit{b}} as the sets of weights matrices and biases vectors. \textit{Filled circles} are the true stellar parameters and CCs, and \textit{empty circles} represent the observed ones. In this representation, the connection between the stellar parameters and CCs is through multilayer perceptrons (dashed grey lines). In our case, we used 6 NNs for most experiments, one for each CC (Figure \ref{fig_mom0}). The main reason for this decision was to avoid multimodality in the NNs parameter space (later explained in Section \ref{resultsanddis}).

Following \cite{Moya2022} and \cite{2019Delgado} we decided to work with 7 chemical abundances; 5 \(\alpha\)-elements (Al, Mg, Si, Ti, Zn) and 2 s-process ones (Y and Sr) to build our CCs, leading to 10 potential combinations between the two groups. Among these, 5 ratios were formed with Y ([Si/Y], [Mg/Y], [Ti/Y], [Zn/Y], and [Al/Y]), while one ratio was formed with Sr ([Mg/Sr]). This decision was motivated by the fact that \textit{Y} is generally easier to obtain and often yields more precise values compared to \textit{Sr}. Besides that, the remaining combinations were not independent of the ones used.

Once the hierarchical Bayesian model is defined we can distinguish two computational stages: 1) the training phase (this stage runs the bayesian inference using MCMC algorithms and uses only the training set defined in section \ref{datasams}), where we use labeled data and derive posterior distributions for the NNs coefficients (\textbf{\textit{W}} and \textbf{\textit{b}}), the stellar parameters and the hyperparameters \textit{D} and \textit{$\Omega$}; and 2) the prediction stage, wherein the model is used to estimate ages for stars not present in the training set.

During the training stage, we sample 4x328 parameters (the 4 stellar variables of each of the 328 stars present in the training set), the hyperparameters and the NNs parameters, which in our case are going to be $ 6 \cdot \textit{\(N_{HLs}\)} \cdot \textit{\(N_{Nodes}\)} $  (where we are assuming an equal number of nodes per HL and in all 6 NNs used). Even though this was the main NN configuration during our work, different architectures were tested, and their proper definitions will be introduced in the corresponding sections, \ref{resultsanddis} and \ref{appendixx}. We distinguish $\hat{\theta}$ for the \textit{observed} stellar parameters and $\theta$ for the \textit{true} or the ones from the dataset. The same distinction applies to the CC vectors (\textit{observed} $\hat{c_{i}}$, \textit{true} $c_{i}$).

The distributions of probabilities of the stellar parameters inferred during the sampling of the training phase are used in the NNs (along with the \textit{\textbf{W}} and \textit{\textbf{b}} distributions) to produce the true CCs values. We assume that stars present a known and independent covariance matrix from each other and that there is not a direct correlation between stellar parameters and CCs. This approach lets us construct individual likelihoods for both groups of parameters (CCs and stellar attributes) in a simpler way by multiplying the probabilities of each observation. Then we can define the joint likelihood function of CCs and stellar parameters by multiplying the individual likelihoods as follows,

\begin{equation}
     \mathcal{L} = \prod_{i=1}^{328} p(\hat{c}_i | \theta_i, \textbf{\textit{W}}, \textbf{\textit{b}} ) \cdot p(\hat{\theta}_i | \theta_i),
    \label{eq-lik}
\end{equation}

\noindent where \textbf{\textit{W}} and \textbf{\textit{b}} represent the set of all matrices and vectors used in all 6 NNs used during the experiment, respectively. Applying Bayes' rule we can finally express the posterior distributions of parameters as

\begin{equation}
    p(\textbf{\textit{W}}, \textbf{\textit{b}}, \Theta|\hat{c}, \hat{\Theta}) \propto \mathcal{L} \cdot \pi(\Theta, D, \Omega) \cdot \pi(\textbf{\textit{W}}) \cdot \pi(\textbf{\textit{b}}),   
    \label{eq-prob2}
\end{equation}

\noindent where

\begin{equation}
    \pi(\Theta, D, \Omega)= p(\Theta | D, \Omega) \cdot \pi(\Omega) \cdot \pi(D).    
    \label{eq-prob3}
\end{equation}

Equation \eqref{eq-prob2} represents the posterior distribution of the parameters $\textbf{\textit{W}}$, $\textbf{\textit{b}}$ given the observed values of the chemical abundance ratios $\hat{c}$ and of the stellar parameters $\hat{\Theta}$. It is proportional to the product of the likelihood function $\mathcal{L}$, which captures the probability of the observed data given the parameters, and the prior distributions: $\pi(\Theta, D, \Omega)$, which accounts for the hierarchical structure of the hyperpriors $D$ and $\Omega$ needed to model $\Theta$, along with $\pi(\textbf{\textit{W}})$ and $\pi(\textbf{\textit{b}})$, representing the priors for $\textbf{\textit{W}}$ and $\textbf{\textit{b}}$, respectively. In Equation \eqref{eq-prob3}, the joint prior distribution is expressed as the product of the conditional prior, reflecting $\Theta$'s dependence on $D$ and $\Omega$, and the hyperprior distributions $\pi(\Omega)$ and $\pi(D)$, which account for uncertainties in the higher-level parameters.

In our approach, we firstly opted for utilizing  Gaussian distributions with varying centering for the initialization of the different weight matrices and bias vectors priors. Through iterative experimentation, we observed that aligning all BNNs priors to 0 significantly improved the predictive performance. Besides that, we decided to use diagonal matrices as their covariances matrices to enforce the independence of parameters for convenience and flexibility during experimentation. In this context, the standard deviation \(\sigma\), which serves as a measure of how the distribution spreads around its mean, is used as a uniform value across all elements of the diagonal covariance matrices applied to both \textbf{\textit{W}} and \textbf{\textit{b}}. Then, this parameter \(\sigma\), emerged as a critical factor during the training phase \citep[e.g.][]{2021Fortuin,2021Noci}. In our work we didn't account for correlations between BNNs and assumed their independence. We can summarize the initialization of our inner BNNs parameters with

\begin{equation}
    \pi(w_{ji}) =  \mathcal{N}( w_{ji} | 0, \sigma^{2}), 
    \label{priorw}
\end{equation}

\begin{equation}
     \pi(b_{i}) =  \mathcal{N}(b_{i}| 0, \sigma^{2}). 
    \label{priorb}
\end{equation}

In Equations \ref{priorw} and \ref{priorb} $w_{ji}$ and $b_{i}$ represent any of the elements of \textbf{\textit{W}} and \textbf{\textit{b}}, respectively.

We employed a non-informative multivariate Gaussian prior to incorporate the correlations between $T_\mathrm{eff}$, $t$, and [Fe/H], whereas we assigned an independent prior for  $\log$ g. The rationale behind this choice is that this parameter showed a low correlation with the other stellar parameters in the training set. This approach also lets us reduce the dimensionality of the hyperparameters \textit{D} and \textit{$\Omega$} by one, improving our computational times.

In our HBM we did not consider the necessity of defining a joint correlation between the two groups of parameters, the CCs, and stellar ones.  The relationship between each CC and the stellar variables is the result of the BNNs.

The multivariate Gaussian distribution used as prior for the stellar parameters was centred on the mean of the observed values. The covariance matrix of $T_\mathrm{eff}$, $t$, and [Fe/H] was disassembled into a scale matrix and a correlation matrix, a detailed exposition of which is available in \cite{Gelman06}:

\begin{equation}
    \Sigma = D \cdot \Omega \cdot D ,
\end{equation}

\noindent where \textit{D} is a diagonal matrix with a scale for each stellar parameter, and \textit{$\Omega$} is the correlation matrix. Following \cite{Moya2022} we defined a Lewandowski-Kurowicka-Joe (LKJ) prior \citep{LEWANDOWSKI2009} with shape parameter $\eta = 3$ for the correlation matrix, representing the equivalent to a uniform distribution on correlations. Finally, we defined a Cauchy prior on the scales centered at 0 with $\gamma = 2$. Once the posterior distributions of the true stellar parameters, \textbf{\textit{W}} and \textbf{\textit{b}} are inferred we can start with the age prediction stage.

\subsection{Prediction stage}

Let \( \theta' \) be the set of true stellar parameters excluding the age ( $\log$ g, $T_\mathrm{eff}$, and [Fe/H]), and \( \hat{\theta}' \) the vector of their observed values. The posterior samples of \( \Sigma \) obtained during the training phase were used to define the prior stellar parameters. These samples are employed to evaluate the likelihood term \( p(\hat{c}, \hat{\theta}' | t, \theta', \textbf{\textit{W}}, \textbf{\textit{b}}) \) to finally produce the distributions of \textit{t} for the stars in the test set.

As previously outlined in this paper, our approach is designed to be as independent as possible from manual modeling. This is the primary reason for our investigation into the automation capabilities and data adaptability of ML. Consequently, we decided to employ BNNs, which have the potential to automatically capture intricate data patterns. In our case, we employed normal priors centered at 0 for our BNNs  \textbf{\textit{W}} and \textbf{\textit{b}} parameters. While these priors depend on some parameter selection and subsequent fine-tuning experimentation, our model could potentially be used for other MS stars or even different stellar populations if we follow the same methodology for parameter adaptation.

In Section \ref{resultsanddis}, we present experiments in which the training phase, implemented via MCMC sampling, generally required between 1.5 and 2 days. However, certain configurations, including the best-performing one, extended the training duration to as much as 3.5 days. In contrast, the prediction time is primarily dictated by the number of samples used. For instance, processing 8,000 samples (organized into four chains of 2,000 samples each) for a single star typically takes about 20 minutes on the CPUs specified in subsection \ref{ref: sms} (this time varies significantly on CPU models).

\section{Results and discussion}\label{resultsanddis}

In this section, we present a comprehensive analysis of the outcomes derived from the HBMs employed in our study. The culmination of extensive computations and model training is encapsulated in a series of succinct and informative tables and graphics presented in the following paragraphs. Herein, we provide an overview of the experimental setup and the results from the BNN architecture designs we used.

\subsection{Experimental setup}

In the following paragraphs, we introduce the distributions of the datasets (described in Section. \ref{datasams}), examine convergence during the inference stage, and outline the evaluation metrics for age prediction tasks. This approach enables a comprehensive evaluation of our architecture efficiency compared to established methodologies.

\begin{itemize}

    \item \textbf{Distributions:} Figure \ref{fig:enter-labelnn2} presents the distributions of stellar parameters from all stars used in this work (a total of 328 for the training set and 23 for the test). These figures exhibit remarkable similarities between the training and test sets, notwithstanding the relatively limited size of the test set.
    
    \begin{figure}
    \centering
    \includegraphics[width=0.6\columnwidth]{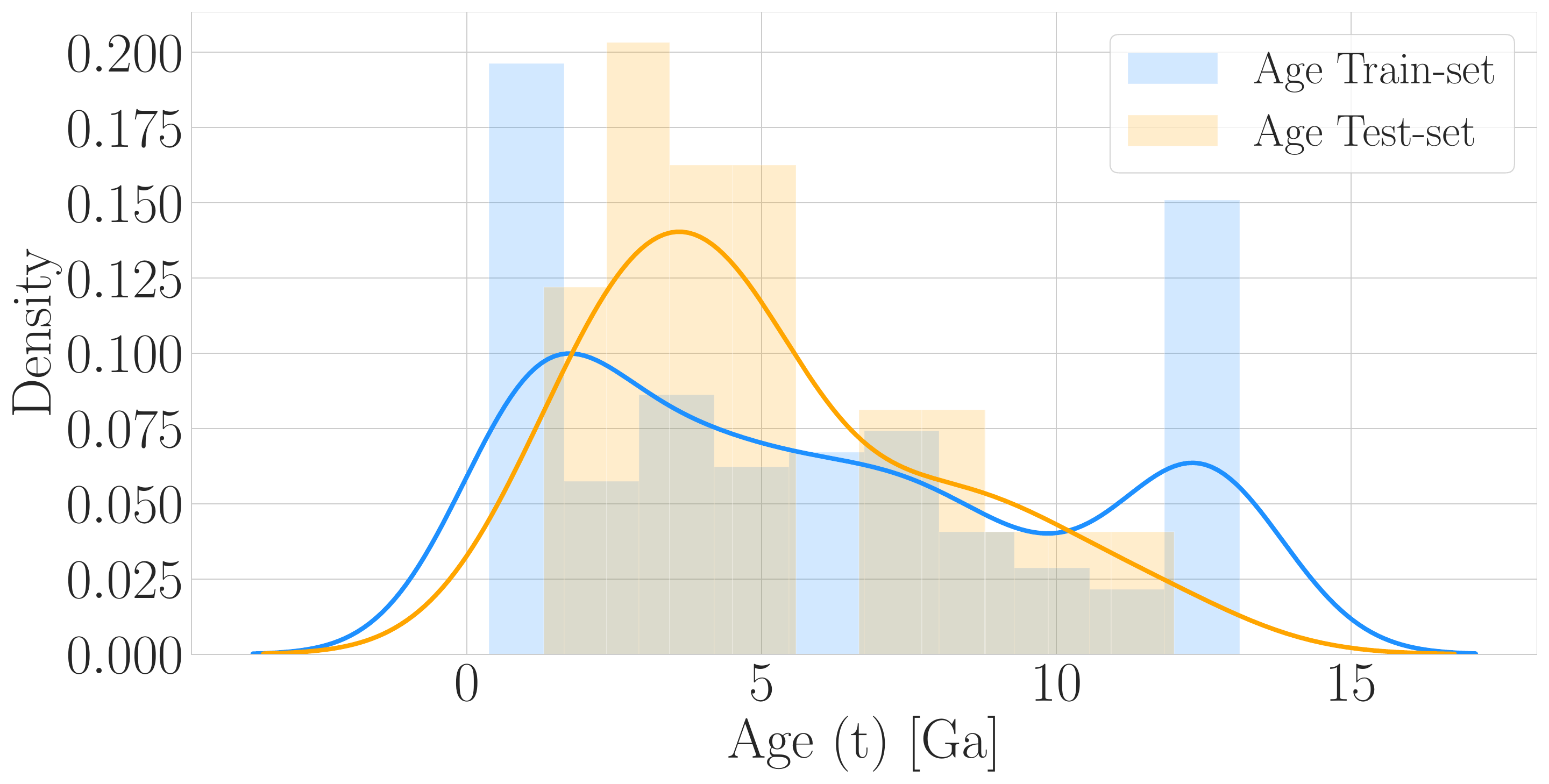}
    \includegraphics[width=0.6\columnwidth]{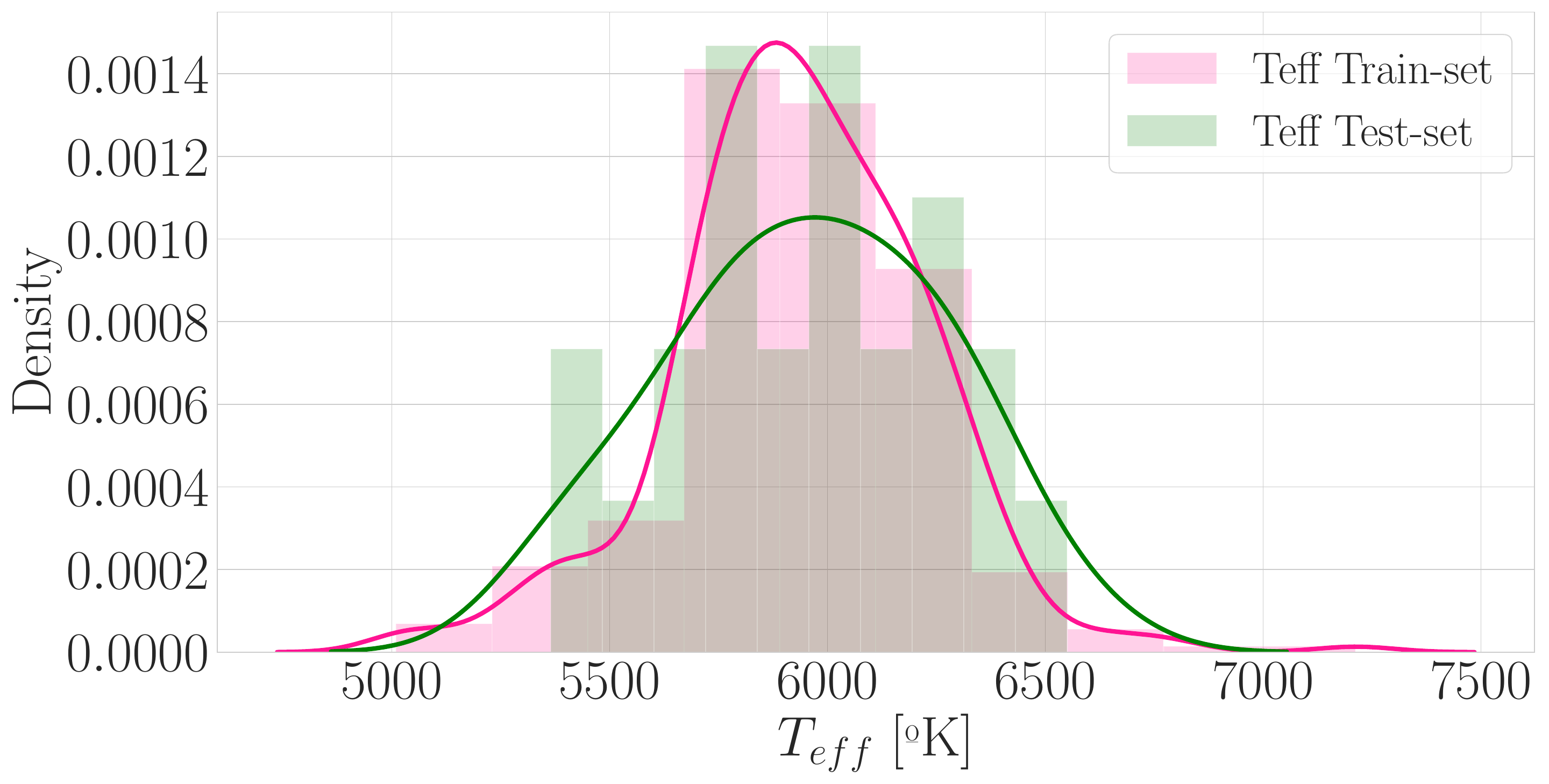}
    \includegraphics[width=0.6\columnwidth]{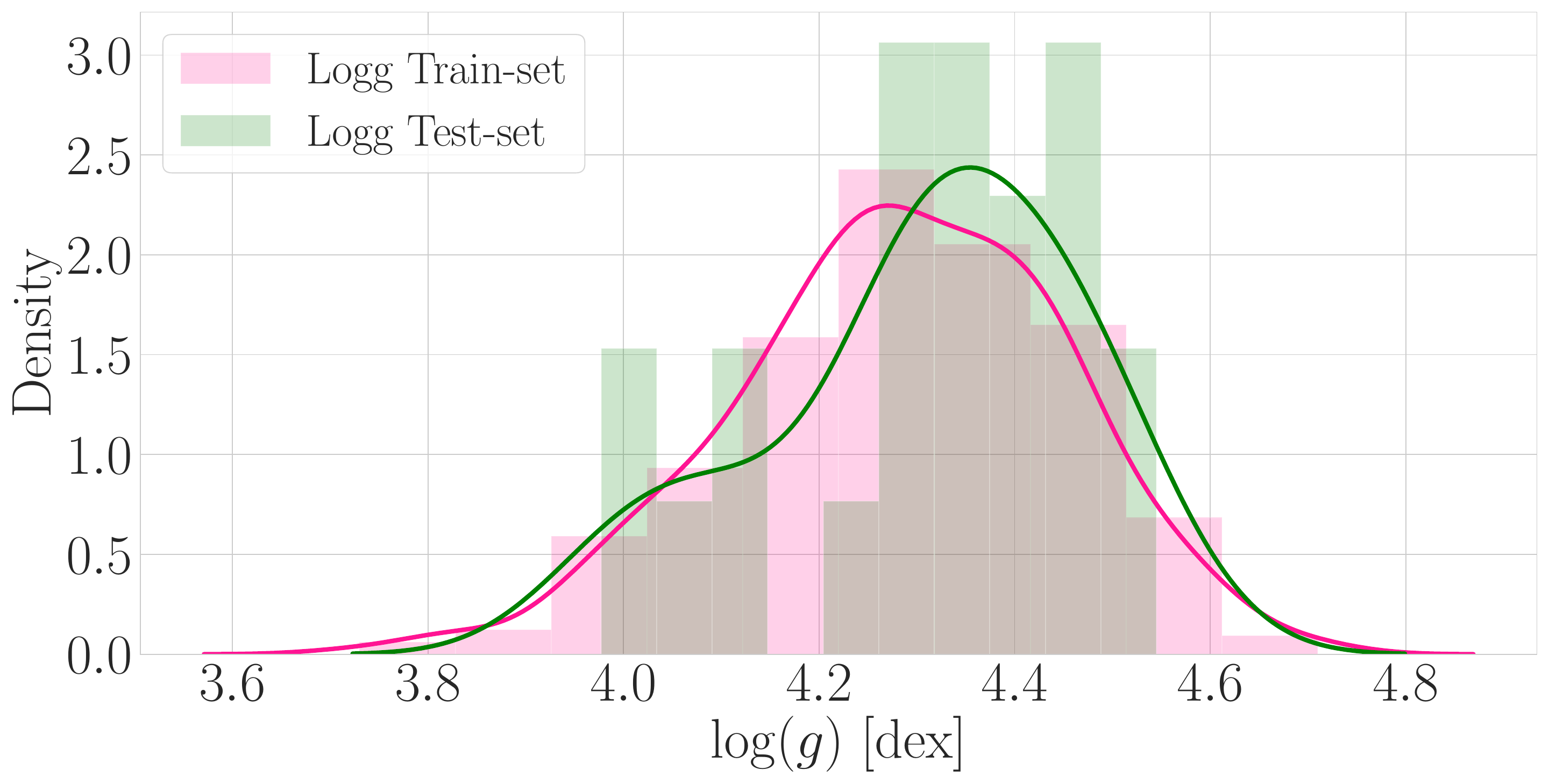}
    \includegraphics[width=0.6\columnwidth]{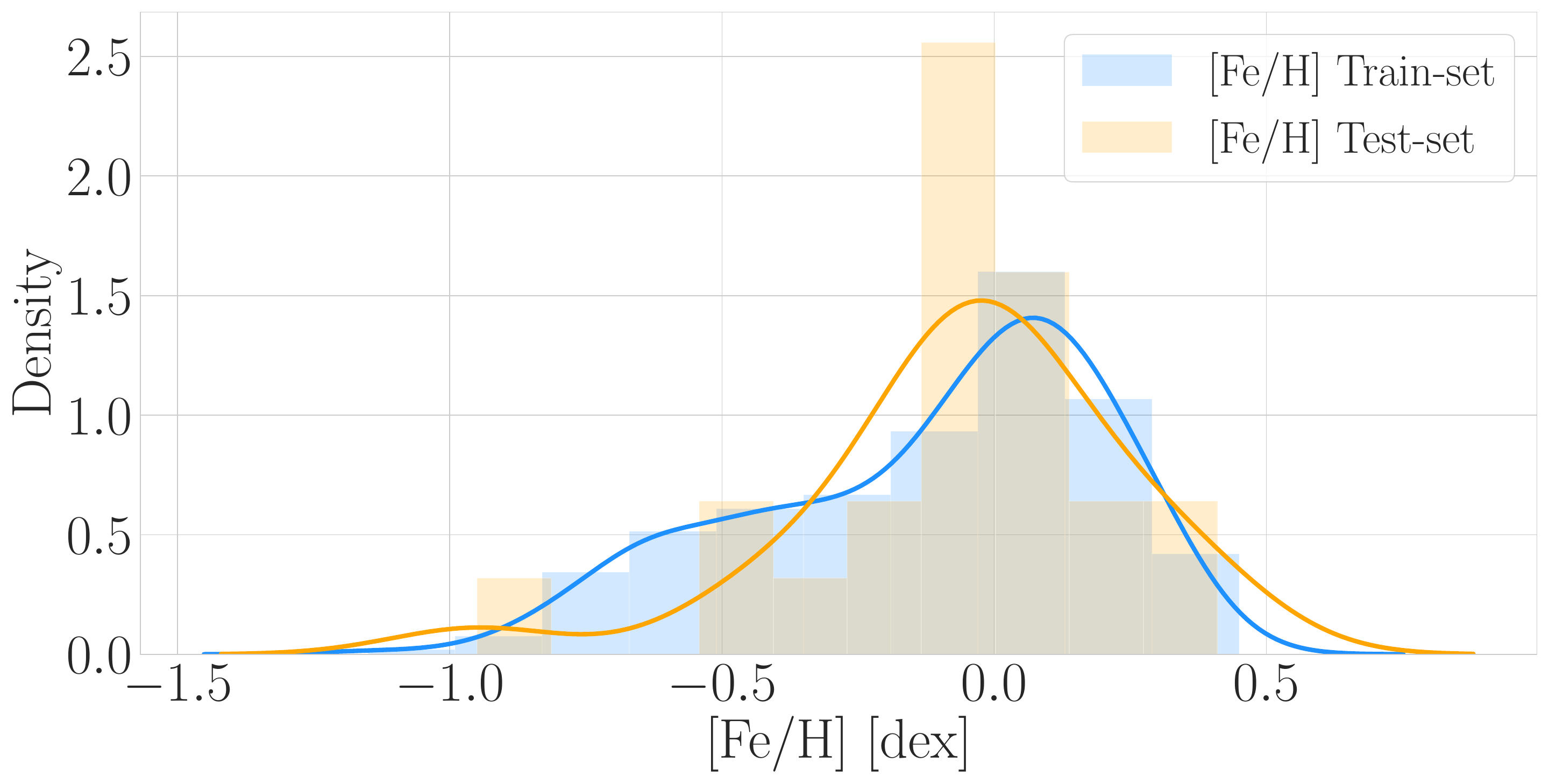}
    \caption{Distribution of the stellar parameters from the training and testing sets used in this work. \textit{From Top to Bottom:} Age, effective temperature, surface gravity, and metallicity.}
    \label{fig:enter-labelnn2}
    \end{figure}

    \item \textbf{Inference convergence:} We assess the convergence of the inference process using the NUTS-MCMC sampler heuristics \citep{2011Hoffman}. This involves analyzing two key parameters: the Gelman-Rubin statistic ($\hat{R}$) and the effective sample size ($N_{\rm eff}$) \citep{neal1993, 1992Geyer}. The $\hat{R}$ statistic measures the convergence of multiple chains, with values close to 1 indicating convergence.  $N_{\rm eff}$ represents the effective number of independent samples and is used to ensure that a sufficient number of samples are obtained for accurate inference. Then the target values for these heuristics are a high \( N_{\text{eff}} \), which indicates low autocorrelation and enhances the informativeness of the samples, and an \( \hat{R} \) value close to one, which signifies well-mixed and converged samples, suggesting reliable parameter estimates.

    In our analysis, we chose a strict convergence criterion, setting the $\hat{R}$ value threshold at 1.05. Additionally, if the $\hat{R}$ condition is satisfied, we also require that the effective sample size, $N_{\text{eff}}$, be at least 20\% of the total number of samples \citep{Gelman06}. Any $N_{\text{eff}}$ below this threshold indicates insufficient effective sampling, leading to rejection of the results. However, a small subset of experiments that meet these criteria may still be unreliable. This situation arises because, in the result tables that follow, we present $\hat{R}$ as an average of the individual $\hat{R}$ values for model parameters. While this approach provides a general overview, it can obscure instances where specific $\hat{R}$ values deviate significantly from 1. To ensure quality, each set of results was carefully reviewed, and instances where this issue occurs are noted in the relevant sections.

    \item \textbf{Age prediction metrics:} For evaluating the performance of our model in predicting stellar ages in the test set, we use the mean absolute error (MAE) metric: 

    \begin{equation}
        \text{MAE} = \frac{1}{n} \sum_{i=1}^{n} \left| t_i - \hat{t}_i \right|,
    \end{equation}
    
     where \( n \) is the number of stars in the test set, \( t_i \) is the target value of the age for star $i$, and \(\hat{t}_i \) is the predicted one. Additionally, we report the results as mean values of the posterior samples along with the standard deviation, representing the uncertainty in our predictions.

    \item \textbf{Baselines:} We compare the results obtained from our model with those reported by \cite{Moya2022} in their multilinear regression (MLR) approach. We provide an overview of their methodology and highlight any similarities or differences between their results and ours. This comparison serves as a benchmark for assessing the effectiveness of our model in relation to existing state-of-the-art approaches.

\end{itemize}

The experimentation aims to shed light on several key aspects related to BNNs applied to stellar characterization. In order to assess which architectural configurations yield the most accurate and reliable results,  we evaluate the performance of different BNN designs, exploring multiple network architectures (other experiments, presented in \ref{appendixx},  were run with single BNN configurations). This exploration offers valuable guidance on parameter-tuning strategies to enhance model performance and predictive accuracy.

Furthermore, the experimentation provides an assessment of the potential of BNNs to compete with classical approaches in this domain \citep[e.g.][]{Moya2022}. By comparing the performance of BNNs against traditional methods our study aims to gauge the extent to which BNNs can replicate or improve upon the predictive capabilities of established techniques. This comparative analysis serves to showcase the versatility and effectiveness of BNNs in capturing complex patterns and relationships within the data, potentially surpassing the limitations of conventional approaches.

\subsection{Multiple BNNs with single output}

Our initial approach involved using a single neural network with six outputs. However, this approach encountered multimodality issues, illustrated in Figure \ref{fig:rotated_figure} and discussed in \ref{appendixx}. To address these issues and improve the generalization capabilities of the architecture, we adopted a more conservative strategy by using a separate neural network for each CC. This approach simplifies the relationships the networks need to learn and reduces the complexity of the parameter space.

The initial phase of model architecture exploration involved a stepwise progression in complexity, with detailed configurations outlined in Table \ref{tabapp1}. This exploration commenced by experimenting with a single hidden layer, varying the number of nodes (3, 6, 10, 15, 20, 25). For prior specification of \textbf{\textit{W}} and \textbf{\textit{b}}, the investigation began with narrow standard deviations centered at 0, progressively expanding their value until 1, informed by prior experimentation.
    
The effective sample size and $\hat{R}$ diagnostics are of insufficient quality with 3 nodes but improve significantly starting from 6 nodes onward. Generally in BNNs the parameter space typically grows in complexity as the model's architecture becomes more intricate \citep[e.g.][]{2020Jospin}. However, using six neurons smooths the initially multimodal distribution of weights and biases observed in the three-node configuration, yielding a shape that more closely resembles a Gaussian distribution. This pattern persists across all our experiments involving architectures with more than three nodes, suggesting that this level of complexity is sufficient for the parameters to converge into a monomodal distribution.

Although the exact mechanism behind this behavior is not fully understood, we hypothesize that it could be due to the small age uncertainties of the dataset. Also, the observed multimodality may re-emerge with significantly larger architectures \citep[e.g.][]{2020Jospin}. In our case, the 6-node configuration appears to represent a threshold of complexity, where the number of nodes is sufficient to effectively map the parameter space. At this point, the nodes seem to specialize, focusing on individual modes to explain the data accurately while avoiding unnecessary complexity \citep[e.g.][]{cybenko1989, hornik1991,csaji2001}.

\begin{table}
    \centering
    
    \begin{tabular}{ccccc}
        \toprule
        \textbf{Nodes} & \textbf{\boldmath{$\pi$} [\boldmath{$\sigma$}]} & \textbf{MAE Age} & \textbf{$\hat{R}$} & \textbf{ $N_{\rm eff}$} \\
        \midrule
        3   & 0.25  & 0.99  & 1.156 & 3306 \\
        3   & 0.5   & 0.95  & 1.166 & 2042 \\
        3   & 1     & 0.87  & 1.838 & 1615 \\
        6   & 0.25  & 0.95  & 1.001 & 4960 \\
        6   & 0.5   & 0.85  & 1.003 & 2794 \\
        6   & 1     & 1.14  & 1.013 & 1711 \\
        10  & 0.25  & 0.92  & 1.000 & 8537 \\
        10  & 0.5   & 0.98  & 1.001 & 3942 \\
        10  & 1     & 1.17  & 1.001 & 3433 \\
        15  & 0.25  & 0.90  & 1.000 & 7180 \\
        20  & 0.25  & 0.89  & 1.000 & 5892 \\
        25  & 0.25  & 0.93  & 1.000 & 6955 \\
        \bottomrule
    \end{tabular}
    \caption*{\textbf{Fixed parameters}: 1 hidden layer NN, priors on data and chains centered on 0 (\textit{\textbf{W}} and \textit{\textbf{b}}), NUTS adapt delta parameter 0.95, NUTS max tree depth 13, 4 chains with 3000 samples (1000 as algorithm burn-in).}
    \vspace{0.3cm}
    \caption{Summary of experimental results and parameter configuration. \textit{First column:} number of nodes of the NN architecture. \textit{Second column:} standard deviations used to define the Gaussian priors for inner NN parameters. \textit{Third column:} age mean absolute errors. \textit{Fourth column:} $\hat{R}$ values for convergence diagnostics. \textit{Fifth column:} effective sample sizes.}
    \label{tabapp1}
\end{table}

We used a wider prior of $\sigma$ = 1  as a weaker prior belief about the values of the weights and biases, allowing the model to explore a broader range of parameter values. On the other hand, a narrower prior, like $\sigma$ = 0.25 imposes stronger constraints on the parameters, effectively acting as a stronger form of regularisation \citep[e.g.][]{mackay1992,bishop2006} (having also a direct impact on computational times). Constraining BNNs parameters too tightly through narrow priors may overly restrict the posterior distribution, limiting its ability to explore meaningful regions of the parameter space. This can result in underestimating the true posterior uncertainty and missing important solutions or modes that could otherwise explain the data effectively. Further investigation into the influence of prior width revealed that restrictive standard deviations generally yield improved predictions. Using a value of 0.25 in the standard deviation for the priors yields very good heuristics in 10, 15, 20, and 25 nodes configurations, all of them with MAE \( \approx \) 0.9. Configurations like the 6-nodes one, also give comparable predictions, but not that good heuristics.

Moreover, a prior initialization distant from the posterior can impede chain convergence, resulting in a diminished count of effective samples. Conversely, Table \ref{tabapp1}  presents the discussed results, predominantly indicating high-quality heuristics, suggesting that a greater number of nodes with precise and restrictive priors generally enhance predictive performance.

\begin{table}
    \centering
    
    \begin{tabular}{ccccc}
        \toprule
        \textbf{Nodes} & \textbf{ \boldmath{$\pi$} [ \boldmath{$\sigma$} ]} & \textbf{MAE Age} & \textbf{$\hat{R}$} & \textbf{ $N_{\rm eff}$} \\
        \midrule
        3   & 0.25  & 1.23 & 1.070 & 3979 \\
        3   & 0.5   & 0.89 & 1.037 & 2117 \\
        3   & 1     & 1.22 & 1.153 & 1199 \\
        6   & 0.25  & 1.12 & 1.009 & 5316 \\
        6   & 0.5   & 0.93 & 1.011 & 6220 \\
        6   & 1     & 1.64 & 1.005 & 3445 \\
        10  & 0.25  & 1.12 & 1.003 & 7335 \\
        \bottomrule
    \end{tabular}
    \caption*{\textbf{Fixed parameters}: 2 hidden layers NN, priors on data and chains centered on 0 (\textit{\textbf{W}} and \textit{\textbf{b}}), NUTS adapt delta parameter 0.95, NUTS max tree depth 13, 4 chains with 3000 samples (1000 as algorithm burn-in).}
    \vspace{0.3cm}
    \caption{Summary of experimental results and parameter configuration. \textit{First column:} number of nodes of the NN architecture. \textit{Second column:} standard deviations used to define the Gaussian priors for inner NN parameters. \textit{Third column:} age mean absolute errors. \textit{Fourth column:} $\hat{R}$ values for convergence diagnostics. \textit{Fifth column:} effective sample sizes.}
    \label{tab:your_label}
\end{table}

Subsequent experimentation delved deeper into the architecture of BNNs. While deeper networks have the potential to capture more complex relationships in the data, they also introduce higher complexity and challenges related to inference, such as increased computational cost and difficulties in sampling from complex posterior distributions \citep[e.g.][]{2015Blundell}. From these investigations, we observed that an increased number of layers does not necessarily yield improved outcomes, as shown in Table \ref{tab:your_label}. Contrary to the insights gained from node variations in earlier experiments, heuristics deteriorated with higher layer counts, reaching a point where results became unreliable, being this one of the scenarios where some of the model $\hat{R}$ parameters significantly deviate from 1. Despite launching additional experiments with varied standard deviations for initialization or larger numbers of nodes, the errors remained non-comparable to those observed in the single hidden layer experiments. Moreover, as discussed in previous sections, our training dataset comprises only 328 stars, presenting an opportunity to leverage simpler models.

The outcomes derived from systematic experimentation revealed instances where certain internal parameters of the BNN posterior distribution deviated significantly from their prior initialization. Consequently, a customized and better-centered initialization method was devised to address this issue, results are summarized in Table \ref{tabapp2}. In this approach, a variational-based technique, ADVI (Automatic Differentiation Variational Inference) from PyStan, was employed to obtain an initial approximation of the mean values for our target parameters (\textbf{\textit{W}} and \textbf{\textit{b}}). In a subsequent and independent stage, these values were input into the MCMC-NUTS sampler as the mean value for the BNN priors, and a range of models was computed by varying the standard deviation width. The results were comparable to non-informative initializations, with some cases exhibiting worse performances. The primary advantage of this approach lay in achieving slightly improved computational times, though this aspect was not a focal parameter of interest in our study.

\begin{table}
    \centering
    
    \small 
    \begin{tabular}{ccccc}
        \toprule
        \textbf{Nodes} & \textbf{\boldmath{$\pi$} [\boldmath{$\sigma$}] (D, C)} & \textbf{MAE} & \textbf{$\hat{R}$} & \textbf{ $N_{\rm eff}$} \\
        \midrule
        3 & 0.25, 0.25 & 1.01 & 1.167 & 2347 \\
        3 & 0.5, 0.25  & 0.99 & 1.153 & 1464 \\
        6 & 0.25, 0.1   & 1.01 & 1.000 & 5429 \\
        6 & 0.5, 0.1    & 1.14 & 1.004 & 1919 \\
        6 & 0.25, 0.25  & 0.88 & 1.002 & 3726 \\
        6 & 0.5, 0.25   & 0.92 & 1.006 & 2235 \\
        6 & 0.25, 0.5   & 0.88 & 1.003 & 3256 \\
        6 & 0.5, 0.5    & 0.95 & 1.006 & 1856 \\    
        \bottomrule
    \end{tabular}
    \caption*{\textbf{Fixed parameters}: 1 hidden layer NN, priors on data and ADVI centered on 0 (\textit{\textbf{W}} and \textit{\textbf{b}}), priors on NUTS chains centered on ADVI posteriors (\textit{\textbf{W}} and \textit{\textbf{b}}), NUTS adapt delta parameter 0.95, NUTS max tree depth 13, 4 chains with 3000 samples (1000 as algorithm burn-in).}
    \vspace{0.3cm}
    \caption{Summary of experimental results and parameter configuration. \textit{First column:} number of nodes of the NN architecture. \textit{Second column:} standard deviations used to define the Gaussian priors for inner NN parameters (D: Data and ADVI initialization, C: Chains). \textit{Third column:} age mean absolute errors. \textit{Fourth column:} $\hat{R}$ values for convergence diagnostics. \textit{Fifth column:} effective sample sizes.}
    \label{tabapp2}
\end{table}

Building upon the insights gained from the prior experimental iterations, we sought to enhance the optimal results obtained with a single hidden layer comprising 6 nodes and employing restrictive prior initialization (see Table \ref{tabapp1}). The subsequent batches of experimentation were devised to systematically evaluate the impact of the number of samples and the maximum tree depth (MTD) of the NUTS on model performance.

Doubling the posterior sample size yielded a marginal improvement in the heuristics of the 6-node architecture, with no discernible adverse effects on MAEs. To systematically compare these outcomes, additional experiments were conducted with a shallower tree depth (for NUTS), results are summarized in Table \ref{tabapp4}, to analyze the impact of the MTD. With an MTD set to 10, the sampler reached 100\% sample saturation, indicating that while achieving favorable heuristics and results, the sampler faced more stringent requirements for realistic parameter space exploration. Altering this parameter to 12 partially resolved the saturation issue (some of the inner parameters of our architecture need deeper MTD), but the results exhibited a decline in quality. This set of experiments suggests that a shallower MTD may be employed in tandem with extended training to alleviate the impact of less accurate exploration. Even though time is not the most significant constraint in our study, this could represent a strong limitation for larger datasets. Conversely, elevating the MTD did not lead to a direct enhancement in MAEs but positively influenced heuristics by increasing the count of effective samples. Higher tree depths improve posterior exploration, particularly for complex distributions, but exponentially increase the computational cost per iteration.

\begin{table}
    \centering
    
    \begin{tabular}{ccccc}
        \toprule
        \textbf{Samples} & \textbf{MTD} & \textbf{MAE Age} & \textbf{$\hat{R}$} & \textbf{ $N_{\rm eff}$} \\
        \midrule
        3000 (1000)  & 10  & 0.92 & 1.010 & 827 \\
        3000 (1000)  & 12  & 1.65 & 1.011 & 2334 \\
        6000 (2000)  & 10  & 1.08 & 1.014 & 1130 \\
        6000 (2000)  & 12  & 1.02 & 1.001 & 5474 \\
        6000 (2000)  & 13  & 0.88 & 1.005 & 3487 \\
        6000 (2000)  & 14  & 1.04 & 1.004 & 6741 \\
        \bottomrule
    \end{tabular}
    \caption*{\textbf{Fixed parameters}: 1 hidden layer with 6 nodes NN, priors on data and chains centered on 0 (\textit{\textbf{W}} and \textit{\textbf{b}}), NUTS adapt delta parameter 0.95, 4 chains with 3000 samples (1000 as algorithm burn-in),  Gaussian priors for inner NN parameters are fixed to $\mu = 0$ and $\sigma = 0.5$.}
    \vspace{0.3cm}
    \caption{Summary of experimental results and parameter configuration. \textit{First column:} number of nodes of the NN architecture. \textit{Second column:} Max tree depth NUTS algorithm parameter. \textit{Third column:} age mean absolute errors. \textit{Fourth column:} $\hat{R}$ values for convergence diagnostics. \textit{Fifth column:} effective sample sizes.}
    \label{tabapp4}
\end{table}

Increasing sample sizes is useful if the correct MTD ensures the sampler can properly explore the posterior. However, once the posterior is well-sampled and diagnostics indicate convergence, additional samples may lead to little improvement or could even worsen the results. Based on this insight, we chose to extend our investigation of the MTD. We focused on configurations with the most promising heuristics from Table \ref{tabapp1} and expanded the MTD to 14. Our analysis reveals an overall enhancement in heuristics, particularly beneficial for configurations comprising only 6 nodes. These findings are summarized in Table \ref{tabapp5}, where it is evident that the majority of MAE values approximate 0.9. This underscores the potential for achieving a balance between architectural complexity, outcomes, and computational efficiency with simpler architectures.

\begin{table}
    \centering
    
    \begin{tabular}{ccccc}
        \toprule
        \textbf{Nodes} & \textbf{ \boldmath{$\pi$} [ \boldmath{$\sigma$} ]} & \textbf{MAE Age} & \textbf{$\hat{R}$} & \textbf{ $N_{\rm eff}$} \\
        \midrule
        6   & 0.25  & 0.90 & 1.000 & 5916 \\
        6   & 0.5   & 0.93 & 1.003 & 2940 \\
        8   & 0.25  & 0.96 & 1.002 & 5709 \\
        8   & 0.5   & 0.91 & 1.002 & 3119 \\
        10  & 0.25  & 0.98 & 1.001 & 7260 \\
        15  & 0.25  & 0.96 & 1.000 & 9300 \\
        20  & 0.25  & 0.91 & 1.000 & 6618 \\
        25  & 0.25  & 0.93 & 1.000 & 7575 \\
        \bottomrule
    \end{tabular}
    \caption*{\textbf{Fixed parameters}: 1 hidden layer NN, priors on data and chains centered on 0 (\textit{\textbf{W}} and \textit{\textbf{b}}), NUTS adapt delta parameter 0.95, NUTS max tree depth 14, 4 chains with 3000 samples (1000 as algorithm burn-in).}
    \vspace{0.3cm}
    \caption{Summary of experimental results and parameter configuration. \textit{First column:} number of nodes of the NN architecture. \textit{Second column:} standard deviations used to define the Gaussian priors for inner NN parameters. \textit{Third column:} age mean absolute errors. \textit{Fourth column:} $\hat{R}$ values for convergence diagnostics. \textit{Fifth column:} effective sample sizes.}
    \label{tabapp5}
\end{table}

As concluding experiments we systematically investigate the impact of the parameter adapt delta (AD) of the NUTS-HMC algorithm, a critical variable influencing the acceptance ratio of target samples. A higher AD results in a more conservative step size in the sampler, thereby reducing the likelihood of numerical errors during the integration process. Consequently, the sampler takes smaller steps, potentially enhancing the robustness and accuracy of the samples at the expense of increased computational requirements per effective sample. The data depicted in Table \ref{tabapp6} closely align with those observed in Table \ref{tabapp5}. Also, the quality of our heuristics improved due to the adoption of a more restrictive sampling approach, ensuring a balanced acceptance rate that enhances both efficiency and reliability in posterior exploration. 

In Table \ref{tabapp7}, we amalgamated the insights garnered from the current experimentation, incorporating longer sampling durations, deeper MTD exploration, and more restrictive AD settings, particularly targeting configurations comprising 6 nodes with varying prior ranges. This batch yielded some of the most promising heuristics. As a succinct conclusion drawn from our current experimentation, we present the following observations:

\begin{itemize}
  
  \item \textit{Architectures and nodes}: Higher layers didn't always improve outcomes due to increased complexity and inference challenges. Single-layer architectures with a higher count of neurons can simplify the parameter space and still capture complex patterns for our problem.
  
  \item \textit{Impact of prior width}: Wider priors allowed broader exploration, while narrower priors acted as stronger regularization. Restrictive standard deviations generally improve predictions but seem to increase sensitivity in heuristics.
  
  \item \textit{Initialization}: Distant initialization hindered convergence, while precise priors enhanced performance. A VI-based initialization technique, meaning a better centering for our case, slightly reduced computational times during our experimentation.
  
  \item \textit{Sample size and MTD}: Doubling samples slightly improved heuristics, while higher tree depth increased effective samples.
  
  \item \textit{Acceptance Denial parameter}: Higher values restricted acceptance, affecting sample acceptance ratio and the robustness of the sampling, at the cost of increased computational times.

\end{itemize}

\begin{table}
    \centering
    
    \begin{tabular}{ccccc}
        \toprule
        \textbf{Nodes} & \textbf{ \boldmath{$\pi$} [ \boldmath{$\sigma$} ]} & \textbf{MAE Age} & \textbf{$\hat{R}$} & \textbf{ $N_{\rm eff}$} \\
        \midrule
        6   & 0.25  & 0.83 & 1.000 & 6411 \\
        6   & 0.5   & 0.92 & 1.002 & 3104 \\
        10  & 0.25  & 1.05 & 1.000 & 7527 \\
        15  & 0.25  & 1.00 & 1.000 & 7931 \\
        20  & 0.25  & 0.86 & 1.000 & 7623 \\
        \bottomrule
    \end{tabular}
    \caption*{\textit{Fixed parameters}: 1 hidden layer NN, priors on data and chains centered on 0 (\textit{\textbf{W}} and \textit{\textbf{b}}), NUTS adapt delta parameter 0.99, NUTS max tree depth 13, 4 chains with 3000 samples (1000 as algorithm burn-in).}
    \vspace{0.3cm}
    \caption{Summary of experimental results and parameter configuration. \textit{First column:} number of nodes of the NN architecture. \textit{Second column:} standard deviations used to define the Gaussian priors for inner NN parameters. \textit{Third column:} age mean absolute errors. \textit{Fourth column:} $\hat{R}$ values for convergence diagnostics. \textit{Fifth column:} effective sample sizes.}
    \label{tabapp6}
\end{table}

\begin{table}
    \centering
    
    \begin{tabular}{ccccc}
        \toprule
        \textbf{Nodes} & \textbf{ \boldmath{$\pi$} [ \boldmath{$\sigma$} ]} & \textbf{MAE Age} & \textbf{$\hat{R}$} & \textbf{ $N_{\rm eff}$} \\
        \midrule
        6   & 0.1   & 1.25 & 1.000 & 24061 \\
        6   & 0.25  & 1.05 & 1.001 & 8802 \\
        6   & 0.5   & 0.90 & 1.002 & 5873 \\
        \bottomrule
    \end{tabular}
    \caption*{\textit{Fixed parameters}: 1 hidden layer NN, priors on data and chains centered on 0 (\textit{\textbf{W}} and \textit{\textbf{b}}), NUTS adapt delta parameter 0.99, NUTS max tree depth 14, 4 chains with 6000 samples (2000 as algorithm burn-in).}
    \vspace{0.3cm}
    \caption{Summary of experimental results and parameter configuration. \textit{First column:} number of nodes of the NN architecture. \textit{Second column:} standard deviations used to define the Gaussian priors for inner NN parameters. \textit{Third column:} age mean absolute errors. \textit{Fourth column:} $\hat{R}$ values for convergence diagnostics. \textit{Fifth column:} effective sample sizes.}
    \label{tabapp7}
\end{table}

\subsection{Best architecture and comparison with the state of the art}

In this section, we compare the performance of our best model architecture with the results reported in \cite{Moya2022}, providing a rationale for our design choices. Based on insights from prior experiments and the outcomes summarized in Table \ref{tabapp7}, we identified $\sigma = 0.25$ and $\sigma = 0.5$ as potential optimal parameters. To ensure robust statistical validation, we conducted the prediction stage 10 independent times, allowing us to evaluate the consistency and reliability of the results.

\begin{figure}
    \centering
    \includegraphics[width=0.8\linewidth]{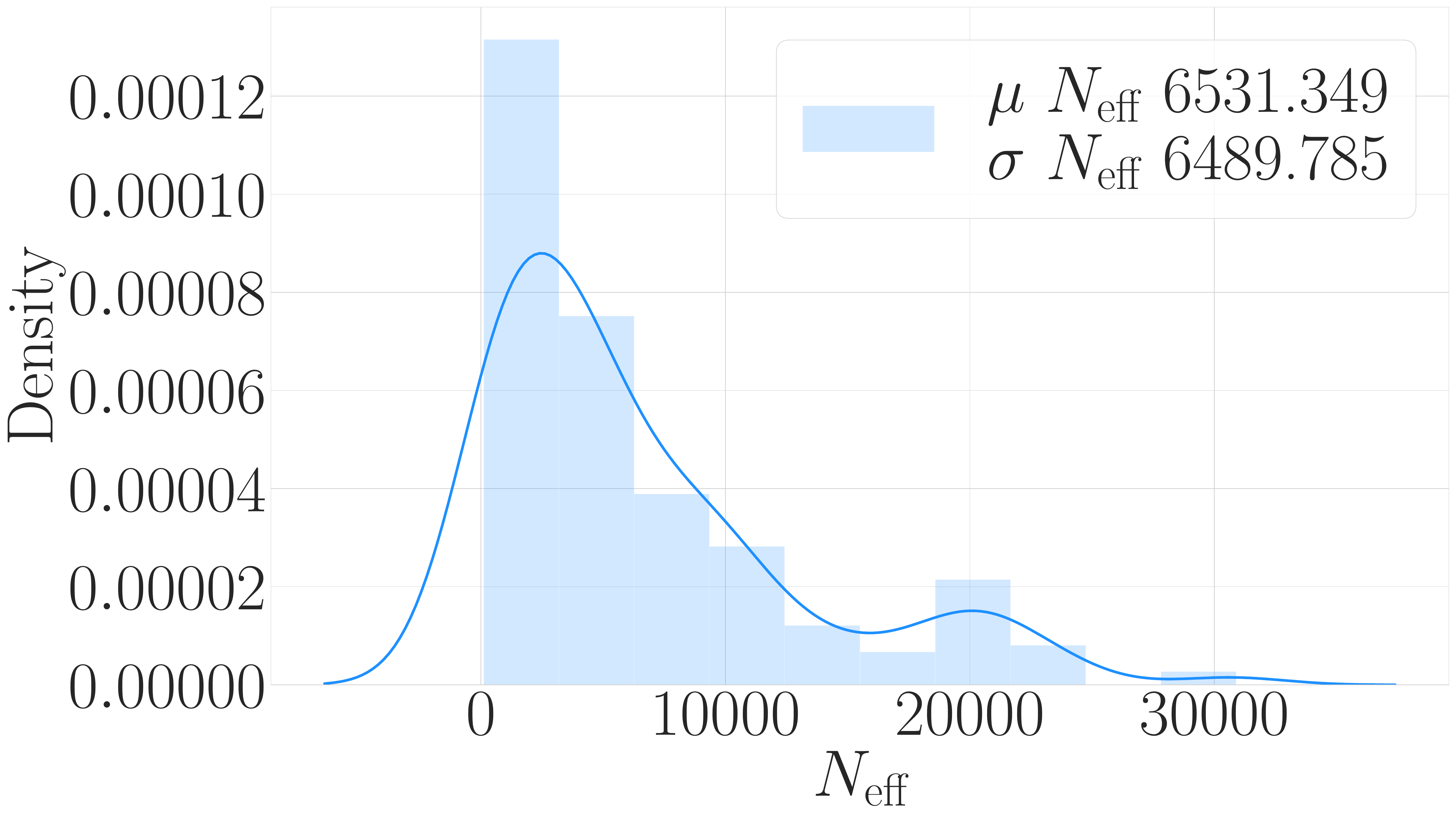}
    \includegraphics[width=0.8\linewidth]{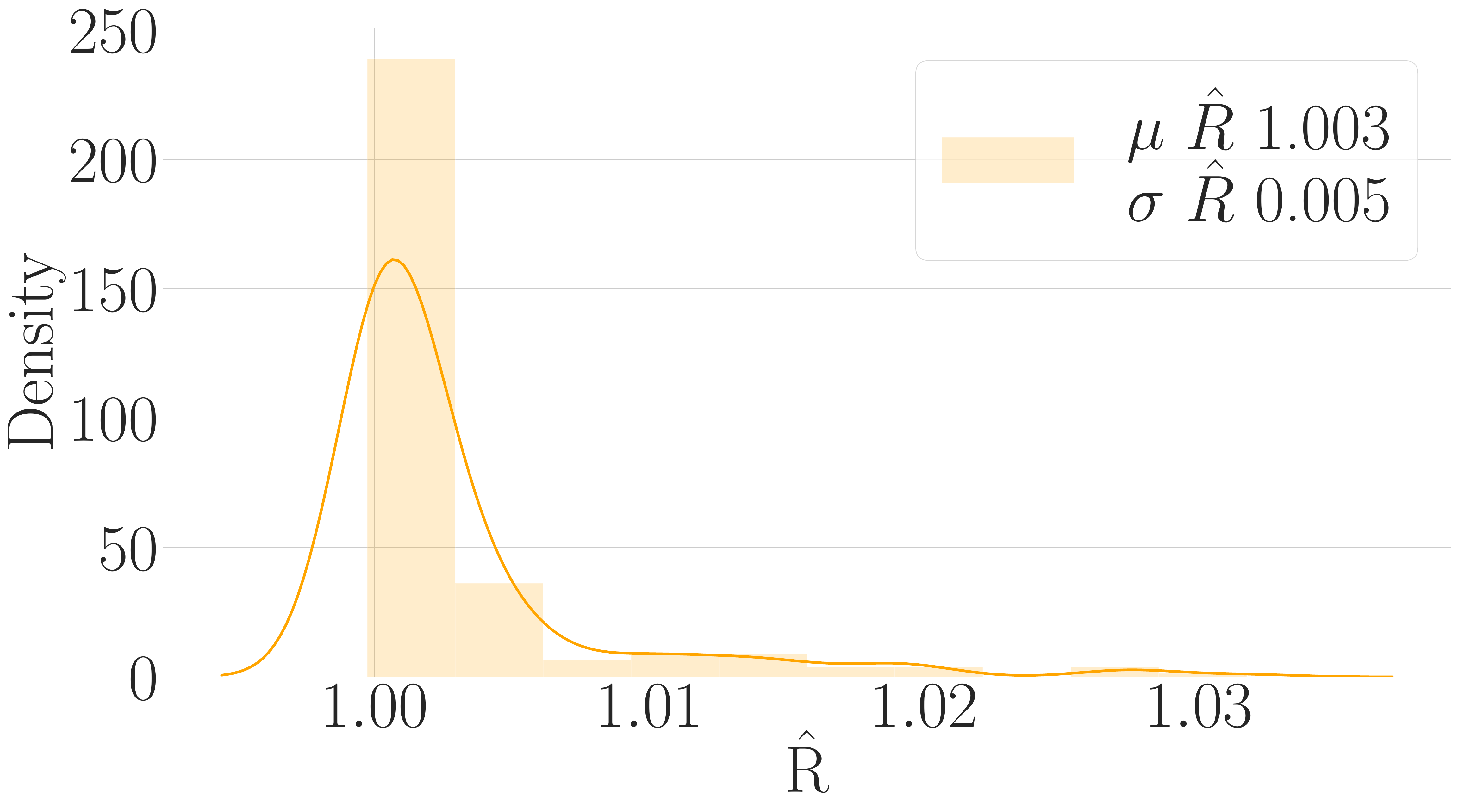}
    \caption{From the best experiment configuration of the work, Table \ref{tabappBNN}. \textit{Top:} Distribution of the probabilities of the MCMC-NUTS sampling heuristic of convergence  $N_{\rm eff}$ (Number of effective samples). \textit{Bottom: }Distribution of the probabilities of the MCMC-NUTS sampling heuristic of convergence $\hat{R}$ (Gelman-Rubin statistic).}
    \label{fig:enter-labelappen}
\end{figure}

In summary, the optimal architecture identified consists of a single HL with six nodes  (utilizing six BNNs, each corresponding to a specific CC). The parameters yielding the best results are presented in Table \ref{tabapp8} and illustrated in Figure \ref{fig:enter-labelappen}. Even though $\sigma = 0.25$ tended to achieve better results during our experimentation, we selected $\sigma = 0.5$ as the prior initialization width for the winning architecture, as it provides the best balance among computational cost, robustness, simplicity, heuristics, and performance outcomes. This was achieved by using 6000 samples per chain, where 2000 samples were allocated for burn-in and preconditioning of the sampler (with a total of 4 chains for parameter space exploration). The parameters MTD and AD, which are crucial for sampling, were set to 14 and 0.99, respectively. The winning configuration parameters are summarized in Table \ref{tabappBNN}.

\begin{table} 
    \centering
    
    \begin{tabular}{ccccc}
        \toprule
        \textbf{Nodes} & \textbf{ \boldmath{$\pi$} [ \boldmath{$\sigma$} ]} & \textbf{MAE Age} & \textbf{$\hat{R}$} & \textbf{ $N_{\rm eff}$} \\
        \midrule
        6   & 0.25  & 0.91 & 1.005 & 6927 \\
        6   & 0.5   & 0.92 & 1.003 & 6531 \\
        \bottomrule
    \end{tabular}
    \caption*{\textit{Fixed parameters}: 1 hidden layer NN, priors on data and chains centered on 0 (\textit{\textbf{W}} and \textit{\textbf{b}}), NUTS adapt delta parameter 0.99, NUTS max tree depth 14, 4 chains with 6000 samples (2000 as algorithm burn-in).}
    \vspace{0.3cm}
    \caption{Summary of final experimental results and parameter configuration. Best designs from Table \ref{tabapp7} (second and third row) [Average of 10 prediction cicles]. \textit{First column:} number of nodes of the NN architecture. \textit{Second column:} standard deviations used to define the Gaussian priors for inner NN parameters. \textit{Third column:} age mean absolute errors. \textit{Fourth column:} $\hat{R}$ values for convergence diagnostics. \textit{Fifth column:} effective sample sizes.}
    \label{tabapp8}
\end{table}

\begin{figure}
	\centering 
	\includegraphics[width= 0.8\columnwidth]{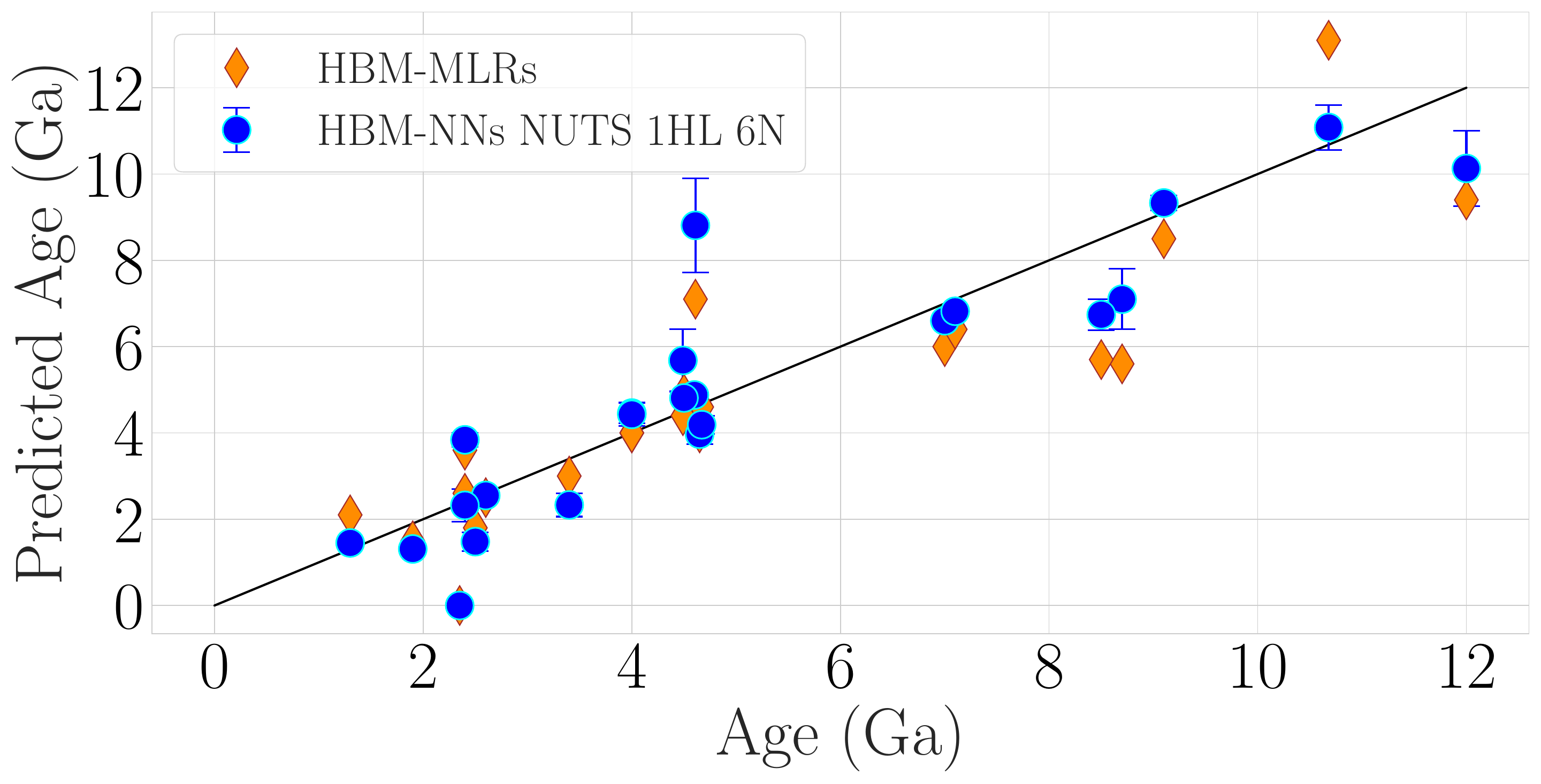}	
	\includegraphics[width= 0.8\columnwidth]{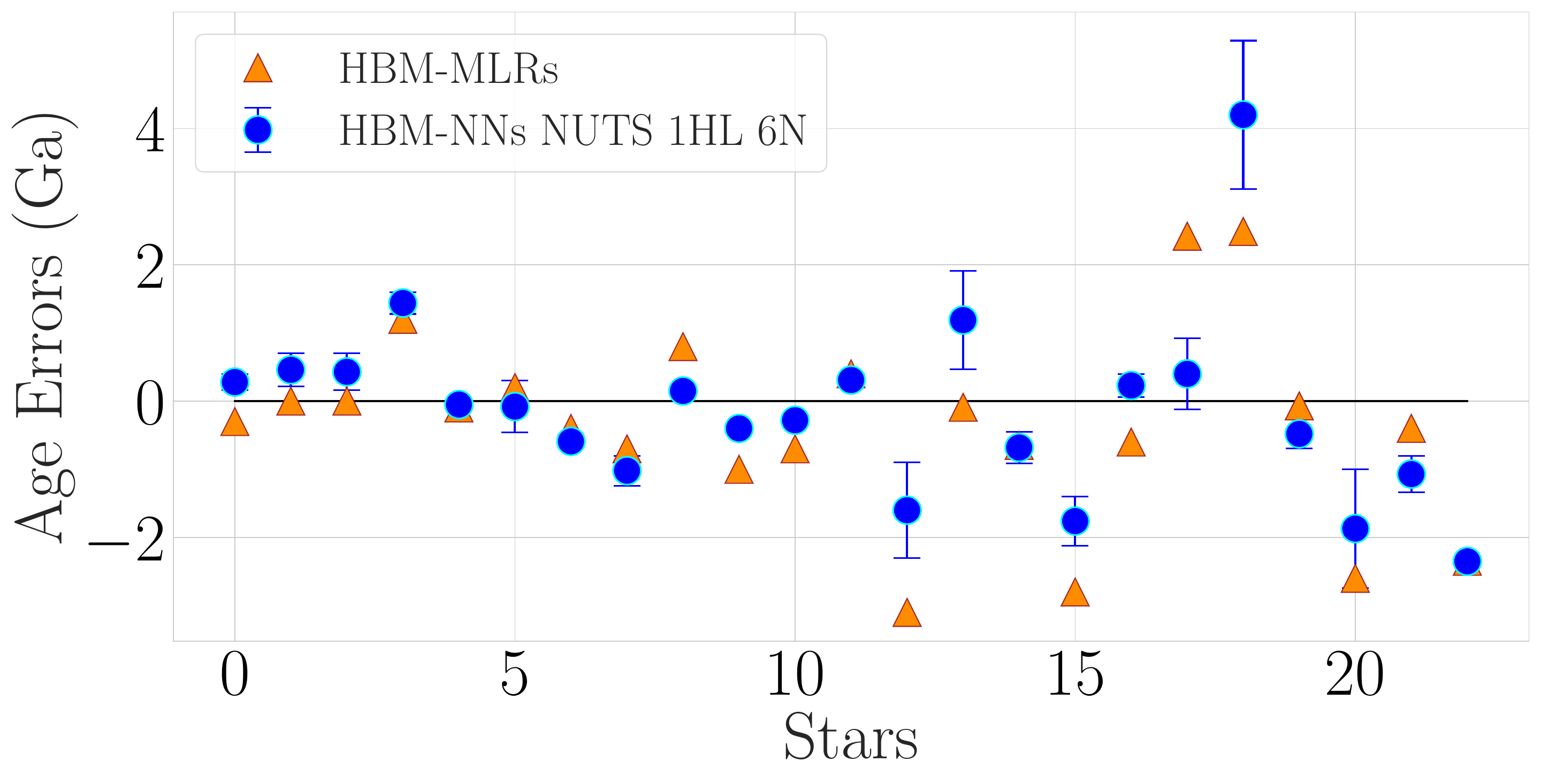}	
	\caption{Hierarchical Bayesian model result comparison. (\textit{HBM-MLRs:}) architecture whose probability relationships are modeled by multi-linear regressions, from \cite{Moya2022}. (\textit{HBM-NNs NUTS 1HL 6N:}) \textbf{[Best]} architecture whose probability relationships are modeled by NNs. The architecture consists of 1 Hidden Layer with 6 nodes and single-output (x6 NNs), sampled with MCMC-NUTS algorithm. (\textit{Top}:) Age prediction for the 23 test stars. (\textit{Bottom}:) Age errors for the 23 test stars.} 
	\label{fig_mom02}%
\end{figure}

This configuration achieved a MAE of 0.92 after ten training runs. The heuristics ( $N_{\rm eff}$ and $\hat{R}$) for this architecture are plotted in Figure \ref{fig:enter-labelappen}, showing an average $\hat{R}$ value of 1.003 with approximately 6500 effective samples per parameter. The final predictions of this architecture are shown in Figure \ref{fig_mom02}, which compares our results with the state-of-the-art HBM-MLR detailed in \cite{Moya2022} for the same test set. Our results are comparable to theirs, and our predictions align closely with their findings. Notably, for most predictions, we achieved small uncertainties compared to their mean value, with only one of the 23 stars exhibiting a larger uncertainty. This discrepancy could be due to the chemical peculiarities of that star or poor representation in the training set, which may have biased our prediction.

\begin{table}
    \centering    
    \begin{tabular}{cc}
        \toprule
        \textbf{Parameter} & \textbf{Value} \\
        \midrule
        HLs & 1 \\
        Nodes & 6 \\
        MTD & 14 \\
        AD & 0.99 \\
        \(\mu\) and \(\sigma\) priors & [0, 0.5] \\
        Samples & 6000 (2000) \\
        Chains & 4 \\
        $\hat{R}$ & 1.003 \\
        MAE Age & 0.92 \\
        \bottomrule
    \end{tabular}
    \caption{Summary of the parameter configuration for the best BNN architecture obtained during the experimentation.}
    \label{tabappBNN}
\end{table}

It is important to note that comparable results were achievable with less restrictive parameters and the use of wider neural networks, which also yielded good outcomes. However, we demonstrated that with appropriate fine-tuning, good results could be achieved using simpler architectures.

In this work, we present an approach for stellar dating that achieves results comparable to those reported in previous studies \citep{Moya2022}. The primary contribution of our approach lies in its ability to leverage neural networks to model complex relationships within the dataset without the need for manually designed empirical relations. This flexibility not only simplifies the modeling process but also reduces the risk of oversimplification inherent in manual modeling, which could otherwise lead to suboptimal models and unreliable uncertainty estimates. Furthermore, our model demonstrates versatility and adaptability, making it suitable for application to a wide range of stellar types and astrophysical problems. For instance, it can be employed to deepen our understanding of the evolution of stellar ages as a function of galactocentric radius within the Milky Way \citep[][]{2022Viscasillas}. Additionally, the model could be used for advancing our knowledge of the intricate relationships between stellar parameters in the context of magneto-gyrochronology, potentially refining the methods used to estimate stellar ages based on rotational and magnetic properties \citep[e.g.][]{2023Mathur2}. This applicability underscores the model's potential as a powerful tool for addressing fundamental questions in stellar astrophysics.

\section{Summary and conclusions}

\begin{figure*}
    \centering
    \rotatebox{90}{\includegraphics[width=15cm,height=12cm]{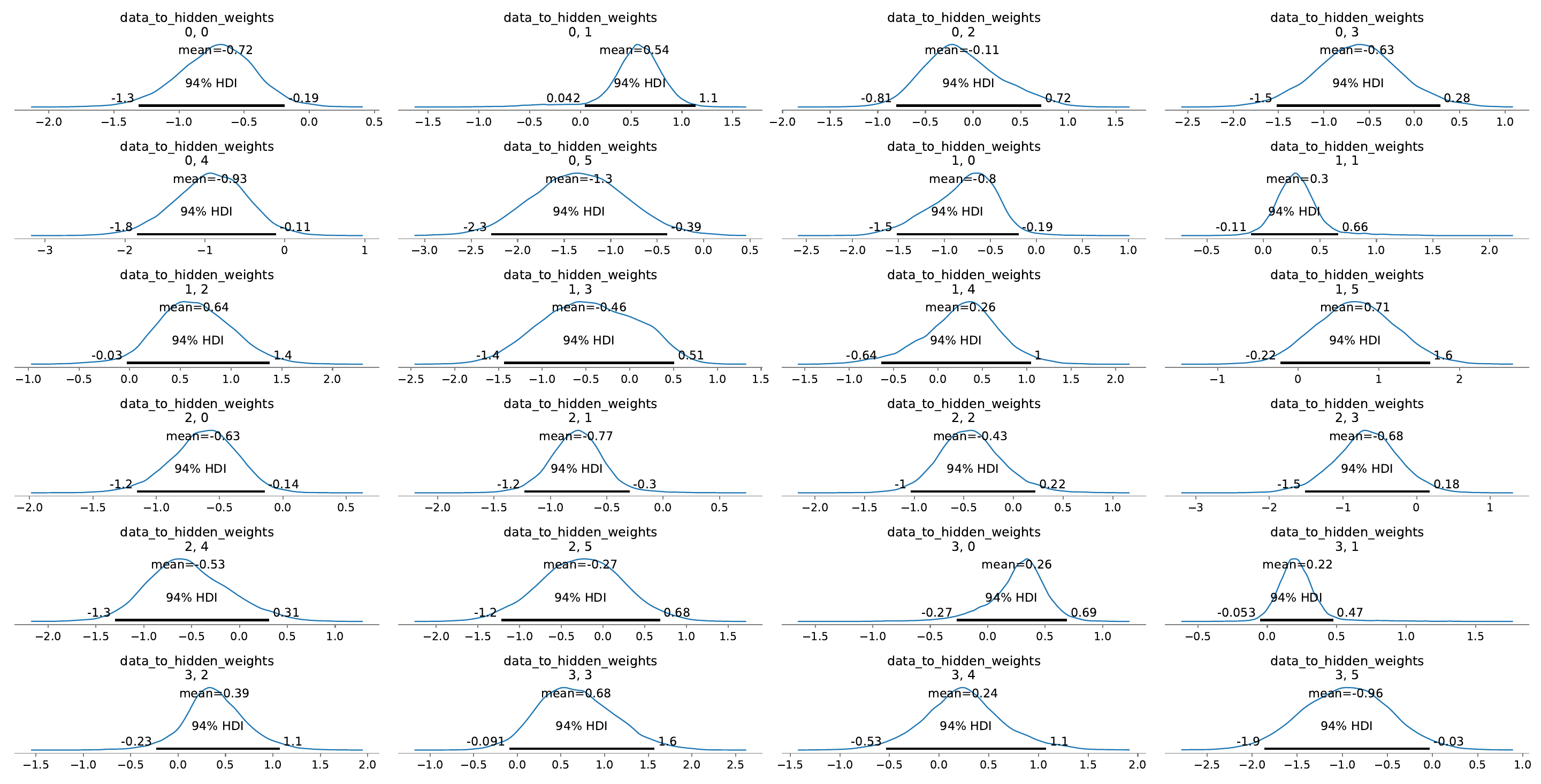}}
    \caption{Example of posterior distributions for a weight matrix in one hidden layer, six nodes neural network applied to a model with a single output, single CC (in this case \textit{[Mg/Y]}) prediction, (for better illustration, see figure \ref{fig_mom0}). The figure shows the distributions of each weight (\textit{$W_{ji}$}) in the initial weight matrix, which represents the connections between the input data and the hidden layer for the best experiment configuration (second row of Table \ref{tabapp8}). Each distribution includes its mean and the 94\% Highest Density Interval (HDI). The distributions are organized from left to right and top to bottom, corresponding to each of the four stellar parameters and their connections to the six nodes in the hidden layer.}
    \label{multibest}
\end{figure*}

The outcomes presented in this paper affirm the successful formulation of an effective and robust methodology for uncertainty quantification, specifically applied in the context of stellar dating. Optimal results were attained through the adoption of single-layer architectures. While it is feasible to achieve comparable outcomes using an HBM employing a solitary BNN with six outputs (corresponding to each CC), such models often necessitate the utilization of deeper and wider neural networks, thereby having the potential to exhibit pronounced multi-modality. While the incorporation of multi-modal samplers could potentially address this challenge, our choice was to focus on the reliability and versatility of the well-established Stan language, which provides robust sampling techniques while allowing for customization and adaptability.

As illustrated in Figure \ref{multibest}, the posterior distributions of our BNNs consistently exhibit Gaussian-like shapes, suggesting the potential utilization of VI samplers, such as ADVI, in scenarios prioritizing computational speed. While the results obtained with VI algorithms are comparable during experimentation (mainly in multi-output NN) and, in some architectures, improvement, we deliberately chose MCMC approaches in this study to prioritize a more conservative approach. This preference is reflected in the selection of an MTD sufficiently high to avoid chain saturation and restrictive ADs (e.g. 0.95 or 0.99). Conversely, the number of samples, set at 3000, proved to be adequate, with a marginal enhancement in heuristics observed with additional samples, particularly advantageous with a lower node count.

Our experimental findings underscore the effectiveness of employing generic initializations for BNN parameters in our context. Assumptions such as Gaussian distributions and centering priors at zero were made during the selection process. These assumptions facilitated the exploration of optimal standard deviations. Generally, initializing priors with small standard deviations (0.25 or less), close to zero, yielded improved heuristics. This effect was more pronounced with a higher node count.

In summary, single hidden layer architectures (with single output) with a node count ranging between 6-25 (optimal heuristics/results ratio at 6 nodes, superior heuristics with higher counts), coupled with stringent restrictions on the prior initialization of weights and biases (standard deviation 0.25-0.5), MTD set at 13-14, and AD at 0.99, led to our most successful outcomes. This configuration enabled the presentation of a statistical-machine learning hybrid model that achieved a MAE of less than 1 Ga in our testing sample.  We present a machine learning-based tool for stellar dating that achieves results comparable to prior studies while eliminating the need for manually designed empirical relationships. This approach streamlines modeling, reduces oversimplification risks, and ensures robust predictions with reliable uncertainty estimates. Its versatility also allows us to apply this approach to various datasets, including different stellar types, opening an opportunity for improvement in this, and potentially other fields.

 \section{Future lines}

Future work will focus on leveraging the developed algorithms to investigate the distribution of stellar ages as a function of galactocentric radius. This analysis holds the promise of providing valuable insights into fundamental processes of galactic evolution, such as the Milky Way's formation history and the role of radial migration \citep[e.g.][]{2022Viscasillas}. By integrating precise stellar age estimates with spatial distribution data, this extension aims to deliver a comprehensive characterization of age variation across the galaxy.

Furthermore, we plan to explore the integration of gyrochronology and magnetic activity indices for stellar age determination. Building on recent work \citep{2023Mathur2}, which first combined these indices through linear relationships. While this approach demonstrated promise, it may also introduce errors and biases due to its reliance on simplistic models. By employing our more flexible framework, we aim to mitigate these limitations and develop a robust methodology for combining gyrochronology and magnetic activity indices in a unified age estimation model.

This approach will not only refine stellar age determinations but also create a versatile toolkit for stellar dating, encompassing a range of complementary techniques. Such advancements are expected to enhance our understanding of the evolution of stellar populations and the dynamic history of the Milky Way.

On the other hand, future research in BNNs and uncertainty quantification remains focused on addressing the persistent challenge of multimodality \citep{wiese2023efficient}. Enhancing inference methods to better capture and navigate multimodal distributions is a crucial aspect of this ongoing investigation. The current target aims to develop innovative algorithms that can efficiently handle the complex parameter spaces inherent in neural networks, seeking a balance between computational efficiency and the accurate representation of uncertainty \citep[e.g.][]{2016Liu2,2020Wilson,2020Durasov}. Furthermore, there is a notable emphasis on integrating domain knowledge and expert insights into the Bayesian framework, providing more informed priors that can contribute to robust and interpretable uncertainty estimates \citep[e.g.][]{2021Fortuin}. Besides that, extending the application of BNNs to diverse domains is a key frontier.

In addition, establishing standardized benchmarks and improving current evaluation metrics will be crucial for evaluating the performance of BNNs across different applications, to ensure that progress in the field is measurable and reproducible. In summary, the dynamic landscape of BNN research still revolves around tackling multimodality, refining inference methods for better scalability, and broadening the applicability of these models to address complex real-world challenges.

\section*{Acknowledgements}

A.M. acknowledges funding support from Grant PID2019-107061GB-C65 funded by MCIN/AEI/10.13039/501100011033, and from Generalitat Valenciana in the frame of the GenT Project CIDEGENT/2020/036. V.T. acknowledges funding support from Generalitat Valenciana in the frame of the INVESTIGO Project INVEST/2022/56. I also extend my thanks to Alberto Torralba, for his expert advice and assistance.

\nocite{*}
\bibliography{references}

\begin{thebibliography}{105}
\expandafter\ifx\csname natexlab\endcsname\relax\def\natexlab#1{#1}\fi
\providecommand{\url}[1]{\texttt{#1}}
\providecommand{\href}[2]{#2}
\providecommand{\path}[1]{#1}
\providecommand{\DOIprefix}{doi:}
\providecommand{\ArXivprefix}{arXiv:}
\providecommand{\URLprefix}{URL: }
\providecommand{\Pubmedprefix}{pmid:}
\providecommand{\doi}[1]{\href{http://dx.doi.org/#1}{\path{#1}}}
\providecommand{\Pubmed}[1]{\href{pmid:#1}{\path{#1}}}
\providecommand{\bibinfo}[2]{#2}
\ifx\xfnm\relax \def\xfnm[#1]{\unskip,\space#1}\fi
\bibitem[{{Adibekyan} et~al.(2011){Adibekyan}, {Santos}, {Sousa} and {Israelian}}]{2011Adibekyan}
\bibinfo{author}{{Adibekyan}, V.Z.}, \bibinfo{author}{{Santos}, N.C.}, \bibinfo{author}{{Sousa}, S.G.}, \bibinfo{author}{{Israelian}, G.}, \bibinfo{year}{2011}.
\newblock \bibinfo{title}{{A new {\ensuremath{\alpha}}-enhanced super-solar metallicity population}}.
\newblock \bibinfo{journal}{\aap} \bibinfo{volume}{535}, \bibinfo{pages}{L11}.
\newblock \DOIprefix\doi{10.1051/0004-6361/201118240}, \href{http://arxiv.org/abs/1111.4936}{{\tt arXiv:1111.4936}}.
\bibitem[{{Adibekyan} et~al.(2012){Adibekyan}, {Sousa}, {Santos}, {Delgado Mena}, {Gonz{\'a}lez Hern{\'a}ndez}, {Israelian}, {Mayor} and {Khachatryan}}]{2012Adibekyan}
\bibinfo{author}{{Adibekyan}, V.Z.}, \bibinfo{author}{{Sousa}, S.G.}, \bibinfo{author}{{Santos}, N.C.}, \bibinfo{author}{{Delgado Mena}, E.}, \bibinfo{author}{{Gonz{\'a}lez Hern{\'a}ndez}, J.I.}, \bibinfo{author}{{Israelian}, G.}, \bibinfo{author}{{Mayor}, M.}, \bibinfo{author}{{Khachatryan}, G.}, \bibinfo{year}{2012}.
\newblock \bibinfo{title}{{Chemical abundances of 1111 FGK stars from the HARPS GTO planet search program. Galactic stellar populations and planets}}.
\newblock \bibinfo{journal}{\aap} \bibinfo{volume}{545}, \bibinfo{pages}{A32}.
\newblock \DOIprefix\doi{10.1051/0004-6361/201219401}, \href{http://arxiv.org/abs/1207.2388}{{\tt arXiv:1207.2388}}.
\bibitem[{Aguirre et~al.(2017)Aguirre, Lund, Antia, Ball, Basu, Christensen-Dalsgaard, Lebreton, Reese, Verma, Casagrande, Justesen, Mosumgaard, Chaplin, Bedding, Davies, Handberg, Houdek, Huber, Kjeldsen, Latham, White, Coelho, Miglio and Rendle}]{Aguirre_2017}
\bibinfo{author}{Aguirre, V.S.}, \bibinfo{author}{Lund, M.N.}, \bibinfo{author}{Antia, H.M.}, \bibinfo{author}{Ball, W.H.}, \bibinfo{author}{Basu, S.}, \bibinfo{author}{Christensen-Dalsgaard, J.}, \bibinfo{author}{Lebreton, Y.}, \bibinfo{author}{Reese, D.R.}, \bibinfo{author}{Verma, K.}, \bibinfo{author}{Casagrande, L.}, \bibinfo{author}{Justesen, A.B.}, \bibinfo{author}{Mosumgaard, J.R.}, \bibinfo{author}{Chaplin, W.J.}, \bibinfo{author}{Bedding, T.R.}, \bibinfo{author}{Davies, G.R.}, \bibinfo{author}{Handberg, R.}, \bibinfo{author}{Houdek, G.}, \bibinfo{author}{Huber, D.}, \bibinfo{author}{Kjeldsen, H.}, \bibinfo{author}{Latham, D.W.}, \bibinfo{author}{White, T.R.}, \bibinfo{author}{Coelho, H.R.}, \bibinfo{author}{Miglio, A.}, \bibinfo{author}{Rendle, B.}, \bibinfo{year}{2017}.
\newblock \bibinfo{title}{Standing on the shoulders of dwarfs: the kepler asteroseismic legacy sample. ii. radii, masses, and ages}.
\newblock \bibinfo{journal}{The Astrophysical Journal} \bibinfo{volume}{835}, \bibinfo{pages}{173}.
\newblock \URLprefix \url{https://dx.doi.org/10.3847/1538-4357/835/2/173}, \DOIprefix\doi{10.3847/1538-4357/835/2/173}.
\bibitem[{{Almeida-Fernandes} and {Rocha-Pinto}(2018)}]{2018Almeida}
\bibinfo{author}{{Almeida-Fernandes}, F.}, \bibinfo{author}{{Rocha-Pinto}, H.J.}, \bibinfo{year}{2018}.
\newblock \bibinfo{title}{{A method to estimate stellar ages from kinematical data}}.
\newblock \bibinfo{journal}{\mnras} \bibinfo{volume}{476}, \bibinfo{pages}{184--197}.
\newblock \DOIprefix\doi{10.1093/mnras/sty119}, \href{http://arxiv.org/abs/1801.04046}{{\tt arXiv:1801.04046}}.
\bibitem[{{Ancona} et~al.(2017){Ancona}, {Ceolini}, {{\"O}ztireli} and {Gross}}]{2017Ancona}
\bibinfo{author}{{Ancona}, M.}, \bibinfo{author}{{Ceolini}, E.}, \bibinfo{author}{{{\"O}ztireli}, C.}, \bibinfo{author}{{Gross}, M.}, \bibinfo{year}{2017}.
\newblock \bibinfo{title}{{Towards better understanding of gradient-based attribution methods for Deep Neural Networks}}.
\newblock \bibinfo{journal}{arXiv e-prints} , \bibinfo{pages}{arXiv:1711.06104}\DOIprefix\doi{10.48550/arXiv.1711.06104}, \href{http://arxiv.org/abs/1711.06104}{{\tt arXiv:1711.06104}}.
\bibitem[{Angus et~al.(2015)Angus, Aigrain, Foreman-Mackey and McQuillan}]{2015Angus}
\bibinfo{author}{Angus, R.}, \bibinfo{author}{Aigrain, S.}, \bibinfo{author}{Foreman-Mackey, D.}, \bibinfo{author}{McQuillan, A.}, \bibinfo{year}{2015}.
\newblock \bibinfo{title}{Calibrating gyrochronology using kepler asteroseismic targets}.
\newblock \bibinfo{journal}{Monthly Notices of the Royal Astronomical Society} \bibinfo{volume}{450}, \bibinfo{pages}{1787–1798}.
\newblock \URLprefix \url{http://dx.doi.org/10.1093/mnras/stv423}, \DOIprefix\doi{10.1093/mnras/stv423}.
\bibitem[{{Angus} et~al.(2019){Angus}, {Morton}, {Brewer}, {Bedell}, {Foreman-Mackey}, {Curtis} and {Hogg}}]{2019Angus}
\bibinfo{author}{{Angus}, R.}, \bibinfo{author}{{Morton}, T.D.}, \bibinfo{author}{{Brewer}, J.}, \bibinfo{author}{{Bedell}, M.}, \bibinfo{author}{{Foreman-Mackey}, D.}, \bibinfo{author}{{Curtis}, J.}, \bibinfo{author}{{Hogg}, D.W.}, \bibinfo{year}{2019}.
\newblock \bibinfo{title}{{An age-dating method that combines stellar evolution models with gyrochronology}}, in: \bibinfo{booktitle}{American Astronomical Society Meeting Abstracts \#233}, p. \bibinfo{pages}{235.09}.
\bibitem[{Arbel et~al.(2023)Arbel, Pitas, Vladimirova and Fortuin}]{arbel2023primer}
\bibinfo{author}{Arbel, J.}, \bibinfo{author}{Pitas, K.}, \bibinfo{author}{Vladimirova, M.}, \bibinfo{author}{Fortuin, V.}, \bibinfo{year}{2023}.
\newblock \bibinfo{title}{A primer on bayesian neural networks: Review and debates}.
\newblock \href{http://arxiv.org/abs/2309.16314}{{\tt arXiv:2309.16314}}.
\bibitem[{{Barnes}(2003)}]{2003Barnes}
\bibinfo{author}{{Barnes}, S.A.}, \bibinfo{year}{2003}.
\newblock \bibinfo{title}{{On the Rotational Evolution of Solar- and Late-Type Stars, Its Magnetic Origins, and the Possibility of Stellar Gyrochronology}}.
\newblock \bibinfo{journal}{\aj} \bibinfo{volume}{586}, \bibinfo{pages}{464--479}.
\newblock \DOIprefix\doi{10.1086/367639}, \href{http://arxiv.org/abs/astro-ph/0303631}{{\tt arXiv:astro-ph/0303631}}.
\bibitem[{Basu(2020)}]{2020basu}
\bibinfo{author}{Basu, V.}, \bibinfo{year}{2020}.
\newblock \bibinfo{title}{Prediction of stellar age with the help of extra-trees regressor in machine learning}, in: \bibinfo{booktitle}{Proceedings of the International Conference on Innovative Computing \& Communications (ICICC) 2020}, p.~\bibinfo{pages}{.}
\newblock \bibinfo{note}{Available at SSRN: \url{https://ssrn.com/abstract=3563397} or \url{http://dx.doi.org/10.2139/ssrn.3563397}}.
\bibitem[{Bishop(2006)}]{bishop2006}
\bibinfo{author}{Bishop, C.M.}, \bibinfo{year}{2006}.
\newblock \bibinfo{title}{Pattern Recognition and Machine Learning}.
\newblock \bibinfo{publisher}{Springer-Verlag New York, Inc.}, \bibinfo{address}{Secaucus, NJ, USA}.
\bibitem[{{Blundell} et~al.(2015){Blundell}, {Cornebise}, {Kavukcuoglu} and {Wierstra}}]{2015Blundell}
\bibinfo{author}{{Blundell}, C.}, \bibinfo{author}{{Cornebise}, J.}, \bibinfo{author}{{Kavukcuoglu}, K.}, \bibinfo{author}{{Wierstra}, D.}, \bibinfo{year}{2015}.
\newblock \bibinfo{title}{{Weight Uncertainty in Neural Networks}}.
\newblock \bibinfo{journal}{arXiv e-prints} , \bibinfo{pages}{arXiv:1505.05424}\DOIprefix\doi{10.48550/arXiv.1505.05424}, \href{http://arxiv.org/abs/1505.05424}{{\tt arXiv:1505.05424}}.
\bibitem[{{Brandt} and {Huang}(2015)}]{2015Brandt}
\bibinfo{author}{{Brandt}, T.D.}, \bibinfo{author}{{Huang}, C.X.}, \bibinfo{year}{2015}.
\newblock \bibinfo{title}{{Bayesian Ages for Early-type Stars from Isochrones Including Rotation, and a Possible Old Age for the Hyades}}.
\newblock \bibinfo{journal}{\aj} \bibinfo{volume}{807}, \bibinfo{pages}{58}.
\newblock \DOIprefix\doi{10.1088/0004-637X/807/1/58}, \href{http://arxiv.org/abs/1501.04404}{{\tt arXiv:1501.04404}}.
\bibitem[{{Bressan} et~al.(2012){Bressan}, {Marigo}, {Girardi}, {Salasnich}, {Dal Cero}, {Rubele} and {Nanni}}]{2012Bressan}
\bibinfo{author}{{Bressan}, A.}, \bibinfo{author}{{Marigo}, P.}, \bibinfo{author}{{Girardi}, L.}, \bibinfo{author}{{Salasnich}, B.}, \bibinfo{author}{{Dal Cero}, C.}, \bibinfo{author}{{Rubele}, S.}, \bibinfo{author}{{Nanni}, A.}, \bibinfo{year}{2012}.
\newblock \bibinfo{title}{{PARSEC: stellar tracks and isochrones with the PAdova and TRieste Stellar Evolution Code}}.
\newblock \bibinfo{journal}{\mnras} \bibinfo{volume}{427}, \bibinfo{pages}{127--145}.
\newblock \DOIprefix\doi{10.1111/j.1365-2966.2012.21948.x}, \href{http://arxiv.org/abs/1208.4498}{{\tt arXiv:1208.4498}}.
\bibitem[{Bu et~al.(2020)Bu, Yerra, Xie, Pan, Gang and Wu}]{2020bu}
\bibinfo{author}{Bu, Y.}, \bibinfo{author}{Yerra, B.k.}, \bibinfo{author}{Xie, J.}, \bibinfo{author}{Pan, J.}, \bibinfo{author}{Gang, Z.}, \bibinfo{author}{Wu, Y.}, \bibinfo{year}{2020}.
\newblock \bibinfo{title}{Estimation of stellar ages and masses using gaussian process regression}.
\newblock \bibinfo{journal}{The Astrophysical Journal Supplement Series} \bibinfo{volume}{249}, \bibinfo{pages}{7}.
\newblock \DOIprefix\doi{10.3847/1538-4365/ab8bcd}.
\bibitem[{Carpenter et~al.(2017)Carpenter, Gelman, Hoffman, Lee, Goodrich, Betancourt, Brubaker, Guo, Li and Riddell}]{Carpenter2017}
\bibinfo{author}{Carpenter, B.}, \bibinfo{author}{Gelman, A.}, \bibinfo{author}{Hoffman, M.D.}, \bibinfo{author}{Lee, D.}, \bibinfo{author}{Goodrich, B.}, \bibinfo{author}{Betancourt, M.}, \bibinfo{author}{Brubaker, M.}, \bibinfo{author}{Guo, J.}, \bibinfo{author}{Li, P.}, \bibinfo{author}{Riddell, A.}, \bibinfo{year}{2017}.
\newblock \bibinfo{title}{Stan: A probabilistic programming language}.
\newblock \bibinfo{journal}{Journal of Statistical Software} \bibinfo{volume}{76}, \bibinfo{pages}{1--32}.
\newblock \URLprefix \url{https://doi.org/10.18637/jss.v076.i01}, \DOIprefix\doi{10.18637/jss.v076.i01}.
\bibitem[{{Ciuc{\u{a}}} et~al.(2021){Ciuc{\u{a}}}, {Kawata}, {Miglio}, {Davies} and {Grand}}]{2021Ioana}
\bibinfo{author}{{Ciuc{\u{a}}}, I.}, \bibinfo{author}{{Kawata}, D.}, \bibinfo{author}{{Miglio}, A.}, \bibinfo{author}{{Davies}, G.R.}, \bibinfo{author}{{Grand}, R.J.J.}, \bibinfo{year}{2021}.
\newblock \bibinfo{title}{{Unveiling the distinct formation pathways of the inner and outer discs of the Milky Way with Bayesian Machine Learning}}.
\newblock \bibinfo{journal}{\mnras} \bibinfo{volume}{503}, \bibinfo{pages}{2814--2824}.
\newblock \DOIprefix\doi{10.1093/mnras/stab639}, \href{http://arxiv.org/abs/2003.03316}{{\tt arXiv:2003.03316}}.
\bibitem[{Cs{\'a}ji(2001)}]{csaji2001}
\bibinfo{author}{Cs{\'a}ji, B.C.}, \bibinfo{year}{2001}.
\newblock \bibinfo{title}{Approximation with artificial neural networks}.
\newblock Ph.D. thesis. Faculty of Sciences, E{\"o}tv{\"o}s Lor{\'a}nd University, Hungary.
\bibitem[{Cybenko(1989)}]{cybenko1989}
\bibinfo{author}{Cybenko, G.}, \bibinfo{year}{1989}.
\newblock \bibinfo{title}{Approximations by superpositions of sigmoidal functions}.
\newblock \bibinfo{journal}{Mathematics of Control, Signals, and Systems} \bibinfo{volume}{2}, \bibinfo{pages}{303--314}.
\newblock \DOIprefix\doi{10.1007/BF02551274}.
\bibitem[{{da Silva} et~al.(2006){da Silva}, {Girardi}, {Pasquini}, {Setiawan}, {von der L{\"u}he}, {de Medeiros}, {Hatzes}, {D{\"o}llinger} and {Weiss}}]{2006Silva}
\bibinfo{author}{{da Silva}, L.}, \bibinfo{author}{{Girardi}, L.}, \bibinfo{author}{{Pasquini}, L.}, \bibinfo{author}{{Setiawan}, J.}, \bibinfo{author}{{von der L{\"u}he}, O.}, \bibinfo{author}{{de Medeiros}, J.R.}, \bibinfo{author}{{Hatzes}, A.}, \bibinfo{author}{{D{\"o}llinger}, M.P.}, \bibinfo{author}{{Weiss}, A.}, \bibinfo{year}{2006}.
\newblock \bibinfo{title}{{Basic physical parameters of a selected sample of evolved stars}}.
\newblock \bibinfo{journal}{\aap} \bibinfo{volume}{458}, \bibinfo{pages}{609--623}.
\newblock \DOIprefix\doi{10.1051/0004-6361:20065105}, \href{http://arxiv.org/abs/astro-ph/0608160}{{\tt arXiv:astro-ph/0608160}}.
\bibitem[{{da Silva} et~al.(2012){da Silva}, {Porto de Mello}, {Milone}, {da Silva}, {Ribeiro} and {Rocha-Pinto}}]{2012Silva}
\bibinfo{author}{{da Silva}, R.}, \bibinfo{author}{{Porto de Mello}, G.F.}, \bibinfo{author}{{Milone}, A.C.}, \bibinfo{author}{{da Silva}, L.}, \bibinfo{author}{{Ribeiro}, L.S.}, \bibinfo{author}{{Rocha-Pinto}, H.J.}, \bibinfo{year}{2012}.
\newblock \bibinfo{title}{{Accurate and homogeneous abundance patterns in solar-type stars of the solar neighbourhood: a chemo-chronological analysis}}.
\newblock \bibinfo{journal}{\aap} \bibinfo{volume}{542}, \bibinfo{pages}{A84}.
\newblock \DOIprefix\doi{10.1051/0004-6361/201118751}, \href{http://arxiv.org/abs/1204.4433}{{\tt arXiv:1204.4433}}.
\bibitem[{Davidian(2003)}]{2002Stephen}
\bibinfo{author}{Davidian, M.}, \bibinfo{year}{2003}.
\newblock \bibinfo{title}{Hierarchical linear models: Applications and data analysis methods (2nd ed.). stephen w. raudenbush and anthony s. bryk}.
\newblock \bibinfo{journal}{Journal of the American Statistical Association} \bibinfo{volume}{98}, \bibinfo{pages}{767--768}.
\newblock \DOIprefix\doi{10.2307/30045307}.
\bibitem[{{Delgado Mena} et~al.(2019a){Delgado Mena}, {Moya}, {Adibekyan}, {Tsantaki}, {Gonzalez Hernandez}, {Israelian}, {Davies}, {Chaplin}, {Sousa}, {Ferreira} and {Santos}}]{2019Delgado2}
\bibinfo{author}{{Delgado Mena}, E.}, \bibinfo{author}{{Moya}, A.}, \bibinfo{author}{{Adibekyan}, V.}, \bibinfo{author}{{Tsantaki}, M.}, \bibinfo{author}{{Gonzalez Hernandez}, J.I.}, \bibinfo{author}{{Israelian}, G.}, \bibinfo{author}{{Davies}, G.R.}, \bibinfo{author}{{Chaplin}, W.J.}, \bibinfo{author}{{Sousa}, S.G.}, \bibinfo{author}{{Ferreira}, A.}, \bibinfo{author}{{Santos}, N.C.}, \bibinfo{year}{2019}a.
\newblock \bibinfo{title}{{VizieR Online Data Catalog: Masses and ages of 1059 HARPS-GTO stars (Delgado Mena+, 2019)}}.
\newblock \bibinfo{journal}{\aa} \DOIprefix\doi{10.26093/cds/vizier.36240078}.
\bibitem[{{Delgado Mena} et~al.(2019b){Delgado Mena}, {Moya}, {Adibekyan}, {Tsantaki}, {Gonz{\'a}lez Hern{\'a}ndez}, {Israelian}, {Davies}, {Chaplin}, {Sousa}, {Ferreira} and {Santos}}]{2019Delgado}
\bibinfo{author}{{Delgado Mena}, E.}, \bibinfo{author}{{Moya}, A.}, \bibinfo{author}{{Adibekyan}, V.}, \bibinfo{author}{{Tsantaki}, M.}, \bibinfo{author}{{Gonz{\'a}lez Hern{\'a}ndez}, J.I.}, \bibinfo{author}{{Israelian}, G.}, \bibinfo{author}{{Davies}, G.R.}, \bibinfo{author}{{Chaplin}, W.J.}, \bibinfo{author}{{Sousa}, S.G.}, \bibinfo{author}{{Ferreira}, A.C.S.}, \bibinfo{author}{{Santos}, N.C.}, \bibinfo{year}{2019}b.
\newblock \bibinfo{title}{{Abundance to age ratios in the HARPS-GTO sample with Gaia DR2. Chemical clocks for a range of [Fe/H]}}.
\newblock \bibinfo{journal}{\aap} \bibinfo{volume}{624}, \bibinfo{pages}{A78}.
\newblock \DOIprefix\doi{10.1051/0004-6361/201834783}, \href{http://arxiv.org/abs/1902.02127}{{\tt arXiv:1902.02127}}.
\bibitem[{{Delgado Mena} et~al.(2017){Delgado Mena}, {Tsantaki}, {Adibekyan}, {Sousa}, {Santos}, {Gonz{\'a}lez Hern{\'a}ndez} and {Israelian}}]{2017Delgado}
\bibinfo{author}{{Delgado Mena}, E.}, \bibinfo{author}{{Tsantaki}, M.}, \bibinfo{author}{{Adibekyan}, V.Z.}, \bibinfo{author}{{Sousa}, S.G.}, \bibinfo{author}{{Santos}, N.C.}, \bibinfo{author}{{Gonz{\'a}lez Hern{\'a}ndez}, J.I.}, \bibinfo{author}{{Israelian}, G.}, \bibinfo{year}{2017}.
\newblock \bibinfo{title}{{Chemical abundances of 1111 FGK stars from the HARPS GTO planet search program. II. Cu, Zn, Sr, Y, Zr, Ba, Ce, Nd, and Eu}}.
\newblock \bibinfo{journal}{\aap} \bibinfo{volume}{606}, \bibinfo{pages}{A94}.
\newblock \DOIprefix\doi{10.1051/0004-6361/201730535}, \href{http://arxiv.org/abs/1705.04349}{{\tt arXiv:1705.04349}}.
\bibitem[{{Dong} et~al.(2017){Dong}, {Su}, {Zhu} and {Bao}}]{2017Dong}
\bibinfo{author}{{Dong}, Y.}, \bibinfo{author}{{Su}, H.}, \bibinfo{author}{{Zhu}, J.}, \bibinfo{author}{{Bao}, F.}, \bibinfo{year}{2017}.
\newblock \bibinfo{title}{{Towards Interpretable Deep Neural Networks by Leveraging Adversarial Examples}}.
\newblock \bibinfo{journal}{arXiv e-prints} , \bibinfo{pages}{arXiv:1708.05493}\DOIprefix\doi{10.48550/arXiv.1708.05493}, \href{http://arxiv.org/abs/1708.05493}{{\tt arXiv:1708.05493}}.
\bibitem[{Dotter et~al.(2017)Dotter, Conroy, Cargile and Asplund}]{Dotter_2017}
\bibinfo{author}{Dotter, A.}, \bibinfo{author}{Conroy, C.}, \bibinfo{author}{Cargile, P.}, \bibinfo{author}{Asplund, M.}, \bibinfo{year}{2017}.
\newblock \bibinfo{title}{The influence of atomic diffusion on stellar ages and chemical tagging}.
\newblock \bibinfo{journal}{The Astrophysical Journal} \bibinfo{volume}{840}, \bibinfo{pages}{99}.
\newblock \URLprefix \url{https://dx.doi.org/10.3847/1538-4357/aa6d10}, \DOIprefix\doi{10.3847/1538-4357/aa6d10}.
\bibitem[{{Durasov} et~al.(2020){Durasov}, {Bagautdinov}, {Baque} and {Fua}}]{2020Durasov}
\bibinfo{author}{{Durasov}, N.}, \bibinfo{author}{{Bagautdinov}, T.}, \bibinfo{author}{{Baque}, P.}, \bibinfo{author}{{Fua}, P.}, \bibinfo{year}{2020}.
\newblock \bibinfo{title}{{Masksembles for Uncertainty Estimation}}.
\newblock \bibinfo{journal}{arXiv e-prints} , \bibinfo{pages}{arXiv:2012.08334}\DOIprefix\doi{10.48550/arXiv.2012.08334}, \href{http://arxiv.org/abs/2012.08334}{{\tt arXiv:2012.08334}}.
\bibitem[{{Fortuin}(2021)}]{2021Fortuin}
\bibinfo{author}{{Fortuin}, V.}, \bibinfo{year}{2021}.
\newblock \bibinfo{title}{{Priors in Bayesian Deep Learning: A Review}}.
\newblock \bibinfo{journal}{arXiv e-prints} , \bibinfo{pages}{arXiv:2105.06868}\DOIprefix\doi{10.48550/arXiv.2105.06868}, \href{http://arxiv.org/abs/2105.06868}{{\tt arXiv:2105.06868}}.
\bibitem[{{Gal} and {Ghahramani}(2015a)}]{2015Gal2}
\bibinfo{author}{{Gal}, Y.}, \bibinfo{author}{{Ghahramani}, Z.}, \bibinfo{year}{2015}a.
\newblock \bibinfo{title}{{Bayesian Convolutional Neural Networks with Bernoulli Approximate Variational Inference}}.
\newblock \bibinfo{journal}{arXiv e-prints} , \bibinfo{pages}{arXiv:1506.02158}\DOIprefix\doi{10.48550/arXiv.1506.02158}, \href{http://arxiv.org/abs/1506.02158}{{\tt arXiv:1506.02158}}.
\bibitem[{{Gal} and {Ghahramani}(2015b)}]{2015Gal}
\bibinfo{author}{{Gal}, Y.}, \bibinfo{author}{{Ghahramani}, Z.}, \bibinfo{year}{2015}b.
\newblock \bibinfo{title}{{Dropout as a Bayesian Approximation: Representing Model Uncertainty in Deep Learning}}.
\newblock \bibinfo{journal}{arXiv e-prints} , \bibinfo{pages}{arXiv:1506.02142}\DOIprefix\doi{10.48550/arXiv.1506.02142}, \href{http://arxiv.org/abs/1506.02142}{{\tt arXiv:1506.02142}}.
\bibitem[{{Gavel} et~al.(2021){Gavel}, {Gruyters}, {Heiter}, {Korn}, {Nordlander}, {Scheutwinkel} and {Richard}}]{2021Gavel}
\bibinfo{author}{{Gavel}, A.}, \bibinfo{author}{{Gruyters}, P.}, \bibinfo{author}{{Heiter}, U.}, \bibinfo{author}{{Korn}, A.J.}, \bibinfo{author}{{Nordlander}, T.}, \bibinfo{author}{{Scheutwinkel}, K.H.}, \bibinfo{author}{{Richard}, O.A.}, \bibinfo{year}{2021}.
\newblock \bibinfo{title}{{Atomic diffusion and mixing in old stars. VII. Abundances of Mg, Ti, and Fe in M 30}}.
\newblock \bibinfo{journal}{\aap} \bibinfo{volume}{652}, \bibinfo{pages}{A75}.
\newblock \DOIprefix\doi{10.1051/0004-6361/202140770}, \href{http://arxiv.org/abs/2110.12391}{{\tt arXiv:2110.12391}}.
\bibitem[{Gelman et~al.(2004)Gelman, Carlin, Stern and Rubin}]{gelmanbda04}
\bibinfo{author}{Gelman, A.}, \bibinfo{author}{Carlin, J.B.}, \bibinfo{author}{Stern, H.S.}, \bibinfo{author}{Rubin, D.B.}, \bibinfo{year}{2004}.
\newblock \bibinfo{title}{Bayesian Data Analysis}.
\newblock \bibinfo{edition}{2nd ed.} ed., \bibinfo{publisher}{Chapman and Hall/CRC}.
\bibitem[{Gelman and Hill(2006)}]{Gelman06}
\bibinfo{author}{Gelman, A.}, \bibinfo{author}{Hill, J.}, \bibinfo{year}{2006}.
\newblock \bibinfo{title}{Data Analysis Using Regression And Multilevel/Hierarchical Models}. volume~\bibinfo{volume}{3}.
\newblock \DOIprefix\doi{10.1017/CBO9780511790942}.
\bibitem[{Gelman et~al.(2015)Gelman, Lee and Guo}]{Gelman17}
\bibinfo{author}{Gelman, A.}, \bibinfo{author}{Lee, D.}, \bibinfo{author}{Guo, J.}, \bibinfo{year}{2015}.
\newblock \bibinfo{title}{Stan: A probabilistic programming language for bayesian inference and optimization}.
\newblock \bibinfo{journal}{Journal of Educational and Behavioral Statistics} \bibinfo{volume}{40}, \bibinfo{pages}{530--543}.
\newblock \URLprefix \url{https://doi.org/10.3102/1076998615606113}, \DOIprefix\doi{10.3102/1076998615606113}, \href{http://arxiv.org/abs/https://doi.org/10.3102/1076998615606113}{{\tt arXiv:https://doi.org/10.3102/1076998615606113}}.
\bibitem[{Gelman and Rubin(1992)}]{Gelman92}
\bibinfo{author}{Gelman, A.}, \bibinfo{author}{Rubin, D.B.}, \bibinfo{year}{1992}.
\newblock \bibinfo{title}{{Inference from Iterative Simulation Using Multiple Sequences}}.
\newblock \bibinfo{journal}{Statistical Science} \bibinfo{volume}{7}, \bibinfo{pages}{457 -- 472}.
\newblock \URLprefix \url{https://doi.org/10.1214/ss/1177011136}, \DOIprefix\doi{10.1214/ss/1177011136}.
\bibitem[{Geyer(1992)}]{1992Geyer}
\bibinfo{author}{Geyer, C.J.}, \bibinfo{year}{1992}.
\newblock \bibinfo{title}{{Practical Markov Chain Monte Carlo}}.
\newblock \bibinfo{journal}{Statistical Science} \bibinfo{volume}{7}, \bibinfo{pages}{473 -- 483}.
\newblock \URLprefix \url{https://doi.org/10.1214/ss/1177011137}, \DOIprefix\doi{10.1214/ss/1177011137}.
\bibitem[{{Goupil} et~al.(2024){Goupil}, {Catala}, {Samadi}, {Belkacem}, {Ouazzani}, {Reese}, {Appourchaux}, {Mathur}, {Cabrera}, {B{\"o}rner}, {Paproth}, {Moedas}, {Verma}, {Lebreton}, {Deal}, {Ballot}, {Chaplin}, {Christensen-Dalsgaard}, {Cunha}, {Lanza}, {Miglio}, {Morel}, {Serenelli}, {Mosser}, {Creevey}, {Moya}, {Garcia}, {Nielsen} and {Hatt}}]{2024Goupil}
\bibinfo{author}{{Goupil}, M.J.}, \bibinfo{author}{{Catala}, C.}, \bibinfo{author}{{Samadi}, R.}, \bibinfo{author}{{Belkacem}, K.}, \bibinfo{author}{{Ouazzani}, R.M.}, \bibinfo{author}{{Reese}, D.R.}, \bibinfo{author}{{Appourchaux}, T.}, \bibinfo{author}{{Mathur}, S.}, \bibinfo{author}{{Cabrera}, J.}, \bibinfo{author}{{B{\"o}rner}, A.}, \bibinfo{author}{{Paproth}, C.}, \bibinfo{author}{{Moedas}, N.}, \bibinfo{author}{{Verma}, K.}, \bibinfo{author}{{Lebreton}, Y.}, \bibinfo{author}{{Deal}, M.}, \bibinfo{author}{{Ballot}, J.}, \bibinfo{author}{{Chaplin}, W.J.}, \bibinfo{author}{{Christensen-Dalsgaard}, J.}, \bibinfo{author}{{Cunha}, M.}, \bibinfo{author}{{Lanza}, A.F.}, \bibinfo{author}{{Miglio}, A.}, \bibinfo{author}{{Morel}, T.}, \bibinfo{author}{{Serenelli}, A.}, \bibinfo{author}{{Mosser}, B.}, \bibinfo{author}{{Creevey}, O.}, \bibinfo{author}{{Moya}, A.}, \bibinfo{author}{{Garcia}, R.A.}, \bibinfo{author}{{Nielsen}, M.B.}, \bibinfo{author}{{Hatt}, E.}, \bibinfo{year}{2024}.
\newblock \bibinfo{title}{{Predicted asteroseismic detection yield for solar-like oscillating stars with PLATO}}.
\newblock \bibinfo{journal}{arXiv e-prints} , \bibinfo{pages}{arXiv:2401.07984}\DOIprefix\doi{10.48550/arXiv.2401.07984}, \href{http://arxiv.org/abs/2401.07984}{{\tt arXiv:2401.07984}}.
\bibitem[{{Gu{\'e}d{\'e}} et~al.(2012){Gu{\'e}d{\'e}}, {Lebreton}, {Babusiaux} and {Haywood}}]{2012Guede}
\bibinfo{author}{{Gu{\'e}d{\'e}}, C.}, \bibinfo{author}{{Lebreton}, Y.}, \bibinfo{author}{{Babusiaux}, C.}, \bibinfo{author}{{Haywood}, M.}, \bibinfo{year}{2012}.
\newblock \bibinfo{title}{{Age dating large samples of stars: Ways toward improved accuracy}}, in: \bibinfo{editor}{{Boissier}, S.}, \bibinfo{editor}{{de Laverny}, P.}, \bibinfo{editor}{{Nardetto}, N.}, \bibinfo{editor}{{Samadi}, R.}, \bibinfo{editor}{{Valls-Gabaud}, D.}, \bibinfo{editor}{{Wozniak}, H.} (Eds.), \bibinfo{booktitle}{SF2A-2012: Proceedings of the Annual meeting of the French Society of Astronomy and Astrophysics}, pp. \bibinfo{pages}{195--198}.
\bibitem[{{Hoffman} and {Gelman}(2011)}]{2011Hoffman}
\bibinfo{author}{{Hoffman}, M.D.}, \bibinfo{author}{{Gelman}, A.}, \bibinfo{year}{2011}.
\newblock \bibinfo{title}{{The No-U-Turn Sampler: Adaptively Setting Path Lengths in Hamiltonian Monte Carlo}}.
\newblock \bibinfo{journal}{arXiv e-prints} , \bibinfo{pages}{arXiv:1111.4246}\DOIprefix\doi{10.48550/arXiv.1111.4246}, \href{http://arxiv.org/abs/1111.4246}{{\tt arXiv:1111.4246}}.
\bibitem[{Hornik(1991)}]{hornik1991}
\bibinfo{author}{Hornik, K.}, \bibinfo{year}{1991}.
\newblock \bibinfo{title}{Approximation capabilities of multilayer feedforward networks}.
\newblock \bibinfo{journal}{Neural Networks} \bibinfo{volume}{4}, \bibinfo{pages}{251--257}.
\newblock \DOIprefix\doi{10.1016/0893-6080(91)90009-T}.
\bibitem[{{Izmailov} et~al.(2021){Izmailov}, {Vikram}, {Hoffman} and {Wilson}}]{2021Izmailov}
\bibinfo{author}{{Izmailov}, P.}, \bibinfo{author}{{Vikram}, S.}, \bibinfo{author}{{Hoffman}, M.D.}, \bibinfo{author}{{Wilson}, A.G.}, \bibinfo{year}{2021}.
\newblock \bibinfo{title}{{What Are Bayesian Neural Network Posteriors Really Like?}}
\newblock \bibinfo{journal}{arXiv e-prints} , \bibinfo{pages}{arXiv:2104.14421}\DOIprefix\doi{10.48550/arXiv.2104.14421}, \href{http://arxiv.org/abs/2104.14421}{{\tt arXiv:2104.14421}}.
\bibitem[{{Jeffery} et~al.(2016){Jeffery}, {von Hippel}, {van Dyk}, {Stenning}, {Robinson}, {Stein} and {Jefferys}}]{2016Jeffery}
\bibinfo{author}{{Jeffery}, E.J.}, \bibinfo{author}{{von Hippel}, T.}, \bibinfo{author}{{van Dyk}, D.A.}, \bibinfo{author}{{Stenning}, D.C.}, \bibinfo{author}{{Robinson}, E.}, \bibinfo{author}{{Stein}, N.}, \bibinfo{author}{{Jefferys}, W.H.}, \bibinfo{year}{2016}.
\newblock \bibinfo{title}{{A Bayesian Analysis of the Ages of Four Open Clusters}}.
\newblock \bibinfo{journal}{\aj} \bibinfo{volume}{828}, \bibinfo{pages}{79}.
\newblock \DOIprefix\doi{10.3847/0004-637X/828/2/79}, \href{http://arxiv.org/abs/1611.00835}{{\tt arXiv:1611.00835}}.
\bibitem[{{Kiman} et~al.(2021){Kiman}, {Faherty}, {Cruz}, {Gagn{\'e}}, {Angus}, {Schmidt}, {Mann}, {Bardalez Gagliuffi} and {Rice}}]{2021Kiman}
\bibinfo{author}{{Kiman}}, \bibinfo{author}{{Faherty}}, \bibinfo{author}{{Cruz}}, \bibinfo{author}{{Gagn{\'e}}}, \bibinfo{author}{{Angus}}, \bibinfo{author}{{Schmidt}}, \bibinfo{author}{{Mann}}, \bibinfo{author}{{Bardalez Gagliuffi}}, \bibinfo{author}{{Rice}}, \bibinfo{year}{2021}.
\newblock \bibinfo{title}{{Age-relations for low-mass stars}}, in: \bibinfo{booktitle}{The 20.5th Cambridge Workshop on Cool Stars, Stellar Systems, and the Sun (CS20.5)}, p. \bibinfo{pages}{264}.
\newblock \DOIprefix\doi{10.5281/zenodo.4567649}.
\bibitem[{{Kiman} et~al.(2020){Kiman}, {Angus}, {Xu}, {Schmidt}, {Gagn{\'e}}, {Cruz}, {Faherty} and {Rice}}]{2020Kiman}
\bibinfo{author}{{Kiman}, R.}, \bibinfo{author}{{Angus}, R.}, \bibinfo{author}{{Xu}, S.}, \bibinfo{author}{{Schmidt}, S.}, \bibinfo{author}{{Gagn{\'e}}, J.}, \bibinfo{author}{{Cruz}, K.}, \bibinfo{author}{{Faherty}, J.}, \bibinfo{author}{{Rice}, E.}, \bibinfo{year}{2020}.
\newblock \bibinfo{title}{{Age-Activity relation for M dwarfs using H{\ensuremath{\alpha}} equivalent widths}}, in: \bibinfo{booktitle}{American Astronomical Society Meeting Abstracts \#235}, p. \bibinfo{pages}{274.16}.
\bibitem[{{Kiman} et~al.(2022){Kiman}, {Xu}, {Faherty}, {Gagn{\'e}}, {Angus}, {Brandt}, {Casewell} and {Cruz}}]{2022Kiman}
\bibinfo{author}{{Kiman}, R.}, \bibinfo{author}{{Xu}, S.}, \bibinfo{author}{{Faherty}, J.K.}, \bibinfo{author}{{Gagn{\'e}}, J.}, \bibinfo{author}{{Angus}, R.}, \bibinfo{author}{{Brandt}, T.D.}, \bibinfo{author}{{Casewell}, S.L.}, \bibinfo{author}{{Cruz}, K.L.}, \bibinfo{year}{2022}.
\newblock \bibinfo{title}{{wdwarfdate: A Python Package to Derive Bayesian Ages of White Dwarfs}}.
\newblock \bibinfo{journal}{\aj} \bibinfo{volume}{164}, \bibinfo{pages}{62}.
\newblock \DOIprefix\doi{10.3847/1538-3881/ac7788}, \href{http://arxiv.org/abs/2206.05388}{{\tt arXiv:2206.05388}}.
\bibitem[{{Kirkpatrick} et~al.(2017){Kirkpatrick}, {Pascanu}, {Rabinowitz}, {Veness}, {Desjardins}, {Rusu}, {Milan}, {Quan}, {Ramalho}, {Grabska-Barwinska}, {Hassabis}, {Clopath}, {Kumaran} and {Hadsell}}]{2017Kirkpatrick}
\bibinfo{author}{{Kirkpatrick}, J.}, \bibinfo{author}{{Pascanu}, R.}, \bibinfo{author}{{Rabinowitz}, N.}, \bibinfo{author}{{Veness}, J.}, \bibinfo{author}{{Desjardins}, G.}, \bibinfo{author}{{Rusu}, A.A.}, \bibinfo{author}{{Milan}, K.}, \bibinfo{author}{{Quan}, J.}, \bibinfo{author}{{Ramalho}, T.}, \bibinfo{author}{{Grabska-Barwinska}, A.}, \bibinfo{author}{{Hassabis}, D.}, \bibinfo{author}{{Clopath}, C.}, \bibinfo{author}{{Kumaran}, D.}, \bibinfo{author}{{Hadsell}, R.}, \bibinfo{year}{2017}.
\newblock \bibinfo{title}{{Overcoming catastrophic forgetting in neural networks}}.
\newblock \bibinfo{journal}{Proceedings of the National Academy of Science} \bibinfo{volume}{114}, \bibinfo{pages}{3521--3526}.
\newblock \DOIprefix\doi{10.1073/pnas.1611835114}, \href{http://arxiv.org/abs/1612.00796}{{\tt arXiv:1612.00796}}.
\bibitem[{Krizhevsky et~al.(2012)Krizhevsky, Sutskever and Hinton}]{Krizhevsky12}
\bibinfo{author}{Krizhevsky, A.}, \bibinfo{author}{Sutskever, I.}, \bibinfo{author}{Hinton, G.E.}, \bibinfo{year}{2012}.
\newblock \bibinfo{title}{ImageNet Classification with Deep Convolutional Neural Networks}. volume~\bibinfo{volume}{25}.
\newblock \bibinfo{publisher}{Curran Associates, Inc.}
\newblock \URLprefix \url{https://proceedings.neurips.cc/paper_files/paper/2012/file/c399862d3b9d6b76c8436e924a68c45b-Paper.pdf}.
\bibitem[{{Kurucz}(1993)}]{1993Kurucz}
\bibinfo{author}{{Kurucz}, R.}, \bibinfo{year}{1993}.
\newblock \bibinfo{title}{{ATLAS9 Stellar Atmosphere Programs and 2 km/s grid.}}
\newblock \bibinfo{journal}{ATLAS9 Stellar Atmosphere Programs and 2 km/s grid. Kurucz CD-ROM No. 13. Cambridge} \bibinfo{volume}{13}.
\bibitem[{{Lakshminarayanan} et~al.(2016){Lakshminarayanan}, {Pritzel} and {Blundell}}]{2016Lakshminarayanan}
\bibinfo{author}{{Lakshminarayanan}, B.}, \bibinfo{author}{{Pritzel}, A.}, \bibinfo{author}{{Blundell}, C.}, \bibinfo{year}{2016}.
\newblock \bibinfo{title}{{Simple and Scalable Predictive Uncertainty Estimation using Deep Ensembles}}.
\newblock \bibinfo{journal}{arXiv e-prints} , \bibinfo{pages}{arXiv:1612.01474}\DOIprefix\doi{10.48550/arXiv.1612.01474}, \href{http://arxiv.org/abs/1612.01474}{{\tt arXiv:1612.01474}}.
\bibitem[{Lecun et~al.(1998)Lecun, Bottou, Bengio and Haffner}]{Lecun98}
\bibinfo{author}{Lecun, Y.}, \bibinfo{author}{Bottou, L.}, \bibinfo{author}{Bengio, Y.}, \bibinfo{author}{Haffner, P.}, \bibinfo{year}{1998}.
\newblock \bibinfo{title}{Gradient-based learning applied to document recognition}.
\newblock \bibinfo{journal}{Proceedings of the IEEE} \bibinfo{volume}{86}, \bibinfo{pages}{2278--2324}.
\newblock \DOIprefix\doi{10.1109/5.726791}.
\bibitem[{Lewandowski et~al.(2009)Lewandowski, Kurowicka and Joe}]{LEWANDOWSKI2009}
\bibinfo{author}{Lewandowski, D.}, \bibinfo{author}{Kurowicka, D.}, \bibinfo{author}{Joe, H.}, \bibinfo{year}{2009}.
\newblock \bibinfo{title}{Generating random correlation matrices based on vines and extended onion method}.
\newblock \bibinfo{journal}{Journal of Multivariate Analysis} \bibinfo{volume}{100}, \bibinfo{pages}{1989--2001}.
\newblock \URLprefix \url{https://www.sciencedirect.com/science/article/pii/S0047259X09000876}, \DOIprefix\doi{https://doi.org/10.1016/j.jmva.2009.04.008}.
\bibitem[{{Lin} et~al.(2018){Lin}, {Dotter}, {Ting} and {Asplund}}]{2018Lin}
\bibinfo{author}{{Lin}, J.}, \bibinfo{author}{{Dotter}, A.}, \bibinfo{author}{{Ting}, Y.S.}, \bibinfo{author}{{Asplund}, M.}, \bibinfo{year}{2018}.
\newblock \bibinfo{title}{{Stellar ages and masses in the solar neighbourhood: Bayesian analysis using spectroscopy and Gaia DR1 parallaxes}}.
\newblock \bibinfo{journal}{\mnras} \bibinfo{volume}{477}, \bibinfo{pages}{2966--2975}.
\newblock \DOIprefix\doi{10.1093/mnras/sty709}, \href{http://arxiv.org/abs/1803.10875}{{\tt arXiv:1803.10875}}.
\bibitem[{Liu et~al.(2016)Liu, Asplund, Yong, Meléndez, Ramírez, Karakas, Carlos and Marino}]{2016Liu}
\bibinfo{author}{Liu, F.}, \bibinfo{author}{Asplund, M.}, \bibinfo{author}{Yong, D.}, \bibinfo{author}{Meléndez, J.}, \bibinfo{author}{Ramírez, I.}, \bibinfo{author}{Karakas, A.I.}, \bibinfo{author}{Carlos, M.}, \bibinfo{author}{Marino, A.F.}, \bibinfo{year}{2016}.
\newblock \bibinfo{title}{{The chemical compositions of solar twins in the open cluster M67}}.
\newblock \bibinfo{journal}{Monthly Notices of the Royal Astronomical Society} \bibinfo{volume}{463}, \bibinfo{pages}{696--704}.
\newblock \URLprefix \url{https://doi.org/10.1093/mnras/stw2045}, \DOIprefix\doi{10.1093/mnras/stw2045}.
\bibitem[{{Liu} and {Wang}(2016)}]{2016Liu2}
\bibinfo{author}{{Liu}, Q.}, \bibinfo{author}{{Wang}, D.}, \bibinfo{year}{2016}.
\newblock \bibinfo{title}{{Stein Variational Gradient Descent: A General Purpose Bayesian Inference Algorithm}}.
\newblock \bibinfo{journal}{arXiv e-prints} , \bibinfo{pages}{arXiv:1608.04471}\DOIprefix\doi{10.48550/arXiv.1608.04471}, \href{http://arxiv.org/abs/1608.04471}{{\tt arXiv:1608.04471}}.
\bibitem[{{Lu} et~al.(2024){Lu}, {Angus}, {Foreman-Mackey} and {Hattori}}]{2024Lu}
\bibinfo{author}{{Lu}, Y.}, \bibinfo{author}{{Angus}, R.}, \bibinfo{author}{{Foreman-Mackey}, D.}, \bibinfo{author}{{Hattori}, S.}, \bibinfo{year}{2024}.
\newblock \bibinfo{title}{{In This Day and Age: An Empirical Gyrochronology Relation for Partially and Fully Convective Single Field Stars}}.
\newblock \bibinfo{journal}{\aj} \bibinfo{volume}{167}, \bibinfo{pages}{159}.
\newblock \DOIprefix\doi{10.3847/1538-3881/ad28b9}, \href{http://arxiv.org/abs/2310.14990}{{\tt arXiv:2310.14990}}.
\bibitem[{MacKay(1992)}]{mackay1992}
\bibinfo{author}{MacKay, D.J.}, \bibinfo{year}{1992}.
\newblock \bibinfo{title}{Bayesian interpolation}.
\newblock \bibinfo{journal}{Neural computation} \bibinfo{volume}{4}, \bibinfo{pages}{415--447}.
\bibitem[{{Mathur}(2013)}]{2013Mathur}
\bibinfo{author}{{Mathur}, S.}, \bibinfo{year}{2013}.
\newblock \bibinfo{title}{{Study of Stellar Magnetic Activity with CoRoT and mbox\{Kepler\} Data}}, in: \bibinfo{editor}{{Shibahashi}, H.}, \bibinfo{editor}{{Lynas-Gray}, A.E.} (Eds.), \bibinfo{booktitle}{Progress in Physics of the Sun and Stars: A New Era in Helio- and Asteroseismology}, p. \bibinfo{pages}{425}.
\newblock \DOIprefix\doi{10.48550/arXiv.1308.0645}, \href{http://arxiv.org/abs/1308.0645}{{\tt arXiv:1308.0645}}.
\bibitem[{{Mathur} et~al.(2023a){Mathur}, {Claytor}, {Santos}, {Garc{\'\i}a}, {Amard}, {Bugnet}, {Corsaro}, {Bonanno}, {Breton}, {Godoy-Rivera}, {Pinsonneault} and {van Saders}}]{2023Mathur}
\bibinfo{author}{{Mathur}, S.}, \bibinfo{author}{{Claytor}, Z.R.}, \bibinfo{author}{{Santos}, {\^A}.R.G.}, \bibinfo{author}{{Garc{\'\i}a}, R.A.}, \bibinfo{author}{{Amard}, L.}, \bibinfo{author}{{Bugnet}, L.}, \bibinfo{author}{{Corsaro}, E.}, \bibinfo{author}{{Bonanno}, A.}, \bibinfo{author}{{Breton}, S.N.}, \bibinfo{author}{{Godoy-Rivera}, D.}, \bibinfo{author}{{Pinsonneault}, M.H.}, \bibinfo{author}{{van Saders}, J.}, \bibinfo{year}{2023}a.
\newblock \bibinfo{title}{{Evolution of rotation and magnetic activity of solar-like stars with age:magneto-(gyro-)chronology}}, in: \bibinfo{booktitle}{PLATO Stellar Science Conference 2023}, p.~\bibinfo{pages}{36}.
\newblock \DOIprefix\doi{10.5281/zenodo.8140785}.
\bibitem[{{Mathur} et~al.(2023b){Mathur}, {Claytor}, {Santos}, {Garc{\'\i}a}, {Amard}, {Bugnet}, {Corsaro}, {Bonanno}, {Breton}, {Godoy-Rivera}, {Pinsonneault} and {van Saders}}]{2023Mathur2}
\bibinfo{author}{{Mathur}, S.}, \bibinfo{author}{{Claytor}, Z.R.}, \bibinfo{author}{{Santos}, {\^A}.R.G.}, \bibinfo{author}{{Garc{\'\i}a}, R.A.}, \bibinfo{author}{{Amard}, L.}, \bibinfo{author}{{Bugnet}, L.}, \bibinfo{author}{{Corsaro}, E.}, \bibinfo{author}{{Bonanno}, A.}, \bibinfo{author}{{Breton}, S.N.}, \bibinfo{author}{{Godoy-Rivera}, D.}, \bibinfo{author}{{Pinsonneault}, M.H.}, \bibinfo{author}{{van Saders}, J.}, \bibinfo{year}{2023}b.
\newblock \bibinfo{title}{{Magnetic Activity Evolution of Solar-like Stars. I. S $_{ph}$-Age Relation Derived from Kepler Observations}}.
\newblock \bibinfo{journal}{\aj} \bibinfo{volume}{952}, \bibinfo{pages}{131}.
\newblock \DOIprefix\doi{10.3847/1538-4357/acd118}, \href{http://arxiv.org/abs/2306.11657}{{\tt arXiv:2306.11657}}.
\bibitem[{McCulloch and Pitts(1943)}]{McCulloch1943}
\bibinfo{author}{McCulloch, W.S.}, \bibinfo{author}{Pitts, W.}, \bibinfo{year}{1943}.
\newblock \bibinfo{title}{A logical calculus of the ideas immanent in nervous activity}.
\newblock \bibinfo{journal}{The bulletin of mathematical biophysics} \bibinfo{volume}{5}, \bibinfo{pages}{115--133}.
\newblock \URLprefix \url{https://doi.org/10.1007/BF02478259}, \DOIprefix\doi{10.1007/BF02478259}.
\bibitem[{{Morel} et~al.(2021){Morel}, {Creevey}, {Montalb{\'a}n}, {Miglio} and {Willett}}]{2021Morel}
\bibinfo{author}{{Morel}, T.}, \bibinfo{author}{{Creevey}, O.L.}, \bibinfo{author}{{Montalb{\'a}n}, J.}, \bibinfo{author}{{Miglio}, A.}, \bibinfo{author}{{Willett}, E.}, \bibinfo{year}{2021}.
\newblock \bibinfo{title}{{Testing abundance-age relations beyond solar analogues with Kepler LEGACY stars}}.
\newblock \bibinfo{journal}{\aap} \bibinfo{volume}{646}, \bibinfo{pages}{A78}.
\newblock \DOIprefix\doi{10.1051/0004-6361/202039212}, \href{http://arxiv.org/abs/2011.04481}{{\tt arXiv:2011.04481}}.
\bibitem[{{Moya} et~al.(2010){Moya}, {Amado}, {Barrado}, {Garc{\'\i}a Hern{\'a}ndez}, {Aberasturi}, {Montesinos} and {Aceituno}}]{2010Moya}
\bibinfo{author}{{Moya}, A.}, \bibinfo{author}{{Amado}, P.J.}, \bibinfo{author}{{Barrado}, D.}, \bibinfo{author}{{Garc{\'\i}a Hern{\'a}ndez}, A.}, \bibinfo{author}{{Aberasturi}, M.}, \bibinfo{author}{{Montesinos}, B.}, \bibinfo{author}{{Aceituno}, F.}, \bibinfo{year}{2010}.
\newblock \bibinfo{title}{{Age determination of the HR8799 planetary system using asteroseismology}}.
\newblock \bibinfo{journal}{\mnras} \bibinfo{volume}{405}, \bibinfo{pages}{L81--L85}.
\newblock \DOIprefix\doi{10.1111/j.1745-3933.2010.00863.x}, \href{http://arxiv.org/abs/1003.5796}{{\tt arXiv:1003.5796}}.
\bibitem[{Moya et~al.(2021)Moya, Recio-Martínez and López-Sastre}]{2021Moya}
\bibinfo{author}{Moya, A.}, \bibinfo{author}{Recio-Martínez, J.}, \bibinfo{author}{López-Sastre, R.J.}, \bibinfo{year}{2021}.
\newblock \bibinfo{title}{Ai for dating stars: a benchmarking study for gyrochronology}, in: \bibinfo{booktitle}{2021 IEEE/CVF Conference on Computer Vision and Pattern Recognition Workshops (CVPRW)}, pp. \bibinfo{pages}{1971--1981}.
\newblock \DOIprefix\doi{10.1109/CVPRW53098.2021.00225}.
\bibitem[{{Moya} et~al.(2022){Moya}, {Sarro}, {Delgado-Mena}, {Chaplin}, {Adibekyan} and {Blanco-Cuaresma}}]{Moya2022}
\bibinfo{author}{{Moya}, A.}, \bibinfo{author}{{Sarro}, L.M.}, \bibinfo{author}{{Delgado-Mena}, E.}, \bibinfo{author}{{Chaplin}, W.J.}, \bibinfo{author}{{Adibekyan}, V.}, \bibinfo{author}{{Blanco-Cuaresma}, S.}, \bibinfo{year}{2022}.
\newblock \bibinfo{title}{{Stellar dating using chemical clocks and Bayesian inference}}.
\newblock \bibinfo{journal}{\aap} \bibinfo{volume}{660}, \bibinfo{pages}{A15}.
\newblock \DOIprefix\doi{10.1051/0004-6361/202141125}, \href{http://arxiv.org/abs/2201.05228}{{\tt arXiv:2201.05228}}.
\bibitem[{Müller and Insua(1998)}]{1998Muller}
\bibinfo{author}{Müller, P.}, \bibinfo{author}{Insua, D.R.}, \bibinfo{year}{1998}.
\newblock \bibinfo{title}{{Issues in Bayesian Analysis of Neural Network Models}}.
\newblock \bibinfo{journal}{Neural Computation} \bibinfo{volume}{10}, \bibinfo{pages}{749--770}.
\newblock \URLprefix \url{https://doi.org/10.1162/089976698300017737}, \DOIprefix\doi{10.1162/089976698300017737}.
\bibitem[{Neal(1993)}]{neal1993}
\bibinfo{author}{Neal, R.M.}, \bibinfo{year}{1993}.
\newblock \bibinfo{title}{Probabilistic Inference Using Markov Chain Monte Carlo Methods}.
\newblock \bibinfo{type}{Technical Report} \bibinfo{number}{CRG-TR-93-1}. Department of Computer Science, University of Toronto.
\bibitem[{Neal(1996)}]{Neal1996}
\bibinfo{author}{Neal, R.M.}, \bibinfo{year}{1996}.
\newblock \bibinfo{title}{Bayesian Learning for Neural Networks}.
\newblock \bibinfo{publisher}{Springer-Verlag}, \bibinfo{address}{Berlin, Heidelberg}.
\bibitem[{Neal(2012)}]{neal2012bayesian}
\bibinfo{author}{Neal, R.M.}, \bibinfo{year}{2012}.
\newblock \bibinfo{title}{Bayesian learning for neural networks}. volume \bibinfo{volume}{118}.
\newblock \bibinfo{publisher}{Springer Science \& Business Media}.
\bibitem[{{Nissen}(2015)}]{2015Nissen}
\bibinfo{author}{{Nissen}, P.E.}, \bibinfo{year}{2015}.
\newblock \bibinfo{title}{{High-precision abundances of elements in solar twin stars. Trends with stellar age and elemental condensation temperature}}.
\newblock \bibinfo{journal}{\aap} \bibinfo{volume}{579}, \bibinfo{pages}{A52}.
\newblock \DOIprefix\doi{10.1051/0004-6361/201526269}, \href{http://arxiv.org/abs/1504.07598}{{\tt arXiv:1504.07598}}.
\bibitem[{{Nissen}(2016)}]{2016Nissen}
\bibinfo{author}{{Nissen}, P.E.}, \bibinfo{year}{2016}.
\newblock \bibinfo{title}{{High-precision abundances of Sc, Mn, Cu, and Ba in solar twins. Trends of element ratios with stellar age}}.
\newblock \bibinfo{journal}{\aap} \bibinfo{volume}{593}, \bibinfo{pages}{A65}.
\newblock \DOIprefix\doi{10.1051/0004-6361/201628888}, \href{http://arxiv.org/abs/1606.08399}{{\tt arXiv:1606.08399}}.
\bibitem[{{Nissen} et~al.(2020){Nissen}, {Christensen-Dalsgaard}, {Mosumgaard}, {Silva Aguirre}, {Spitoni} and {Verma}}]{2020Nissen}
\bibinfo{author}{{Nissen}, P.E.}, \bibinfo{author}{{Christensen-Dalsgaard}, J.}, \bibinfo{author}{{Mosumgaard}, J.R.}, \bibinfo{author}{{Silva Aguirre}, V.}, \bibinfo{author}{{Spitoni}, E.}, \bibinfo{author}{{Verma}, K.}, \bibinfo{year}{2020}.
\newblock \bibinfo{title}{{High-precision abundances of elements in solar-type stars. Evidence of two distinct sequences in abundance-age relations}}.
\newblock \bibinfo{journal}{\aap} \bibinfo{volume}{640}, \bibinfo{pages}{A81}.
\newblock \DOIprefix\doi{10.1051/0004-6361/202038300}, \href{http://arxiv.org/abs/2006.06013}{{\tt arXiv:2006.06013}}.
\bibitem[{{Nissen} et~al.(2017){Nissen}, {Silva Aguirre}, {Christensen-Dalsgaard}, {Collet}, {Grundahl} and {Slumstrup}}]{2017Nissen}
\bibinfo{author}{{Nissen}, P.E.}, \bibinfo{author}{{Silva Aguirre}, V.}, \bibinfo{author}{{Christensen-Dalsgaard}, J.}, \bibinfo{author}{{Collet}, R.}, \bibinfo{author}{{Grundahl}, F.}, \bibinfo{author}{{Slumstrup}, D.}, \bibinfo{year}{2017}.
\newblock \bibinfo{title}{{High-precision abundances of elements in Kepler LEGACY stars. Verification of trends with stellar age}}.
\newblock \bibinfo{journal}{\aap} \bibinfo{volume}{608}, \bibinfo{pages}{A112}.
\newblock \DOIprefix\doi{10.1051/0004-6361/201731845}, \href{http://arxiv.org/abs/1710.03544}{{\tt arXiv:1710.03544}}.
\bibitem[{{Noci} et~al.(2021){Noci}, {Roth}, {Bachmann}, {Nowozin} and {Hofmann}}]{2021Noci}
\bibinfo{author}{{Noci}, L.}, \bibinfo{author}{{Roth}, K.}, \bibinfo{author}{{Bachmann}, G.}, \bibinfo{author}{{Nowozin}, S.}, \bibinfo{author}{{Hofmann}, T.}, \bibinfo{year}{2021}.
\newblock \bibinfo{title}{{Disentangling the Roles of Curation, Data-Augmentation and the Prior in the Cold Posterior Effect}}.
\newblock \bibinfo{journal}{arXiv e-prints} , \bibinfo{pages}{arXiv:2106.06596}\DOIprefix\doi{10.48550/arXiv.2106.06596}, \href{http://arxiv.org/abs/2106.06596}{{\tt arXiv:2106.06596}}.
\bibitem[{Ojha et~al.(2017)Ojha, Abraham and Snášel}]{Varun2017}
\bibinfo{author}{Ojha, V.K.}, \bibinfo{author}{Abraham, A.}, \bibinfo{author}{Snášel, V.}, \bibinfo{year}{2017}.
\newblock \bibinfo{title}{Metaheuristic design of feedforward neural networks: A review of two decades of research}.
\newblock \bibinfo{journal}{Engineering Applications of Artificial Intelligence} \bibinfo{volume}{60}, \bibinfo{pages}{97--116}.
\newblock \URLprefix \url{https://www.sciencedirect.com/science/article/pii/S0952197617300234}, \DOIprefix\doi{https://doi.org/10.1016/j.engappai.2017.01.013}.
\bibitem[{Olivares~Romero(2017)}]{2017olivaresromero}
\bibinfo{author}{Olivares~Romero, J.}, \bibinfo{year}{2017}.
\newblock \bibinfo{title}{{Bayesian hierarchical modelling of young stellar clusters}}.
\newblock \bibinfo{type}{Theses}. {Universit{\'e} Grenoble Alpes ; Universidad nacional de educaci{\'o}n a distancia (Madrid)}.
\newblock \URLprefix \url{https://theses.hal.science/tel-01762982}.
\bibitem[{Papamarkou et~al.(2021)Papamarkou, Hinkle, Young and Womble}]{papamarkou2021challenges}
\bibinfo{author}{Papamarkou, T.}, \bibinfo{author}{Hinkle, J.}, \bibinfo{author}{Young, M.T.}, \bibinfo{author}{Womble, D.}, \bibinfo{year}{2021}.
\newblock \bibinfo{title}{Challenges in markov chain monte carlo for bayesian neural networks}.
\newblock \href{http://arxiv.org/abs/1910.06539}{{\tt arXiv:1910.06539}}.
\bibitem[{Pourzanjani et~al.(2017)Pourzanjani, Jiang and Petzold}]{Pourzanjani2017ImprovingTI}
\bibinfo{author}{Pourzanjani, A.A.}, \bibinfo{author}{Jiang, R.M.}, \bibinfo{author}{Petzold, L.R.}, \bibinfo{year}{2017}.
\newblock \bibinfo{title}{Improving the identifiability of neural networks for bayesian inference}.
\newblock \bibinfo{journal}{Second workshop on Bayesian Deep Learning (NIPS 2017), Long Beach, CA, USA.} \URLprefix \url{https://api.semanticscholar.org/CorpusID:46932278}.
\bibitem[{{Pr{\v{s}}a} et~al.(2016){Pr{\v{s}}a}, {Harmanec}, {Torres}, {Mamajek}, {Asplund}, {Capitaine}, {Christensen-Dalsgaard}, {Depagne}, {Haberreiter}, {Hekker}, {Hilton}, {Kopp}, {Kostov}, {Kurtz}, {Laskar}, {Mason}, {Milone}, {Montgomery}, {Richards}, {Schmutz}, {Schou} and {Stewart}}]{2016Prvsa}
\bibinfo{author}{{Pr{\v{s}}a}, A.}, \bibinfo{author}{{Harmanec}, P.}, \bibinfo{author}{{Torres}, G.}, \bibinfo{author}{{Mamajek}, E.}, \bibinfo{author}{{Asplund}, M.}, \bibinfo{author}{{Capitaine}, N.}, \bibinfo{author}{{Christensen-Dalsgaard}, J.}, \bibinfo{author}{{Depagne}, {\'E}.}, \bibinfo{author}{{Haberreiter}, M.}, \bibinfo{author}{{Hekker}, S.}, \bibinfo{author}{{Hilton}, J.}, \bibinfo{author}{{Kopp}, G.}, \bibinfo{author}{{Kostov}, V.}, \bibinfo{author}{{Kurtz}, D.W.}, \bibinfo{author}{{Laskar}, J.}, \bibinfo{author}{{Mason}, B.D.}, \bibinfo{author}{{Milone}, E.F.}, \bibinfo{author}{{Montgomery}, M.}, \bibinfo{author}{{Richards}, M.}, \bibinfo{author}{{Schmutz}, W.}, \bibinfo{author}{{Schou}, J.}, \bibinfo{author}{{Stewart}, S.G.}, \bibinfo{year}{2016}.
\newblock \bibinfo{title}{{Nominal Values for Selected Solar and Planetary Quantities: IAU 2015 Resolution B3}}.
\newblock \bibinfo{journal}{\aj} \bibinfo{volume}{152}, \bibinfo{pages}{41}.
\newblock \DOIprefix\doi{10.3847/0004-6256/152/2/41}, \href{http://arxiv.org/abs/1605.09788}{{\tt arXiv:1605.09788}}.
\bibitem[{{R{\"a}uker} et~al.(2022){R{\"a}uker}, {Ho}, {Casper} and {Hadfield-Menell}}]{2022Rauker}
\bibinfo{author}{{R{\"a}uker}, T.}, \bibinfo{author}{{Ho}, A.}, \bibinfo{author}{{Casper}, S.}, \bibinfo{author}{{Hadfield-Menell}, D.}, \bibinfo{year}{2022}.
\newblock \bibinfo{title}{{Toward Transparent AI: A Survey on Interpreting the Inner Structures of Deep Neural Networks}}.
\newblock \bibinfo{journal}{arXiv e-prints} , \bibinfo{pages}{arXiv:2207.13243}\DOIprefix\doi{10.48550/arXiv.2207.13243}, \href{http://arxiv.org/abs/2207.13243}{{\tt arXiv:2207.13243}}.
\bibitem[{{Rodrigues} et~al.(2017){Rodrigues}, {Bossini}, {Miglio}, {Girardi}, {Montalb{\'a}n}, {Noels}, {Trabucchi}, {Coelho} and {Marigo}}]{2017Rodrigues}
\bibinfo{author}{{Rodrigues}, T.S.}, \bibinfo{author}{{Bossini}, D.}, \bibinfo{author}{{Miglio}, A.}, \bibinfo{author}{{Girardi}, L.}, \bibinfo{author}{{Montalb{\'a}n}, J.}, \bibinfo{author}{{Noels}, A.}, \bibinfo{author}{{Trabucchi}, M.}, \bibinfo{author}{{Coelho}, H.R.}, \bibinfo{author}{{Marigo}, P.}, \bibinfo{year}{2017}.
\newblock \bibinfo{title}{{Determining stellar parameters of asteroseismic targets: going beyond the use of scaling relations}}.
\newblock \bibinfo{journal}{\mnras} \bibinfo{volume}{467}, \bibinfo{pages}{1433--1448}.
\newblock \DOIprefix\doi{10.1093/mnras/stx120}, \href{http://arxiv.org/abs/1701.04791}{{\tt arXiv:1701.04791}}.
\bibitem[{Rosenblatt(1958)}]{Rosenblatt1958}
\bibinfo{author}{Rosenblatt, F.}, \bibinfo{year}{1958}.
\newblock \bibinfo{title}{The perceptron: A probabilistic model for information storage and organization in the brain}.
\newblock \bibinfo{journal}{Psychological Review} \bibinfo{volume}{65}, \bibinfo{pages}{386--408}.
\newblock \DOIprefix\doi{10.1037/h0042519}.
\bibitem[{{Serenelli} et~al.(2013){Serenelli}, {Bergemann}, {Ruchti} and {Casagrande}}]{2013Serenelli}
\bibinfo{author}{{Serenelli}, A.M.}, \bibinfo{author}{{Bergemann}, M.}, \bibinfo{author}{{Ruchti}, G.}, \bibinfo{author}{{Casagrande}, L.}, \bibinfo{year}{2013}.
\newblock \bibinfo{title}{{Bayesian analysis of ages, masses and distances to cool stars with non-LTE spectroscopic parameters}}.
\newblock \bibinfo{journal}{\mnras} \bibinfo{volume}{429}, \bibinfo{pages}{3645--3657}.
\newblock \DOIprefix\doi{10.1093/mnras/sts648}, \href{http://arxiv.org/abs/1212.4497}{{\tt arXiv:1212.4497}}.
\bibitem[{{Silva Aguirre} et~al.(2015a){Silva Aguirre}, {Davies}, {Basu}, {Christensen-Dalsgaard}, {Creevey}, {Metcalfe}, {Bedding}, {Casagrande}, {Handberg}, {Lund}, {Nissen}, {Chaplin}, {Huber}, {Serenelli}, {Stello}, {Van Eylen}, {Campante}, {Elsworth}, {Gilliland}, {Hekker}, {Karoff}, {Kawaler}, {Kjeldsen} and {Lundkvist}}]{2015Silva}
\bibinfo{author}{{Silva Aguirre}, V.}, \bibinfo{author}{{Davies}, G.R.}, \bibinfo{author}{{Basu}, S.}, \bibinfo{author}{{Christensen-Dalsgaard}, J.}, \bibinfo{author}{{Creevey}, O.}, \bibinfo{author}{{Metcalfe}, T.S.}, \bibinfo{author}{{Bedding}, T.R.}, \bibinfo{author}{{Casagrande}, L.}, \bibinfo{author}{{Handberg}, R.}, \bibinfo{author}{{Lund}, M.N.}, \bibinfo{author}{{Nissen}, P.E.}, \bibinfo{author}{{Chaplin}, W.J.}, \bibinfo{author}{{Huber}, D.}, \bibinfo{author}{{Serenelli}, A.M.}, \bibinfo{author}{{Stello}, D.}, \bibinfo{author}{{Van Eylen}, V.}, \bibinfo{author}{{Campante}, T.L.}, \bibinfo{author}{{Elsworth}, Y.}, \bibinfo{author}{{Gilliland}, R.L.}, \bibinfo{author}{{Hekker}, S.}, \bibinfo{author}{{Karoff}, C.}, \bibinfo{author}{{Kawaler}, S.D.}, \bibinfo{author}{{Kjeldsen}, H.}, \bibinfo{author}{{Lundkvist}, M.S.}, \bibinfo{year}{2015}a.
\newblock \bibinfo{title}{{Ages and fundamental properties of Kepler exoplanet host stars from asteroseismology}}.
\newblock \bibinfo{journal}{\mnras} \bibinfo{volume}{452}, \bibinfo{pages}{2127--2148}.
\newblock \DOIprefix\doi{10.1093/mnras/stv1388}, \href{http://arxiv.org/abs/1504.07992}{{\tt arXiv:1504.07992}}.
\bibitem[{{Silva Aguirre} et~al.(2015b){Silva Aguirre}, {Davies}, {Basu}, {Christensen-Dalsgaard}, {Creevey}, {Metcalfe}, {Bedding}, {Casagrande}, {Handberg}, {Lund}, {Nissen}, {Chaplin}, {Huber}, {Serenelli}, {Stello}, {Van Eylen}, {Campante}, {Elsworth}, {Gilliland}, {Hekker}, {Karoff}, {Kawaler}, {Kjeldsen} and {Lundkvist}}]{2015Silva2}
\bibinfo{author}{{Silva Aguirre}, V.}, \bibinfo{author}{{Davies}, G.R.}, \bibinfo{author}{{Basu}, S.}, \bibinfo{author}{{Christensen-Dalsgaard}, J.}, \bibinfo{author}{{Creevey}, O.}, \bibinfo{author}{{Metcalfe}, T.S.}, \bibinfo{author}{{Bedding}, T.R.}, \bibinfo{author}{{Casagrande}, L.}, \bibinfo{author}{{Handberg}, R.}, \bibinfo{author}{{Lund}, M.N.}, \bibinfo{author}{{Nissen}, P.E.}, \bibinfo{author}{{Chaplin}, W.J.}, \bibinfo{author}{{Huber}, D.}, \bibinfo{author}{{Serenelli}, A.M.}, \bibinfo{author}{{Stello}, D.}, \bibinfo{author}{{Van Eylen}, V.}, \bibinfo{author}{{Campante}, T.L.}, \bibinfo{author}{{Elsworth}, Y.}, \bibinfo{author}{{Gilliland}, R.L.}, \bibinfo{author}{{Hekker}, S.}, \bibinfo{author}{{Karoff}, C.}, \bibinfo{author}{{Kawaler}, S.D.}, \bibinfo{author}{{Kjeldsen}, H.}, \bibinfo{author}{{Lundkvist}, M.S.}, \bibinfo{year}{2015}b.
\newblock \bibinfo{title}{{Ages and fundamental properties of Kepler exoplanet host stars from asteroseismology}}.
\newblock \bibinfo{journal}{\mnras} \bibinfo{volume}{452}, \bibinfo{pages}{2127--2148}.
\newblock \DOIprefix\doi{10.1093/mnras/stv1388}, \href{http://arxiv.org/abs/1504.07992}{{\tt arXiv:1504.07992}}.
\bibitem[{{Silver} et~al.(2017){Silver}, {Hubert}, {Schrittwieser}, {Antonoglou}, {Lai}, {Guez}, {Lanctot}, {Sifre}, {Kumaran}, {Graepel}, {Lillicrap}, {Simonyan} and {Hassabis}}]{Silver17}
\bibinfo{author}{{Silver}, D.}, \bibinfo{author}{{Hubert}, T.}, \bibinfo{author}{{Schrittwieser}, J.}, \bibinfo{author}{{Antonoglou}, I.}, \bibinfo{author}{{Lai}, M.}, \bibinfo{author}{{Guez}, A.}, \bibinfo{author}{{Lanctot}, M.}, \bibinfo{author}{{Sifre}, L.}, \bibinfo{author}{{Kumaran}, D.}, \bibinfo{author}{{Graepel}, T.}, \bibinfo{author}{{Lillicrap}, T.}, \bibinfo{author}{{Simonyan}, K.}, \bibinfo{author}{{Hassabis}, D.}, \bibinfo{year}{2017}.
\newblock \bibinfo{title}{{Mastering Chess and Shogi by Self-Play with a General Reinforcement Learning Algorithm}}.
\newblock \bibinfo{journal}{arXiv e-prints} , \bibinfo{pages}{arXiv:1712.01815}\DOIprefix\doi{10.48550/arXiv.1712.01815}, \href{http://arxiv.org/abs/1712.01815}{{\tt arXiv:1712.01815}}.
\bibitem[{Sneden(1973)}]{sneden1973university}
\bibinfo{author}{Sneden, C.}, \bibinfo{year}{1973}.
\newblock \bibinfo{title}{The University of Texas at Austin}.
\newblock Ph.D. thesis. Ph. D. thesis.
\bibitem[{{Soderblom}(2015a)}]{2015Soderblom}
\bibinfo{author}{{Soderblom}, D.}, \bibinfo{year}{2015}a.
\newblock \bibinfo{title}{{Astrophysics: Stellar clocks}}.
\newblock \bibinfo{journal}{\nat} \bibinfo{volume}{517}, \bibinfo{pages}{557--558}.
\newblock \DOIprefix\doi{10.1038/517557a}.
\bibitem[{{Soderblom}(2010)}]{2010Soderblom}
\bibinfo{author}{{Soderblom}, D.R.}, \bibinfo{year}{2010}.
\newblock \bibinfo{title}{{The Ages of Stars}}.
\newblock \bibinfo{journal}{\araa} \bibinfo{volume}{48}, \bibinfo{pages}{581--629}.
\newblock \DOIprefix\doi{10.1146/annurev-astro-081309-130806}, \href{http://arxiv.org/abs/1003.6074}{{\tt arXiv:1003.6074}}.
\bibitem[{{Soderblom}(2015b)}]{2015Soderblom2}
\bibinfo{author}{{Soderblom}, D.R.}, \bibinfo{year}{2015}b.
\newblock \bibinfo{title}{{Ages of Stars: Methods and Uncertainties}}, in: \bibinfo{booktitle}{Asteroseismology of Stellar Populations in the Milky Way}, p.~\bibinfo{pages}{3}.
\newblock \DOIprefix\doi{10.1007/978-3-319-10993-0_1}, \href{http://arxiv.org/abs/1409.2266}{{\tt arXiv:1409.2266}}.
\bibitem[{{Sousa} et~al.(2007){Sousa}, {Santos}, {Israelian}, {Mayor} and {Monteiro}}]{2007Sousa}
\bibinfo{author}{{Sousa}, S.G.}, \bibinfo{author}{{Santos}, N.C.}, \bibinfo{author}{{Israelian}, G.}, \bibinfo{author}{{Mayor}, M.}, \bibinfo{author}{{Monteiro}, M.J.P.F.G.}, \bibinfo{year}{2007}.
\newblock \bibinfo{title}{{A new code for automatic determination of equivalent widths: Automatic Routine for line Equivalent widths in stellar Spectra (ARES)}}.
\newblock \bibinfo{journal}{\aap} \bibinfo{volume}{469}, \bibinfo{pages}{783--791}.
\newblock \DOIprefix\doi{10.1051/0004-6361:20077288}, \href{http://arxiv.org/abs/astro-ph/0703696}{{\tt arXiv:astro-ph/0703696}}.
\bibitem[{{Sousa} et~al.(2008){Sousa}, {Santos}, {Mayor}, {Udry}, {Casagrande}, {Israelian}, {Pepe}, {Queloz} and {Monteiro}}]{2008Sousa}
\bibinfo{author}{{Sousa}, S.G.}, \bibinfo{author}{{Santos}, N.C.}, \bibinfo{author}{{Mayor}, M.}, \bibinfo{author}{{Udry}, S.}, \bibinfo{author}{{Casagrande}, L.}, \bibinfo{author}{{Israelian}, G.}, \bibinfo{author}{{Pepe}, F.}, \bibinfo{author}{{Queloz}, D.}, \bibinfo{author}{{Monteiro}, M.J.P.F.G.}, \bibinfo{year}{2008}.
\newblock \bibinfo{title}{{Spectroscopic parameters for 451 stars in the HARPS GTO planet search program. Stellar [Fe/H] and the frequency of exo-Neptunes}}.
\newblock \bibinfo{journal}{\aap} \bibinfo{volume}{487}, \bibinfo{pages}{373--381}.
\newblock \DOIprefix\doi{10.1051/0004-6361:200809698}, \href{http://arxiv.org/abs/0805.4826}{{\tt arXiv:0805.4826}}.
\bibitem[{Tishby et~al.(1989)Tishby, Levin and Solla}]{Tishby1989}
\bibinfo{author}{Tishby, N.}, \bibinfo{author}{Levin, E.}, \bibinfo{author}{Solla, S.A.}, \bibinfo{year}{1989}.
\newblock \bibinfo{title}{Consistent inference of probabilities in layered networks: predictions and generalizations}.
\newblock \bibinfo{journal}{International 1989 Joint Conference on Neural Networks} , \bibinfo{pages}{403--409 vol.2}\URLprefix \url{https://api.semanticscholar.org/CorpusID:15012839}.
\bibitem[{{Valentin Jospin} et~al.(2020){Valentin Jospin}, {Buntine}, {Boussaid}, {Laga} and {Bennamoun}}]{2020Jospin}
\bibinfo{author}{{Valentin Jospin}, L.}, \bibinfo{author}{{Buntine}, W.}, \bibinfo{author}{{Boussaid}, F.}, \bibinfo{author}{{Laga}, H.}, \bibinfo{author}{{Bennamoun}, M.}, \bibinfo{year}{2020}.
\newblock \bibinfo{title}{{Hands-on Bayesian Neural Networks -- a Tutorial for Deep Learning Users}}.
\newblock \bibinfo{journal}{arXiv e-prints} , \bibinfo{pages}{arXiv:2007.06823}\DOIprefix\doi{10.48550/arXiv.2007.06823}, \href{http://arxiv.org/abs/2007.06823}{{\tt arXiv:2007.06823}}.
\bibitem[{{Valls-Gabaud}(2014)}]{2014Valls}
\bibinfo{author}{{Valls-Gabaud}, D.}, \bibinfo{year}{2014}.
\newblock \bibinfo{title}{{Bayesian isochrone fitting and stellar ages}}, in: \bibinfo{editor}{{Lebreton}, Y.}, \bibinfo{editor}{{Valls-Gabaud}, D.}, \bibinfo{editor}{{Charbonnel}, C.} (Eds.), \bibinfo{booktitle}{EAS Publications Series}, pp. \bibinfo{pages}{225--265}.
\newblock \DOIprefix\doi{10.1051/eas/1465006}, \href{http://arxiv.org/abs/1607.03000}{{\tt arXiv:1607.03000}}.
\bibitem[{{Van-Lane} et~al.(2023){Van-Lane}, {Speagle} and {Douglas}}]{2023Lane}
\bibinfo{author}{{Van-Lane}, P.}, \bibinfo{author}{{Speagle}, J.S.}, \bibinfo{author}{{Douglas}, S.}, \bibinfo{year}{2023}.
\newblock \bibinfo{title}{{A Novel Application of Conditional Normalizing Flows: Stellar Age Inference with Gyrochronology}}.
\newblock \bibinfo{journal}{arXiv e-prints} , \bibinfo{pages}{arXiv:2307.08753}\DOIprefix\doi{10.48550/arXiv.2307.08753}, \href{http://arxiv.org/abs/2307.08753}{{\tt arXiv:2307.08753}}.
\bibitem[{{Viscasillas V{\'a}zquez} et~al.(2022){Viscasillas V{\'a}zquez}, {Magrini}, {Casali}, {Tautvai{\v{s}}ien{\.{e}}}, {Spina}, {Van der Swaelmen}, {Randich}, {Bensby}, {Bragaglia}, {Friel}, {Feltzing}, {Sacco}, {Turchi}, {Jim{\'e}nez-Esteban}, {D'Orazi}, {Delgado-Mena}, {Mikolaitis}, {Drazdauskas}, {Minkevi{\v{c}}i{\={u}}t{\.{e}}}, {Stonkut{\.{e}}}, {Bagdonas}, {Montes}, {Guiglion}, {Baratella}, {Tabernero}, {Gilmore}, {Alfaro}, {Francois}, {Korn}, {Smiljanic}, {Bergemann}, {Franciosini}, {Gonneau}, {Hourihane}, {Worley} and {Zaggia}}]{2022Viscasillas}
\bibinfo{author}{{Viscasillas V{\'a}zquez}, C.}, \bibinfo{author}{{Magrini}, L.}, \bibinfo{author}{{Casali}, G.}, \bibinfo{author}{{Tautvai{\v{s}}ien{\.{e}}}, G.}, \bibinfo{author}{{Spina}, L.}, \bibinfo{author}{{Van der Swaelmen}, M.}, \bibinfo{author}{{Randich}, S.}, \bibinfo{author}{{Bensby}, T.}, \bibinfo{author}{{Bragaglia}, A.}, \bibinfo{author}{{Friel}, E.}, \bibinfo{author}{{Feltzing}, S.}, \bibinfo{author}{{Sacco}, G.G.}, \bibinfo{author}{{Turchi}, A.}, \bibinfo{author}{{Jim{\'e}nez-Esteban}, F.}, \bibinfo{author}{{D'Orazi}, V.}, \bibinfo{author}{{Delgado-Mena}, E.}, \bibinfo{author}{{Mikolaitis}, {\v{S}}.}, \bibinfo{author}{{Drazdauskas}, A.}, \bibinfo{author}{{Minkevi{\v{c}}i{\={u}}t{\.{e}}}, R.}, \bibinfo{author}{{Stonkut{\.{e}}}, E.}, \bibinfo{author}{{Bagdonas}, V.}, \bibinfo{author}{{Montes}, D.}, \bibinfo{author}{{Guiglion}, G.}, \bibinfo{author}{{Baratella}, M.}, \bibinfo{author}{{Tabernero}, H.M.}, \bibinfo{author}{{Gilmore}, G.}, \bibinfo{author}{{Alfaro}, E.}, \bibinfo{author}{{Francois},
  P.}, \bibinfo{author}{{Korn}, A.}, \bibinfo{author}{{Smiljanic}, R.}, \bibinfo{author}{{Bergemann}, M.}, \bibinfo{author}{{Franciosini}, E.}, \bibinfo{author}{{Gonneau}, A.}, \bibinfo{author}{{Hourihane}, A.}, \bibinfo{author}{{Worley}, C.C.}, \bibinfo{author}{{Zaggia}, S.}, \bibinfo{year}{2022}.
\newblock \bibinfo{title}{{The Gaia-ESO survey: Age-chemical-clock relations spatially resolved in the Galactic disc}}.
\newblock \bibinfo{journal}{\aap} \bibinfo{volume}{660}, \bibinfo{pages}{A135}.
\newblock \DOIprefix\doi{10.1051/0004-6361/202142937}, \href{http://arxiv.org/abs/2202.04863}{{\tt arXiv:2202.04863}}.
\bibitem[{{von Hippel} et~al.(2016){von Hippel}, {van Dyk}, {Stenning}, {Robinson}, {Jeffery}, {Stein}, {Jefferys} and {O'Malley}}]{2016Hippel}
\bibinfo{author}{{von Hippel}, T.}, \bibinfo{author}{{van Dyk}, D.A.}, \bibinfo{author}{{Stenning}, D.C.}, \bibinfo{author}{{Robinson}, E.}, \bibinfo{author}{{Jeffery}, E.}, \bibinfo{author}{{Stein}, N.}, \bibinfo{author}{{Jefferys}, W.H.}, \bibinfo{author}{{O'Malley}, E.}, \bibinfo{year}{2016}.
\newblock \bibinfo{title}{{The Power of Principled Bayesian Methods in the Study of Stellar Evolution}}.
\newblock \bibinfo{journal}{arXiv e-prints} , \bibinfo{pages}{arXiv:1605.02810}\DOIprefix\doi{10.48550/arXiv.1605.02810}, \href{http://arxiv.org/abs/1605.02810}{{\tt arXiv:1605.02810}}.
\bibitem[{Werbos(1974)}]{phdthesis}
\bibinfo{author}{Werbos, P.}, \bibinfo{year}{1974}.
\newblock \bibinfo{title}{Beyond Regression: New Tools for Prediction and Analysis in the Behavioral Science. Thesis (Ph. D.).}
\newblock Ph.D. thesis. Appl. Math. Harvard University.
\bibitem[{Wiese et~al.(2023)Wiese, Wimmer, Papamarkou, Bischl, Günnemann and Rügamer}]{wiese2023efficient}
\bibinfo{author}{Wiese, J.G.}, \bibinfo{author}{Wimmer, L.}, \bibinfo{author}{Papamarkou, T.}, \bibinfo{author}{Bischl, B.}, \bibinfo{author}{Günnemann, S.}, \bibinfo{author}{Rügamer, D.}, \bibinfo{year}{2023}.
\newblock \bibinfo{title}{Towards efficient mcmc sampling in bayesian neural networks by exploiting symmetry}.
\newblock \href{http://arxiv.org/abs/2304.02902}{{\tt arXiv:2304.02902}}.
\bibitem[{{Wilson} and {Izmailov}(2020)}]{2020Wilson}
\bibinfo{author}{{Wilson}, A.G.}, \bibinfo{author}{{Izmailov}, P.}, \bibinfo{year}{2020}.
\newblock \bibinfo{title}{{Bayesian Deep Learning and a Probabilistic Perspective of Generalization}}.
\newblock \bibinfo{journal}{arXiv e-prints} , \bibinfo{pages}{arXiv:2002.08791}\DOIprefix\doi{10.48550/arXiv.2002.08791}, \href{http://arxiv.org/abs/2002.08791}{{\tt arXiv:2002.08791}}.
\bibitem[{Xuan et~al.(2019)Xuan, Lu and Zhang}]{Xuan19}
\bibinfo{author}{Xuan, J.}, \bibinfo{author}{Lu, J.}, \bibinfo{author}{Zhang, G.}, \bibinfo{year}{2019}.
\newblock \bibinfo{title}{A survey on bayesian nonparametric learning}.
\newblock \bibinfo{journal}{ACM Comput. Surv.} \bibinfo{volume}{52}.
\newblock \URLprefix \url{https://doi.org/10.1145/3291044}, \DOIprefix\doi{10.1145/3291044}.
\bibitem[{{Yadav} et~al.(2008){Yadav}, {Bedin}, {Piotto}, {Anderson}, {Cassisi}, {Villanova}, {Platais}, {Pasquini}, {Momany} and {Sagar}}]{2008Yadav}
\bibinfo{author}{{Yadav}, R.K.S.}, \bibinfo{author}{{Bedin}, L.R.}, \bibinfo{author}{{Piotto}, G.}, \bibinfo{author}{{Anderson}, J.}, \bibinfo{author}{{Cassisi}, S.}, \bibinfo{author}{{Villanova}, S.}, \bibinfo{author}{{Platais}, I.}, \bibinfo{author}{{Pasquini}, L.}, \bibinfo{author}{{Momany}, Y.}, \bibinfo{author}{{Sagar}, R.}, \bibinfo{year}{2008}.
\newblock \bibinfo{title}{{Ground-based CCD astrometry with wide-field imagers. II. A star catalog for M 67: WFI@2.2 m MPG/ESO astrometry, FLAMES@VLT radial velocities}}.
\newblock \bibinfo{journal}{\aap} \bibinfo{volume}{484}, \bibinfo{pages}{609--620}.
\newblock \DOIprefix\doi{10.1051/0004-6361:20079245}, \href{http://arxiv.org/abs/0803.0004}{{\tt arXiv:0803.0004}}.
\bibitem[{{Zhao} et~al.(2022){Zhao}, {Saxena} and {Cao}}]{2022Zhao2}
\bibinfo{author}{{Zhao}, Y.}, \bibinfo{author}{{Saxena}, D.}, \bibinfo{author}{{Cao}, J.}, \bibinfo{year}{2022}.
\newblock \bibinfo{title}{{AdaptCL: Adaptive Continual Learning for Tackling Heterogeneity in Sequential Datasets}}.
\newblock \bibinfo{journal}{arXiv e-prints} , \bibinfo{pages}{arXiv:2207.11005}\DOIprefix\doi{10.48550/arXiv.2207.11005}, \href{http://arxiv.org/abs/2207.11005}{{\tt arXiv:2207.11005}}.
\bibitem[{Zhao et~al.(2022)Zhao, Saxena and Cao}]{2022Zhao}
\bibinfo{author}{Zhao, Y.}, \bibinfo{author}{Saxena, D.}, \bibinfo{author}{Cao, J.}, \bibinfo{year}{2022}.
\newblock \bibinfo{title}{Revisiting parameter reuse to overcome catastrophic forgetting in neural networks}.
\newblock \bibinfo{journal}{arXiv e-prints} .

\end{thebibliography}
\bibliographystyle{elsarticle-harv}

\clearpage

\appendix

\section{Appendix Title}\label{appendixx}

\subsection{Single BNN with multiple output} \label{multiout}

This subsection focuses on the performance and behavior of a single BNN with multiple output dimensions. By leveraging the inherent capacity of BNNs to model uncertainty, we analyze the outcomes derived from our experiments. The results highlight key observations regarding prediction accuracy, uncertainty quantification, and the network’s ability to handle multi-output scenarios effectively.

\begin{figure*}[ht]
	\centering 
	\includegraphics[width=0.9\textwidth]{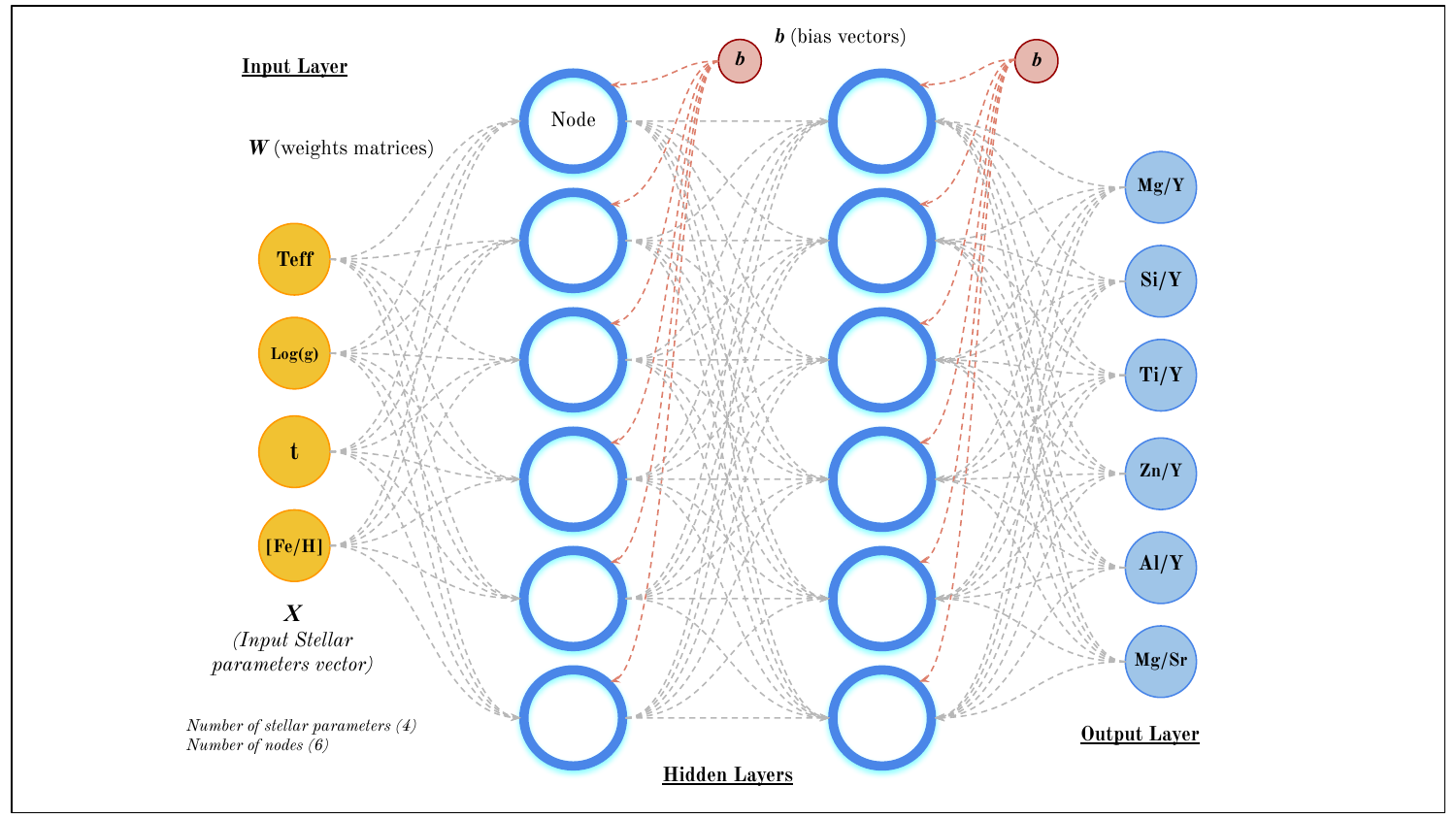}	
	\caption{An example of a multiple-output architecture with two hidden layers, matching the best BNN schematics used for multiple-output configurations during experimentation (single NN for all 6 CCs). \(\textit{\textbf{W}}\) and \(\textit{\textbf{b}}\) represent the weight matrices and bias vectors, \textit{Orange filled circles} are the stellar parameters, ($X$) as input, and \textit{Blue filled circle} are the CCs output. The \textit{Dashed gray and Red lines} illustrate the information flow within the NN (representing the \(\textit{\textbf{W}}\) matrices and \textit{$b$} vectors).} 
	\label{2HLBNN}%
\end{figure*}

The initial version of our HBM-NNs featured a single neural network that outputs six values, each corresponding to an individual CC, Figure \ref{2HLBNN}). This picture depicts the best schematic in multiple output configurations for our problem. Parameters \(\textit{\textbf{W}}\) and \(\textit{\textbf{b}}\) represent the weights matrices and bias vectors, respectively. In Figure \ref{2HLBNN} the dashed gray and red lines depict the information flow within the NN. Additionally, orange-filled circles are the stellar parameters (\textit{X}) as input, and the blue-filled circles are the CC outputs. This design choice aimed to streamline the architecture and reduce the number of parameters entering the Bayesian model.

Increasing the number of HLs and nodes in BNNs enhances their capacity to model intricate and complex patterns in data. However, this increased expressiveness comes at the cost of significantly heightened complexity in the parameter space. A higher number of nodes introduces additional symmetries in how the weights can be arranged, as multiple configurations of weights may represent latent features in the data equally well. This symmetry further compounds the challenge, as the posterior distribution may encompass multiple plausible explanations of the data. Consequently, new modes emerge in the parameter space, contributing to multimodality in the posterior distribution.

The proliferation of modes leads to a parameter space characterized by numerous local optima, complicating the inference process. Such landscapes make posterior exploration more difficult for traditional optimization and sampling methods, as they can become trapped in suboptimal regions. This increased complexity and multimodality result in greater uncertainty in predictions, making them more variable and less stable. 

The inherently multimodal and high-dimensional nature of the parameter space in BNNs \citep[e.g.,][]{1998Muller, bishop2006, 2021Izmailov, arbel2023primer} poses significant challenges for sampling from their posterior distributions. This complexity arises from the interplay of the non-linearities in neural network architectures and the prior assumptions, often resulting in posterior distributions with multiple isolated modes. While MCMC methods are commonly employed for posterior inference, these algorithms can struggle to adequately explore the parameter space. Specifically, MCMC samplers may become trapped in a single mode, failing to traverse energy barriers to explore other regions of the posterior. This limitation can lead to incomplete mixing, slow convergence, and posterior estimates that are biased or unrepresentative of the true distribution. Such inadequacies can compromise the reliability of downstream analyses, producing heuristics or predictions that fail to capture the full uncertainty encoded in the Bayesian framework. Addressing these challenges requires either developing advanced sampling methods tailored to the unique characteristics of BNN posteriors or leveraging variational inference techniques as an alternative.

\begin{figure*}
    \centering
    \rotatebox{90}{\includegraphics[width=15.5cm,height=12cm]{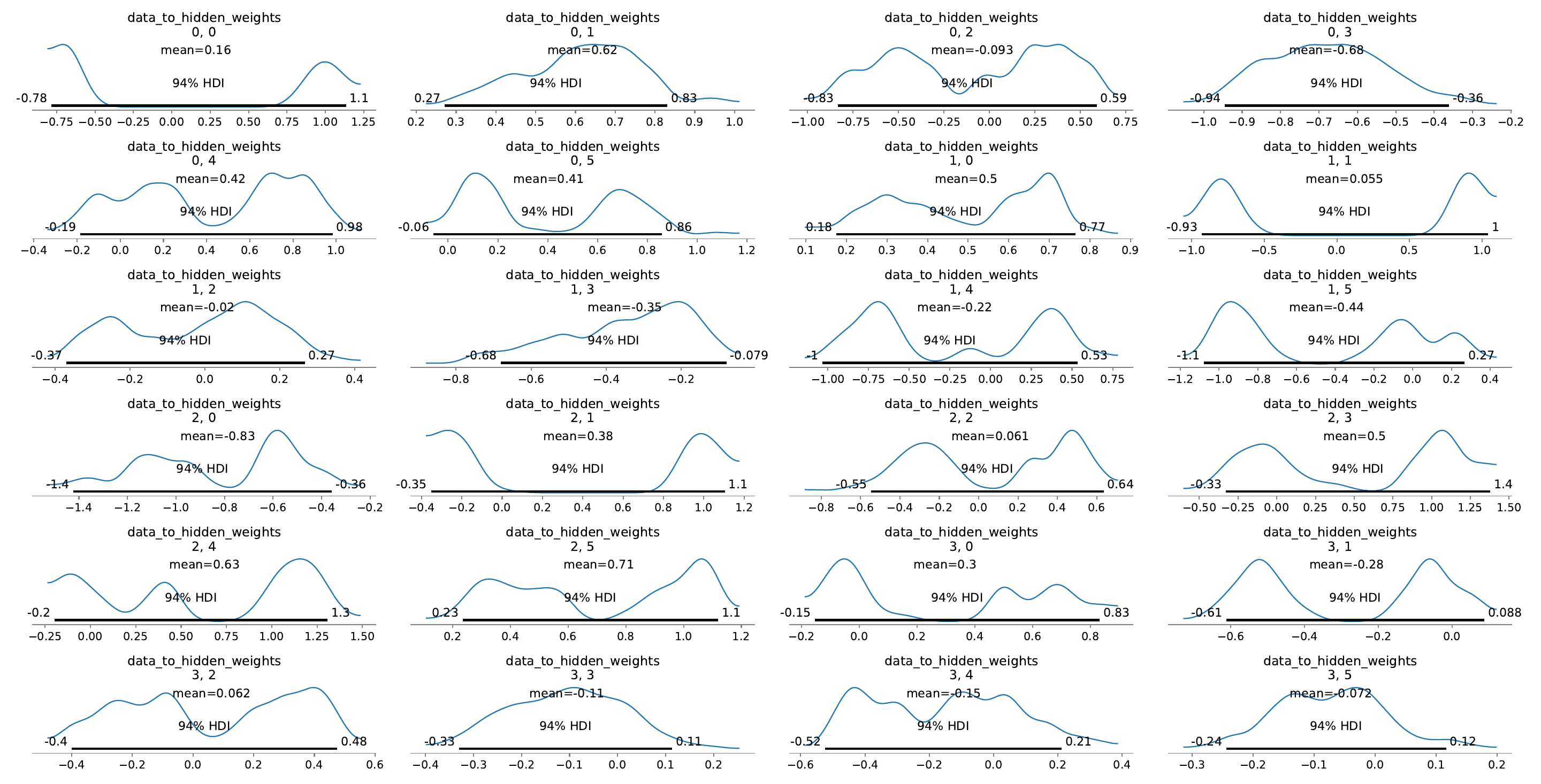}}
    \caption{Example of posterior distributions for a weight matrix in a two hidden layer, six nodes per hidden layer neural network applied to a model with six outputs, 6 CCs prediction at once (for better illustration, see figure \ref{2HLBNN}). The figure shows the distributions of each weight (\textit{$W_{ji}$}) in the initial weight matrix, which represents the connections between the input data and the first hidden layer. Each distribution includes its mean and the 94\% Highest Density Interval (HDI). The distributions are organized from left to right and top to bottom, corresponding to each of the four stellar parameters and their connections to the six nodes in the first hidden layer.}
    \label{fig:rotated_figure}
\end{figure*}

This issue represented a crucial point during our experimentation. Figures \ref{fig:rotated_figure} and \ref{multibest} show an example of the posterior distributions of weights matrices in BNNs used in this work, each illustrating different parameter spaces influenced by the complexity of the network. Both figures depict the posteriors of each element in the first weight matrix, which represents the connections between the input vector and the first HL nodes. Considering the reasons outlined during this work, we aim to have parameter spaces as monomodal as possible in order to have the information concentrated, improving computational times and uncertainty. Even though our first iteration presented complex parameter spaces with multimodality, due to the deeper architecture (Figure \ref{2HLBNN} ), we first opted for these approaches to avoid a higher number of parameters in our Bayesian model.

Initially, to address this multimodality challenge, we implemented various strategies, principally involving the imposition of narrower and centered priors, with a focus on the primary mode of the distributions for the weights and biases. To achieve this, we trained an identical NN architecture as the one in the Bayesian model multiple times, aiming to attain a configuration of \textbf{\textit{W}} and \textbf{\textit{b}} with lower loss. Subsequently, these configurations were introduced as homonyms priors in PyStan.

\begin{figure}
	\centering 
	\includegraphics[width=0.65\textwidth]{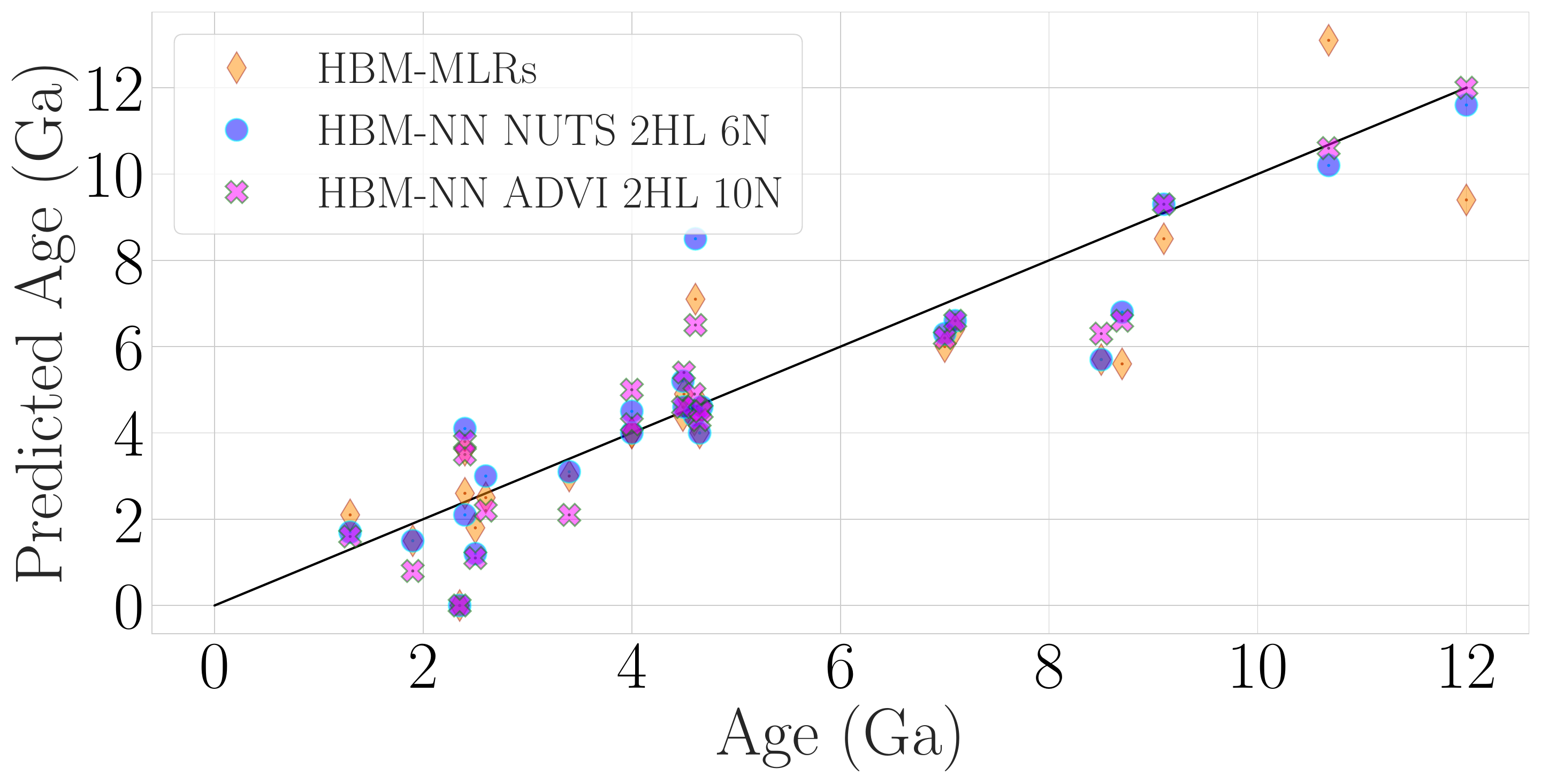}	
	\includegraphics[width=0.65\textwidth]{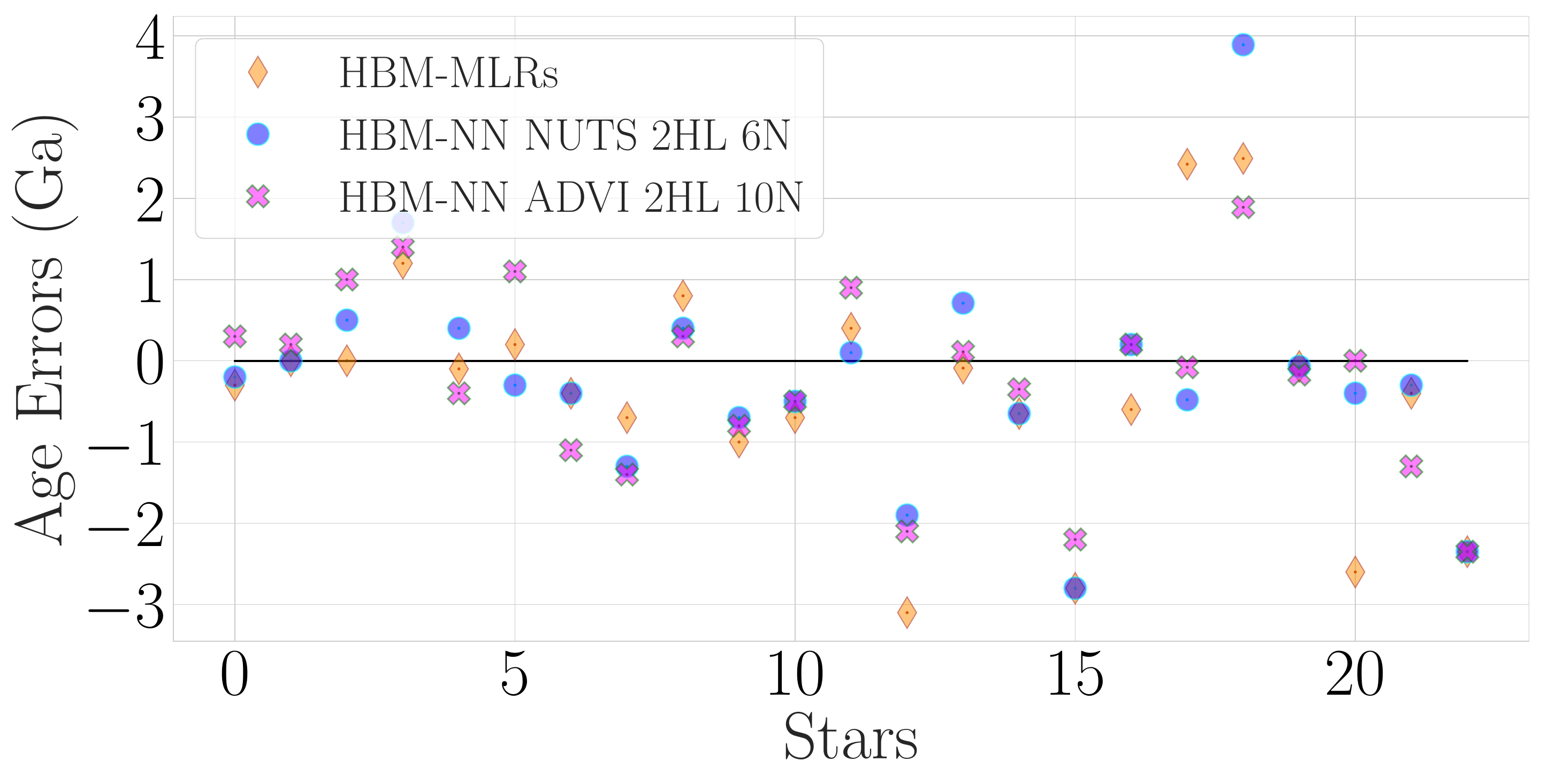}	
	\caption{Hierarchical Bayesian model results comparison. \textit{(HBM-MLRs:}) architecture whose probability relationships are modeled by multi-linear regressions, from \cite{Moya2022}. (\textit{HBM-NN NUTS 2HL 6N:}) architecture whose probability relationships are modeled by one NN. The architecture consists of 2 Hidden Layers with 6 nodes per layer and multiple outputs (6 CCs), sampled with the MCMC-NUTS algorithm. (\textit{HBM-NN ADVI 2HL 10N:}) architecture whose probability relationships are modeled by one NN. The architecture consists of 2 hidden layers with 10 nodes per layer and multiple outputs (6 CCs), sampled with the VI-ADVI algorithm.(\textit{Top}:) Age prediction for the 23 test stars. (\textit{Bottom}:) Age errors for the 23 test stars.}
	\label{fig_mom02s}%
\end{figure}

We determined that the optimal balance between prediction accuracy and computational efficiency was achieved with a setup comprising two hidden layers, each consisting of 6 nodes (Figure \ref{2HLBNN}) for the MCMC-NUTS sampler and 10 for the VI-ADVI-based one.  Leveraging the expressive capacity of NNs, we successfully captured intricate relationships among stellar parameters, yielding results comparable to \cite{Moya2022} (for the same 23 test set stars they achieve a MAE=0.86) while benefitting from automated learning. In Figure \ref{fig_mom02s} we present a comparison of the results obtained by these two configurations. Both architectures achieve similar results, a MAE=0.88 for the NUTS case and 0.87 for the ADVI one.

While this approach yielded reliable results and heuristics, the constrained nature of the priors has the potential to hinder the sampling of other modes, particularly posing risks when encountering distant or disconnected modes in this or other datasets. Moreover, this restrictive approach has a potential adverse impact on computational times.

To achieve more reproducible and stable results we decided to start new experimentation, this time focusing on monomodal distributions that let us concentrate the information. This was the main reason we built a new architecture based on 6 simpler NNs, one for each CC (Figure \ref{fig_mom0}). Illustrating this, Figure \ref{multibest} shows a parameter space with very little multimodality. The posterior distribution in this case is much less complex, suggesting a more deterministic relationship between inputs and output. The near-unimodal landscape indicates a concentrated posterior distribution, meaning that there are fewer alternative hypotheses about the best weight configurations. This simplicity reduces the uncertainty in parameter estimation, leading to a potentially more reproducible solution.

\end{document}